\documentclass[aps,nofootinbib,showpacs,showkeys,11pt]{revtex4-1}
\pdfoutput=1

\usepackage{epsf,epsfig,subfigure,graphicx,amsmath,amssymb,mathtools}
\usepackage{color}
\usepackage{float}
\usepackage{multirow}
\usepackage{nicefrac}
\usepackage{cancel}
\usepackage{verbatim}
\usepackage{setspace}

\newcommand{\dis}[1]{\begin{equation}\begin{split}#1\end{split}\end{equation}}

\newcommand{\cc}[1]{\mathcal{#1}}
\newcommand{\rr}[1]{\mathrm{#1}}

\newcommand{\kk}[1]{\mathfrak{#1}}

\newcommand{\be}{\begin{eqnarray}}
\newcommand{\nn}{\nonumber}
\newcommand{\ee}{\end{eqnarray}}
\newcommand{\ba}{\begin{array}}
\newcommand{\ea}{\end{array}}

\newcommand{\f}[2]{\frac{#1}{#2}}
\newcommand{\nf}[2]{\nicefrac{#1}{#2}}

\newcommand{\N}{\mathcal{\cal N}}

\usepackage{xcolor}

\begin{document}

\title{\large
A New Avenue to Charged Higgs Discovery in Multi-Higgs Models
}

\author{
Radovan Derm\'i\v{s}ek$^1$\footnote{dermisek@indiana.edu},
Jonathan P. Hall$^1$\footnote{halljp@indiana.edu},
Enrico Lunghi$^1$\footnote{elunghi@indiana.edu},
and Seodong Shin$^{1,2}$\footnote{shinseod@indiana.edu}
}
\affiliation{
$^1$Physics Department, Indiana University, Bloomington, IN 47405, USA \\
$^2$CTP and Department of Physics and Astronomy, Seoul National University, Seoul 151-747, Korea \\
}

\begin{abstract}

Current searches for the charged Higgs at the LHC focus only on the $\tau\nu$, $cs$, and $tb$ final states. Instead, we consider the process $pp\to \Phi\to W^\pm H^\mp \to W^+ W^- A$ where $\Phi$ is a heavy neutral Higgs boson, $H^\pm$ is a charged Higgs boson, and $A$ is a light Higgs boson, with mass either below or above the $b\bar{b}$ threshold. The cross-section for this process is typically large when kinematically open since $H^\pm \to W^\pm A$ can be the dominant decay mode of the charged Higgs. The final state we consider has two leptons and missing energy from the doubly leptonic decay of the $W^+ W^-$ and possibly additional jets; it is therefore constrained by existing SM Higgs searches in the $W^+ W^-$ channel. We extract these constraints on the cross-section for this process as a function of the masses of the particles involved. We also apply our results specifically to a type-II two Higgs doublet model with an extra Standard-Model-singlet and obtain new and powerful constraints on $m_{H^\pm}$ and $\tan\beta$. We point out that a slightly modified version of this search, with more dedicated cuts, could be used to possibly discover the charged Higgs, either with existing data or in the future. 

\end{abstract}

\maketitle

%%%%%%%%%%%%%%%%%%%%%%%%%%%%%%%%%%%%%%%%%%%%%%%%%%%%%%%%%%%%%%%%%%%%%%%%%%

\section{Introduction}
\label{sec:intro}
The quest to unveil the mechanism responsible for the breaking of the electroweak symmetry made a huge leap forward with the recent discovery of a scalar particle whose quantum numbers and interactions appear to be compatible, albeit with large uncertainties, with those of the Standard Model (SM) Higgs boson \cite{higgs}. The presence of a fundamental scalar particle renders electroweak physics sensitive to arbitrarily large scales possibly present in a full theory of electroweak, strong, and gravitational interactions. Solutions to this problem usually entail the introduction of new physics just above the electroweak scale. Amongst others, hints that point to the incomplete nature of the SM are the strong empirical evidence for particle dark matter, the baryon-antibaryon asymmetry of the universe, and the pattern of neutrino masses and mixing. Even before addressing these problems it is important to realize that while the structure of currently observed gauge interactions is completely dictated by the SM gauge groups alone the pattern of electroweak symmetry breaking is not. In particular, within the context of a perturbative (Higgs) mechanism there are absolutely no ``symmetry'' reasons for introducing a single doublet (besides the empirical observation that such a choice leads directly to the rather successful Cabibbo-Kobayashi-Maskawa pattern of flavor changing and $CP$ violation). Moreover, it is well known that supersymmetry, one of the most popular extensions of the SM that actually addresses some of the above mentioned problems, requires the introduction of a second Higgs doublet. In view of these observations it is clear that understanding how many fundamental scalars are involved in the electroweak spontaneous symmetry  breaking mechanism is one of the most pressing questions we currently face. In particular, any model with at least two doublets contain at least two charged Higgs boson ($H^\pm$) and at least two extra neutral Higgses. In this paper we investigate a previously overlooked technique that could uncover a charged Higgs from a multi-Higgs scenario.

Direct charged Higgs production in the top-bottom fusion channel typically has cross-sections $\cc{O}(1~{\rm pb})$~\cite{Dittmaier:2009np} and discovery would be fairly difficult in this channel~\cite{Assamagan:2004tt, Lowette:2006bs}. If the charged Higgs mass is lower than the top mass, it is possible to bypass this problem by looking for charged Higgs bosons in top decays ($t\to H^+ b$), taking advantage of the very large $t\bar t$ production cross-section. Moreover, most current experimental studies consider only charged Higgs decays to pairs of fermions ($H^+ \to \tau^+\nu$, $H^+ \to c\bar s$, and $H^+ \to t\bar b$). Under these assumptions ATLAS and CMS were able to place bounds on ${\rm BR} (t\to H^+ b)$ at the 1--5~\% level~\cite{Aad:2012tj,Chatrchyan:2012vca,Aad:2013hla,Flechl:2013wza} for $m_{H^\pm} < m_t$~\footnote{A preliminary result of ATLAS reduces this to $\mathcal{O}(0.1\%)$ \cite{atlaspreliminary}.}. It is well known that the presence of a light neutral Higgs can significantly modify these conclusions. In fact, the $H^+ \to W^+ A$ decay ($A$ being a neutral $CP$-even or -odd Higgs boson) can easily dominate the charged Higgs decay width if it is kinematically allowed and the $A$ has non-vanishing mixing with one of the neutral components of a Higgs doublet. Such a light neutral pseudoscalar Higgs ($A=a_1$) has been looked for by BaBar~\cite{Aubert:2009cp,Aubert:2009cka} in $\Upsilon \to a_1 \gamma \to (\tau\tau,\mu\mu) \gamma$ decays and by ATLAS~\cite{Atlas_lightA} and CMS~\cite{Chatrchyan:2012am} in $pp \to a_1 \to \mu\mu$ direct production. These bounds are easily evaded by assuming that the lightest neutral Higgs $a_1$ has a singlet component. Under this condition, in the context of a type-II two Higgs doublet model (2HDM) with an additional singlet, the ${\rm BR}(t \to b H^+)$ can be as large as $\cc{O}(10~\%)$ for $\tan \beta < 6$ ($\tan\beta$ being the ratio of the vacuum expectation values of the neutral components of the two Higgs doublets) even for $a_1$ as light as 8~GeV~\cite{Dermisek:2012cn}. Trilepton events in $t\bar t$ production can be used to discover at the LHC a charged Higgs produced in top decays and decaying to $W^\pm A$ with as little as 20~${\rm fb}^{-1}$ integrated luminosity at 8 TeV center of mass energy. 

At the LHC the charged Higgs can be alternatively produced in the decay of a heavier neutral Higgs ($\Phi$). Heavy neutral Higgs bosons are dominantly produced in gluon-gluon fusion ($gg$F) with a significant cross-section, leading to sizable charged Higgs production rates. For our somewhat model independent analysis, we ignore possible mass relations amongst the various Higgs bosons as they depend on the exact Lagrangian of the model. In the presence of a light Higgs $A$ the decay $H^+ \to W^+ A$ is mostly dominant for $m_{H^+} < m_t$ and remains comparable to $H^+ \to t\bar b$ otherwise, depending on the values of the various parameters. Note that the $H^+ \to W^+ h_1$ decay (we take $h_1$ be the particle recently discovered at the LHC) vanishes in the limit that $h_1$ is completely SM-like.

\begin{figure}
\includegraphics[width=0.49\linewidth]{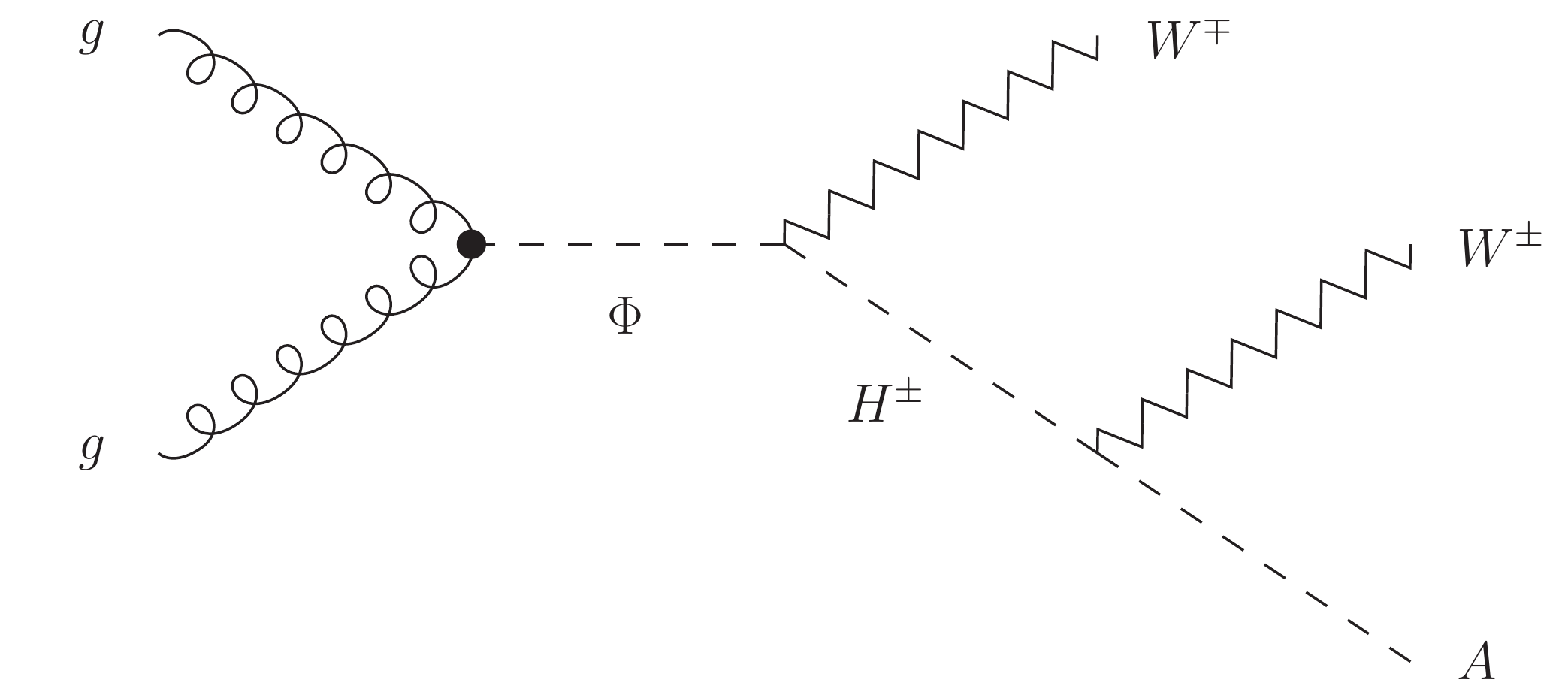}
\includegraphics[width=0.49\linewidth]{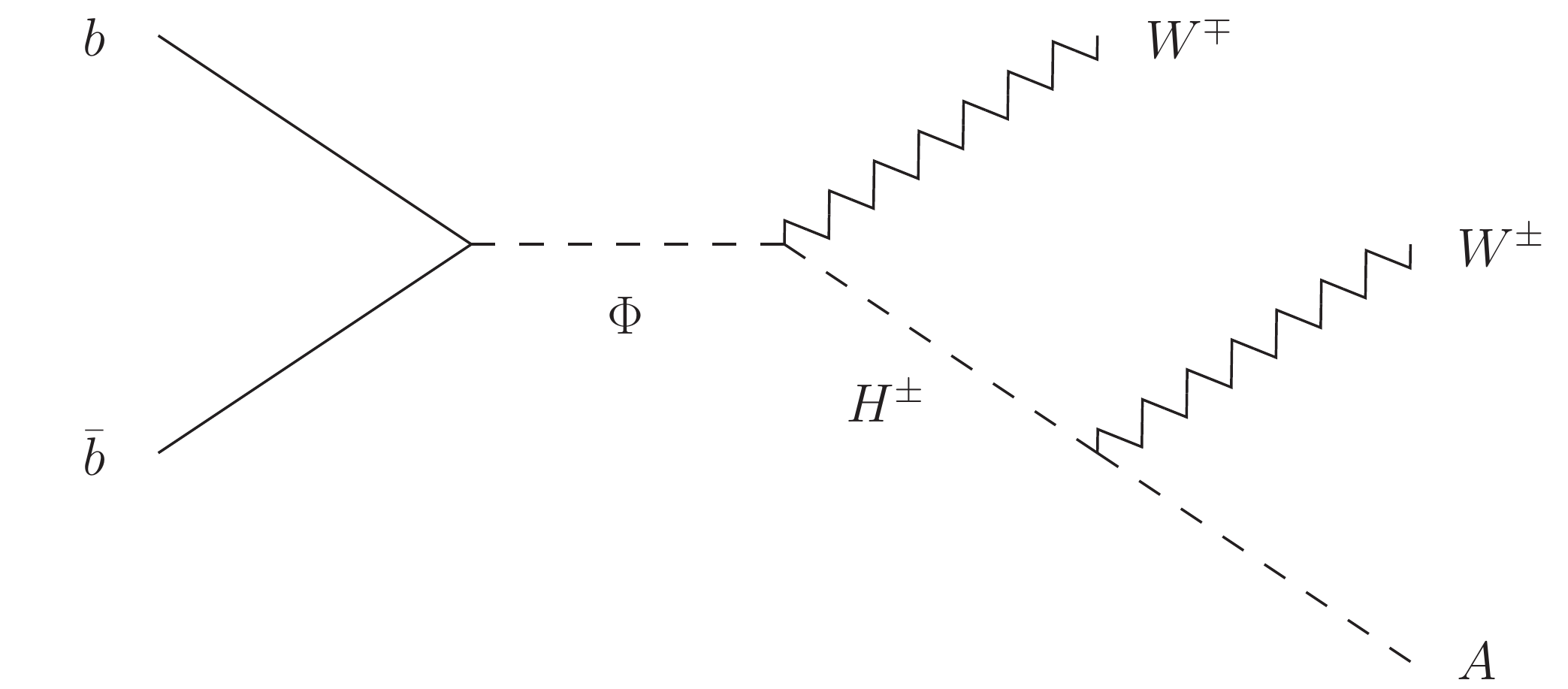}
\vskip3mm
\includegraphics[width=0.49\linewidth]{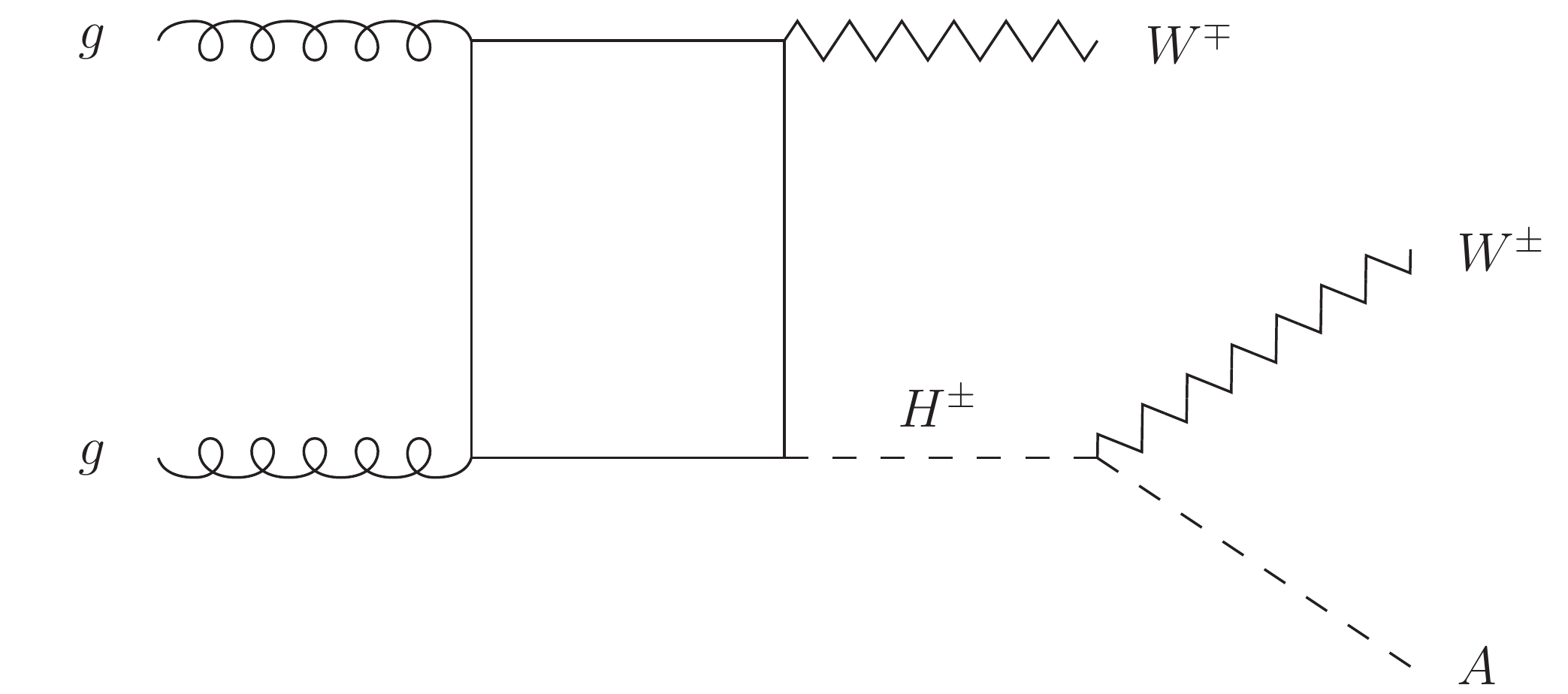}
\includegraphics[width=0.49\linewidth]{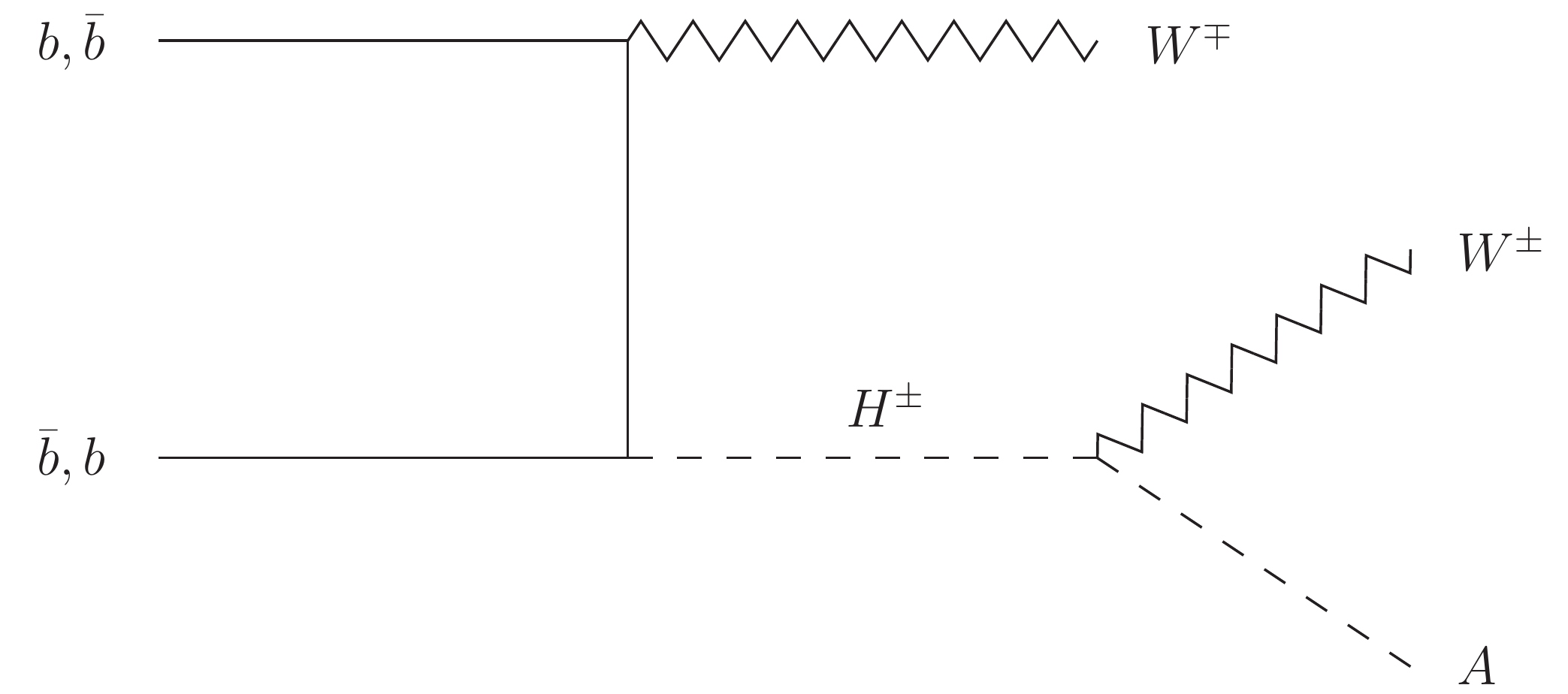}

\caption{
Feynman diagrams for $pp \to W^+ W^- A$. If an intermediate on-shell $\Phi$ is present, the upper diagrams dominate the cross-section. The bottom fusion diagram (upper right) is only sizable at large $\tan\beta$.
\label{fig:process}}
\end{figure}

In this study we consider the process $pp \to \Phi \to H^\pm W^\mp \to W^+ W^- A$ as shown in Fig.~\ref{fig:process}. The constraints we derive are valid for $m_A$ not too far above the $b\bar{b}$ threshold, where the decay $A\to b\bar{b}$ should be dominant (they are also approximately valid below this threshold, as discussed in Sec.~\ref{sec:lim}). At large transverse momentum of the $b\bar b$ pair (transverse momentum relevant for the event selection), the angular separation of the two bottom quarks is small and they are combined into a single jet~
\footnote{
The ATLAS collaboration recently announced the results of a search for a similar process, where the light state $A$ is identified with the 125~GeV $CP$-even Higgs, dominantly decaying into two separable $b$-jets~\cite{Aad:2013dza}. 
They consider the semileptonic decay of the $WW$. This was based on the suggestion put forward in Ref.~\cite{Evans:2012nd}. See also Ref.~\cite{hwsemi} which includes the non-resonant production of $H^\pm W^\mp$.}.
The final state we consider is, therefore, constrained by the standard $h\to WW$ searches by CMS~\cite{cmshww} (with 19.5~fb$^{-1}$ at 8~TeV and 4.9~fb$^{-1}$ at 7~TeV) and ATLAS~\cite{ATLAShWW} (with 20.7~fb$^{-1}$ at 8~TeV and 4.6~fb$^{-1}$ at 7~TeV). We use the data provided in the CMS analysis to place bounds.

The impact of the experimental cuts depends on the kinematics and is controlled by the masses of the three intermediate Higgs bosons only. We therefore derive constraints on the LHC cross-section for the considered process that depend only on the masses of the relevant particles and not on other model-dependent parameters or the $CP$ nature of the neutral Higgs bosons $\Phi$ and $A$. We also apply our results to a $CP$ conserving type-II 2HDM with an additional singlet~\cite{Dermisek:2005ar, Dermisek:2005gg, Dermisek:2007yt, Chang:2008cw}. In this framework the lightest neutral Higgs ($A$) is identified with the lightest $CP$-odd eigenstate $a_1$ and the heavy Higgs ($\Phi$) with the heavy $CP$-odd Higgs $a_2$. To the extent that the $a_2 \to H^+ W^-$ decay dominates over other decays involving Higgs bosons (and this can easily be the case) and decays to other beyond-the-Standard-Model particles our bounds depend only on $m_{a_2}$, $m_{H^\pm}$, $m_{a_1}$, $\tan(\beta)$, and $\vartheta_A$ (the mixing angle in the $CP$-odd sector). A novelty in our analysis is the exclusion of parameter space regions at low $\tan\beta$. The 8 TeV LHC data analyzed so far allow one, using our approach, to probe only a relatively light charged Higgs (roughly below the $tb$ threshold); in the future, regions in parameter space with a heavy charged Higgs will be accessible as well.
We also consider the same scenario but with one of the $CP$-even states ($h_2$) as the heavy neutral state $\Phi$. The types of scenario we consider and constrain can easily be consistent with constraints on the custodial symmetry breaking parameter $\rho = M_W^2 / (M_Z^2 \cos^2 \vartheta_W)$.
%-----------------------------------
%----------------------

The paper is organized as follows. In Sec.~\ref{sec:crossbr} we discuss the production and decay cross-section for our signal. In particular, after introducing the type-II 2HDM + singlet scenario in Sec.~\ref{sec:def} we discuss charged ($H^\pm$) and neutral ($\Phi$) Higgs decays in Secs.~\ref{sec:charged} and \ref{sec:heavy}, the $gg \to \Phi$ production cross-section in Sec.~\ref{sec:production}, and the total cross-section (production times branching ratios) in Sec.~\ref{sec:total}. In Sec.~\ref{sec:lim} we show the upper bound on the total cross-section that we extract from SM Higgs to $WW$ searches. In Sec.~\ref{sec:param} we specialize the previous results to our reference scenario (type-II 2HDM with an additional singlet, $\Phi = a_2$ and $A=a_1$) and present the new exclusion bounds at low $\tan\beta$ that we extract. Finally, in Sec.~\ref{sec:conclusions}, we present our conclusions.

%%%%%%%%%%%%%%%%%%%%%%%%%%%%%%%%%%%%%%%%%%%%%%%%%%%%%%%%%%%%%%%%%%%%%%%%%%
\section{Charged Higgs production and decay}
\label{sec:crossbr}

In the multi-Higgs models containing at least two SU(2) doublets, there can exist a heavy neutral Higgs ($\Phi$) which decays into $H^{\pm} W^{\mp}$. The process is shown in Fig.~\ref{fig:process} with the charged Higgs decaying to a light neutral Higgs $A$ and another $W$ boson. Looking for this process could be the first way the charged Higgs is discovered and its properties measured. This is due to the large value of $\sigma(gg \to \Phi \to W^\mp H^\pm \to W^\mp W^\pm A$) when all particles can be on-shell. In this section, we focus on showing how large such a production cross-section times the branching ratios can be, especially in the context of the type-II 2HDM + singlet scenario. In the following subsections, we show that the branching ratios of $H^\pm \to W^\pm A$ and $\Phi \to H^\pm W^\mp$ can be sizable when kinematics allow and the production cross-section of $\Phi$ is roughly as large as that of the SM Higgs. Our general cross-section constraints depend only on the masses of the particles involved and will be discussed in the next section. For the specific type-II 2HDM + singlet reference scenario we can constrain physical parameters (the masses; $\tan\beta$; and $\vartheta_A$, the mixing angle in the $CP$-odd sector) without specifying the Lagrangian in the Higgs sector and we assume no mass relations among the Higgs bosons states.

%%%%%%%%%%%%%%%%%%%%%%%%%%%%%%%%%%%%%%%%%%%%%%%%%%%%%%%%%%%%
\subsection{Our example reference scenario: the type-II two Higgs doublet model with an additional SM singlet}
\label{sec:def}

Considering the type-II 2HDM with one extra complex singlet scalar
we define the field-space basis by
\be
\left(\ba{c}h\\H\\N\ea\right) &=& \left(\ba{ccc}
\cos(\beta) & \sin(\beta) & 0 \\
-\sin(\beta) & \cos(\beta) & 0 \\
0 & 0 & 1\ea\right)\left(\ba{c}
\sqrt{2}\kk{Re}H_d^0 - v_d\\
\sqrt{2}\kk{Re}H_u^0 - v_u\\
\sqrt{2}\kk{Re}{S} -s\ea\right),\nn\\\\
A_H &=&
\sqrt{2}\left(\cos(\beta)\kk{Im}H_u^0 - \sin(\beta)\kk{Im}H_d^0\right),\nn\\
A_N &=& \sqrt{2}\kk{Im}{S},\nn
\ee
where $S$ is the SM-singlet and $s$ is its possibly non-zero VEV and $\tan\beta = v_u / v_d$.
In this convention, $h$ interacts exactly as a SM Higgs in both gauge and Yukawa interactions; $H$ has no coupling to the gauge boson pairs and interacts with the up-type quarks (down-type quarks and charged leptons) with couplings multiplied by $\cot\beta$ ($\tan\beta$) relative to the SM Higgs couplings.
The orthogonal state to $A_H$ and $A_N$ is the $Z$-boson Goldstone mode.

We define an orthogonal matrix $\cc{U}$ that transforms the $CP$-even field-space basis states into the $CP$-even mass eigenstates
\be
\left(\ba{c}h_1\\h_2\\h_3\ea\right) &=& 
\left(\ba{ccc}
\cc{U}_{1h}&\cc{U}_{1H}&\cc{U}_{1N}\\
\cc{U}_{2h}&\cc{U}_{2H}&\cc{U}_{2N}\\
\cc{U}_{3h}&\cc{U}_{3H}&\cc{U}_{3N}\ea\right)
\left(\ba{c}h\\H\\N\ea\right)
\ee
We define $h_1$ to be the particle recently discovered at the LHC and do not demand that $h_i$ are ordered by mass. The overlap of $h_1$ with the SM-like state $h$ appears to be large. The mass eigenstates $h_2$ and $h_3$ are then approximately superpositions of $H$ and $N$ only. When we consider $h_2$ to be the heavy state produced from $pp$ collisions $\cc{U}_{2H}$, the overlap of $h_2$ and $H$, becomes an important parameter.

We define a mixing angle between the $CP$-odd mass
eigenstates $\vartheta_A$ by
\be
\left(\ba{c}a_1\\a_2\ea\right) &=& \left(\ba{cc}
\cos(\vartheta_A) & \sin(\vartheta_A)\\
-\sin(\vartheta_A) & \cos(\vartheta_A)\ea\right)\left(\ba{c}
A_H\\A_N\ea\right),\label{eq:eigen}
\ee
where $a_1$ is defined to be the lighter state.

The state $a_1$ is identified with $A$ in our process $pp\to \Phi \to W^\mp H^\pm \to W^\mp W^\pm A$. We mainly consider $\Phi$ to be the other $CP$-odd state $a_2$ but also consider the case where it is one of the $CP$-even states, defined to be $h_2$.

When the mass of $a_1$ is below the $b\bar b$ threshold the constraints from the decay $\Upsilon \to a_1 \gamma \to (\tau\tau,\mu\mu)\gamma$ at BaBar and the light scalar search at the LHC ($pp\to a_1\to\mu\mu$) lead to an upper bound on $\cos \vartheta_A \tan\beta$ of about 0.5~\cite{Atlas_lightA,Chatrchyan:2012am,Dermisek:2010mg}. We concentrate on two benchmark $a_1$ masses: 8 and 15~GeV. Our results depend weakly on this mass; therefore, the 8~GeV threshold is representative of masses just below and just above the $b\bar b$ threshold, where the constraint $\cos \vartheta_A \tan\beta\lesssim 0.5$ does and does not apply respectively.

In the parameter region where one of the $CP$-even Higgses $h_{2,3}$ is lighter than 150 GeV, the direct search bounds for light neutral Higgses in associated production $h_i a_1$ ($i=2,3$) at LEP-II can be considered~\cite{lep2}.
The final states can be, for example, $4b$ or $2b2\tau$.
However, even for $h_{2,3}$ light enough for this associated production to be possible, the cross-section is proportional to the doublet component of $a_1$ and is usually small in our scenario. The upper bounds in~\cite{lep2} constrain $\cos^2(\vartheta_A)\,\cc{U}_{iH}^2$ ($i=2,3$) times branching ratios as a function of the masses, but this can easily be small enough to be consistent with the bounds. We therefore ignore the LEP-II constraint throughout this paper.

\begin{figure}
\includegraphics[width=0.6\linewidth]{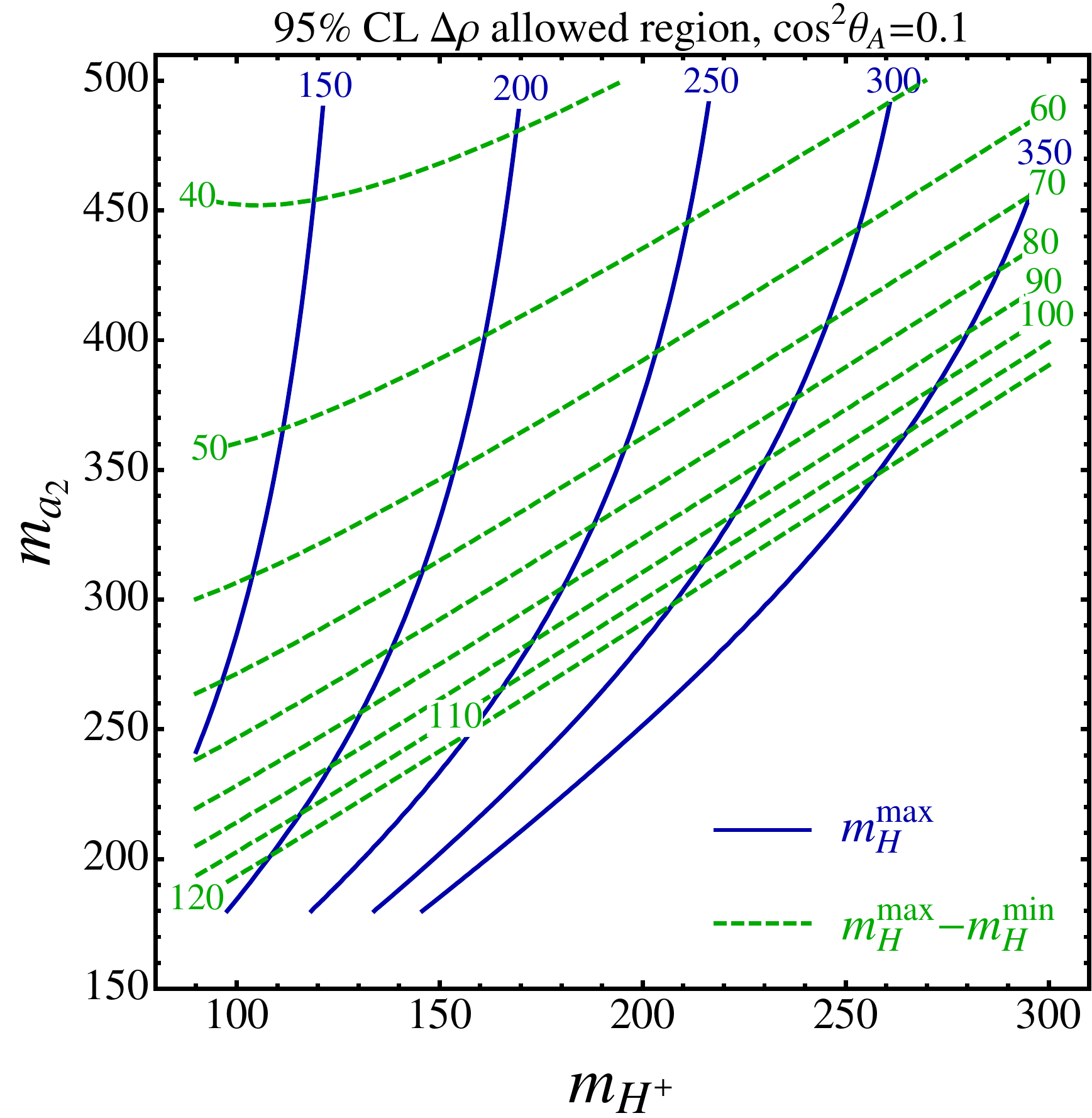}
\caption{Maximum value (solid blue) and range (dashed green) of the extra neutral Higgs ($H$) mass required to satisfy at 95\% C.L. the $\Delta \rho$ constraint \cite{pdg}. This is for the type-II 2HDM + singlet model discussed in the text.}\label{fig:deltarho}
\end{figure}

The masses of the extra neutral and charged Higgs bosons can affect the custodial symmetry breaking parameter $\rho = M_W^2 / (M_Z^2 \cos^2 \vartheta_W)$, where $\vartheta_W$ is the weak mixing angle. Since we are considering extensions of the Higgs sector involving only $SU(2)$ doublets and singlets, contributions to $\Delta \rho \equiv \rho -1$ appear only at loop level. In our type-II 2HDM + singlet reference scenario with $\Phi=a_2$ (mostly doublet), $A=a_1$ (mostly singlet) and the SM--like Higgs boson discovered at the LHC identified with $h_1$, $\Delta\rho$ depends also on the two remaining $CP$-even states $h_{2,3}$. For a simple demonstration of the $\Delta \rho$ constraint, we assume that one of these two states is completely doublet (the field-space basis state $H$ defined above). Then the main contributions to the vacuum polarization of the $W^\pm$ by $H^\pm - a_2$ and $H^\pm - H$ loops need to be cancelled by that of the $Z$ by $H - a_2$ loop. Therefore one can roughly expect the contribution due to the mass difference between $H^\pm$ and $a_2$ can be cancelled by that between $H$ and $a_2$, while making that of the $H^\pm - H$ loop to the $W$ boson small. (In the 2HDM, complete contributions to the oblique parameters are well depicted in the Appendix D of the reference ~\cite{Haber:2010bw}.) In Fig.~\ref{fig:deltarho}, we show the $H$ mass range allowed at 95~\% C.L. by the present determination of $\Delta \rho$~\cite{pdg} for given masses of $a_2$ and $H^\pm$. The solid (blue) contours give the maximum value of $m_H$ required to satisfy the experimental $\Delta \rho$ constraint; the dashed (green) contours show the difference between the maximum and minimum $m_H$ required and are therefore a measure of the (low) fine tuning between $m_H$ and $m_{H^\pm}$ that we require. We find that in the parameter space where our process is dominant ($m_\Phi-m_{H^\pm}\gtrsim M_W$) the contributions to $\Delta\rho$ can easily be compensated by the contributions of other Higgs states, although the fine tuning between $m_{H^+}$ and $m_{H}$ increases as $m_{a_2}$ does. It is quite possible for the $H$ state to remain unconstrained by LHC Higgs searches. Based on this result, we simply ignore the $\Delta \rho$ constraint throughout this paper. We also ignore possible mass relations amongst the various Higgs bosons which depend on the exact details of the Higgs sector Lagrangian.
%-----------------------------------
%----------------------

%%%%%%%%%%%%%%%%%%%%%%%%%%%%%%%%%%%%%%%%%%%%%%%%%%%%%%%%%%%%
\subsection{Charged Higgs decays}
\label{sec:charged}

When the charged Higgs is lighter than the top quark (light charged Higgs), investigating only the usual $\tau\nu$ or $cs$ final states from its decay may not be enough for discovery. This is because the process $H^+ \to W^+ A$, whose decay rate is proportional to $m_{H^+}^3$, can easily dominate over the $\tau^+ \nu$ and $c\bar{s}$ final states. The detailed analysis of the light charged Higgs from the top quark decay in the context of the type-II 2HDM + singlet is shown in Ref.~\cite{Dermisek:2012cn}, where the lightest $CP$ odd neutral Higgs $a_1$ is the particle $A$. The main factors determining the BR($H^+ \to W^+ a_1$) are the SU(2) doublet fraction (at the amplitude level) in $a_1$ ($\cos\vartheta_A$) and $\tan\beta$. According to that analysis, BR($H^+ \to W^+ a_1$) rapidly approaches unity for $m_{H^+} > M_W + m_{a_1}$ even when the light Higgs $a_1$ is highly singlet-like, as long as $\tan\beta$ is small.

%-----------------------
\begin{figure}
\raisebox{-\height}{\includegraphics[width=0.49\linewidth]{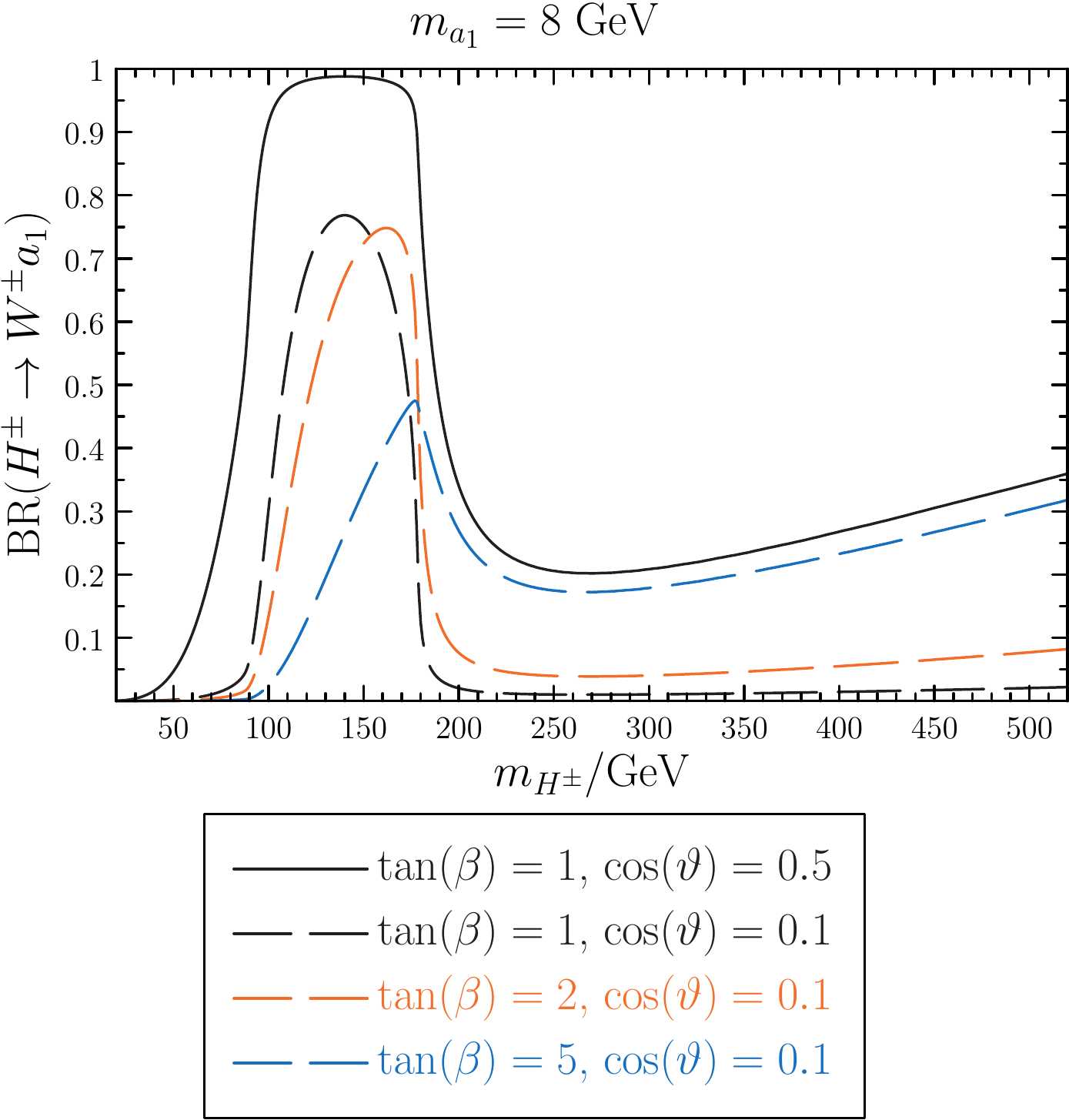}}
\raisebox{-\height}{\includegraphics[width=0.49\linewidth]{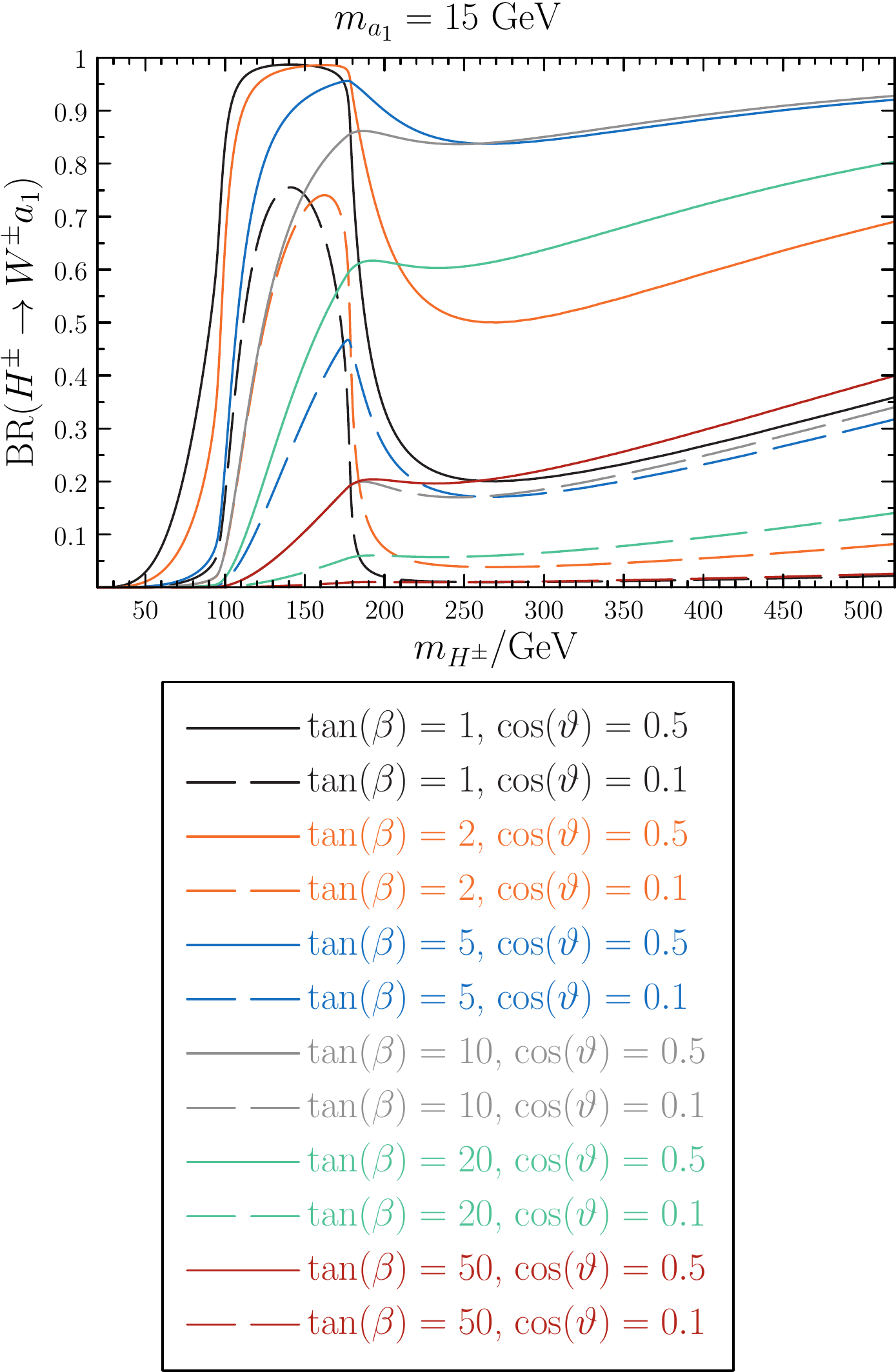}}
\caption{The BR($H^\pm \to W^\pm a_1$). Off-shell tops and $W$s are included. Above threshold the ratio of the $H^+\to W^+ a_1$ and $H^+ \to t\bar{b}$ decay rates is proportional to $\cos^2\vartheta_A \tan^2\beta m_{H^\pm}^2$ for small $\tan\beta$ ($\lesssim 5$). For $m_{a_1} = 8$~GeV, the constraint from the decay $\Upsilon \to a_1 \gamma$ at BABAR and the light scalar search at CMS lead to an upper bound on $\cos \vartheta_A \tan\beta$ of about 0.5~\cite{Atlas_lightA,Chatrchyan:2012am,Dermisek:2010mg}. In this region the black solid line therefore represents the maximum possible branching ratio. For $m_{a_1} = 15$~GeV these bounds do not apply. For $\tan\beta \gg 7$, the $\tan\beta$ dependence of BR($H^+ \to W^+ a_1$) is reversed since the $(m_b / v)^2 \tan^2\beta$ term in $\Gamma(H^+ \to t\bar{b})$ is dominant.
\label{fig:brh}}
\end{figure}
%-------------------

For a charged Higgs heavier than the top quark (heavy charged Higgs), the channel $H^+ \to t\bar{b}$ opens to compete with the process $H^+ \to W^+ A$. In the context of the type-II 2HDM + singlet, we show the dependence of the BR($H^+ \to W^+ a_1$) on $\cos\vartheta_A$ and $\tan\beta$ in Fig. \ref{fig:brh}.
For low $\tan\beta \lesssim 5$, the value of $\Gamma(H^+ \to t\bar{b})$ is dominantly determined by the $(m_t / v)^2 \cot^2\beta$ term, so the BR($H^+ \to W^+ a_1$) increases for larger $\tan\beta$. (See App.~\ref{appen:width} for the detailed formulae.) Above threshold the ratio of the $H^+ \to W^+ a_1$ and $H^+ \to t\bar{b}$ decay rates is proportional to $\cos^2\vartheta_A \tan^2\beta m^2_{H^+}$. For $m_{a_1}=8$~GeV the constraint $\cos \vartheta_A \tan\beta\lesssim 0.5$ applies and hence BR($H^+ \to W^+ a_1$) is at most around 30~\% for $m_{H^+} < 400$~GeV, increasing for larger charged Higgs masses. On the other hand, we do not need to consider this bound when $a_1$ is heavier than about 9~GeV, so in this case the BR($H^+ \to W^+ a_1$) can be larger than 0.5, corresponding to larger values of $\cos\vartheta_A \tan\beta$ when $m_{a_1}$ is set to 15~GeV in Fig.~\ref{fig:brh}.
Far above thresholds and at low $\tan\beta$ we have
\dis{
\frac{\Gamma(H^+\rightarrow W^+ a_1)}{\Gamma(H^+\rightarrow t\bar{b})}
&\rightarrow
\frac{m_{H^\pm}^2 \tan^2\beta\cos^2(\vartheta_A)}{6m_t^2}.}
Consequently, BR($H^\pm \to W^\pm a_1$) can be still larger than 0.5 even after the on-shell $H^+ \to t\bar{b}$ decay opens, as long as $a_1$ is heavier than about 9~GeV.
For large values of $\tan\beta$ ($\gg 7$) the $\tan\beta$ dependence of BR($H^+ \to W^+ a_1$) is reversed since the $(m_b / v)^2 \tan^2\beta$ term in $\Gamma(H^+ \to t\bar{b})$ is dominant.

%%%%%%%%%%%%%%%%%%%%%%%%%%%%%%%%%%%%
\subsection{Heavy neutral Higgs decays}
\label{sec:heavy}

%-------------------------------------------------------------------------------
\begin{figure}
\includegraphics[width=0.8\linewidth]{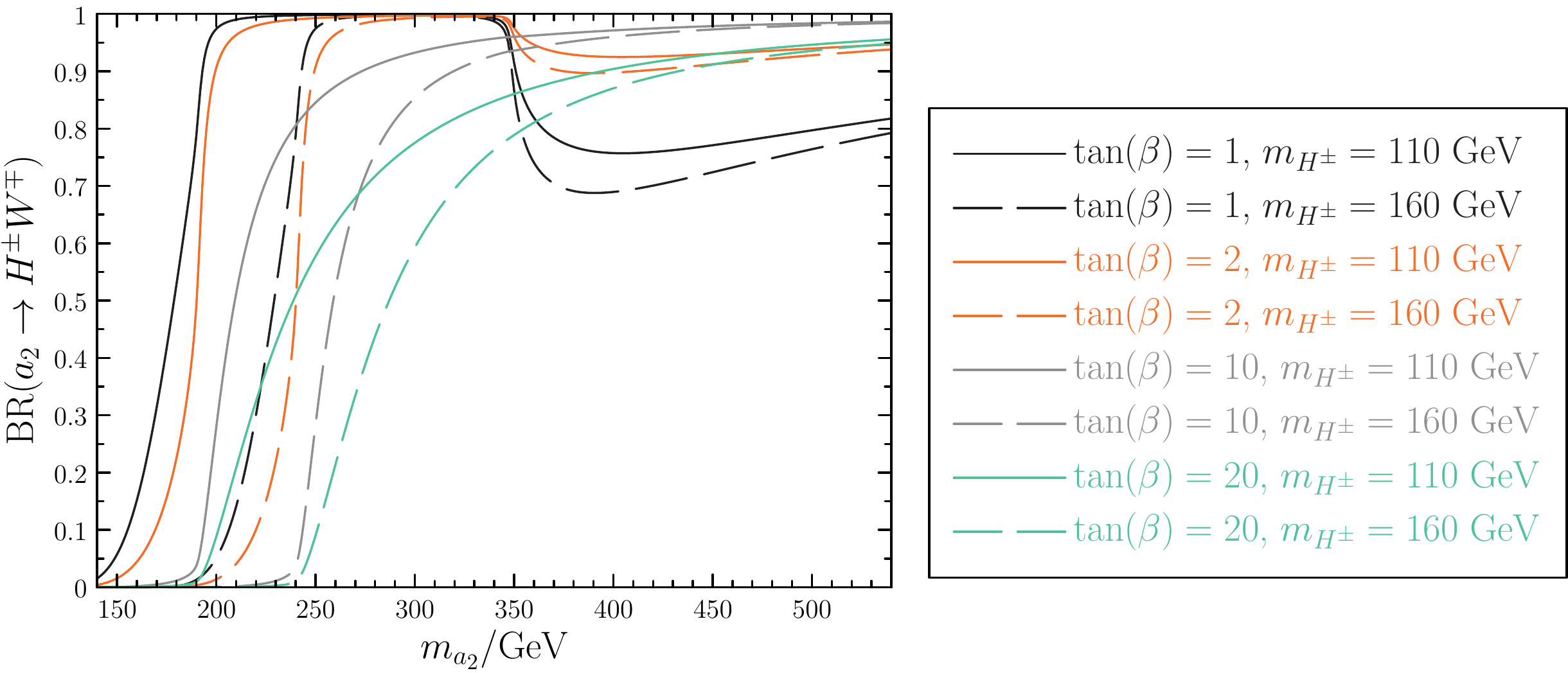}
\caption{The BR($a_2 \to H^\pm W^\mp$). Other than this channel we only consider
decays to fermions, including off-shell tops. In this case all decays take place via the doublet ($A_H$) component of $a_2$ and the $\sin^2\vartheta_A$ dependence cancels out of all branching ratios.
\label{fig:bra2}}
\end{figure}
%---------------------------------------------------------------------------

The $\Phi \rightarrow W^\pm H^\mp$ decay can easily dominate over decays into SM fermions, including top quarks. In the type II 2HDM + singlet scenario we set $A = a_1$ and $\Phi = a_2$ (which is the heavy $CP$-odd Higgs); then Fig.~\ref{fig:bra2} shows how BR($a_2 \rightarrow H^\pm W^\mp$) varies with $\tan\beta$ and the various masses. For small $\tan\beta$, the branching ratio is affected by the partial width $a_2 \to t\bar{t}$, whose rate depends on $\cot^2\beta$. Since we only consider this decay and decays into SM fermions, all taking place via the doublet ($A_H$) component of $a_2$, the $\sin^2(\vartheta_A)$ dependence cancels out of all of the branching ratios of $a_2$. For our reference type-II 2HDM + singlet scenario we assume the possible decays $a_2 \to h_i Z$ and $a_2 \to h_i a_1$ to be subdominant compared to $a_2\to H^\pm W^\mp$, where $h_i$ is a $CP$-even neutral Higgs. This is in order to reduce the number of parameters relevant for determining cross-section times branching ratios in this reference scenario (to be compared to the general bounds on this cross-section times branching ratios that we derive). The processes $a_2 \to h_i a_1$ are model dependent even within the type II 2HDM + singlet scenario. As for the possible decay modes $a_2 \to h_i Z$: the more SM-like the 125~GeV particle discovered at the LHC ($h_1$) is (the more $h_1\sim h$, see Sec.~\ref{sec:def}), the more suppressed the decay to $h_1 Z$ will be. On the other hand the other final states $h_{i>1}Z$ can reduce the relevant BR($a_2\to H^\pm W^\mp$) by up to about 1/3, if we consider that the $\Delta \rho$ constraint requires very approximate mass degeneracy of $H^\pm$ and any state significantly overlapping with $H$ (The width to $ZH$ is equal to the width to $H^+ W^-$ if one ignores the phase-space factor). The results we present (e.g.~the new bound in the $(\tan\beta, m_{H^\pm})$ plane for $m_{a_2} \sim 2 m_t$) are not much affected by the presence of this decay mode and we will neglect it altogether in the following. Far above thresholds we have
\dis{
\frac{\Gamma \left(a_2 \rightarrow H^+ W^- \right) + \Gamma \left(a_2 \rightarrow H^- W^+ \right)}{\Gamma \left( a_2 \rightarrow t\bar{t} \right)} \rightarrow \frac{m_{a_2}^2 \tan^2\beta}{3m_t^2}.
}

Finally let us comment on the possibility of taking $\Phi$ to be the $CP$-even state $h_2$. As can be seen from the results collected in App.~\ref{appen:width}, the decay rates are very similar to those for a $CP$-odd Higgs ($\Phi = a_2$ case). For this case we similarly neglect the two-body decays to $Z a_i$, $a_ia_j$, and $h_1 h_1$. In this case too, the mixing-matrix-element-squared $\mathcal{U}^2_{2H}$ (see Sec.~\ref{sec:def}) dependence cancels out of all branching ratios and appears only in the production cross-section.

%%%%%%%%%%%%%%%%%%%%%%%%%%%%%%%%%%%
\subsection{Heavy neutral Higgs production}
\label{sec:production}

The dominant production mechanism for $h_{\rm SM}$ at the LHC is $gg$F mediated by quark loops, mainly dominated by the top quark loop due to its large Yukawa coupling. The production cross-section of $\Phi$ depends on its modified couplings to up- and down-type quarks. The $A_H$ and $H$ interaction states, defined in Sec.~\ref{sec:def}, have couplings to up-type quarks suppressed by $1/\tan\beta$ and couplings to down-type quarks enhanced by $\tan\beta$. The production of $a_2$ is also modified at leading order since there are different form factors for the scalar and pseudoscalar couplings; $CP$-even Higgs bosons couple to fermions via scalar couplings and $CP$-odd couple via pseudoscalar.

%-------------------------
\begin{figure}[h!]
\includegraphics[width=0.49\linewidth]{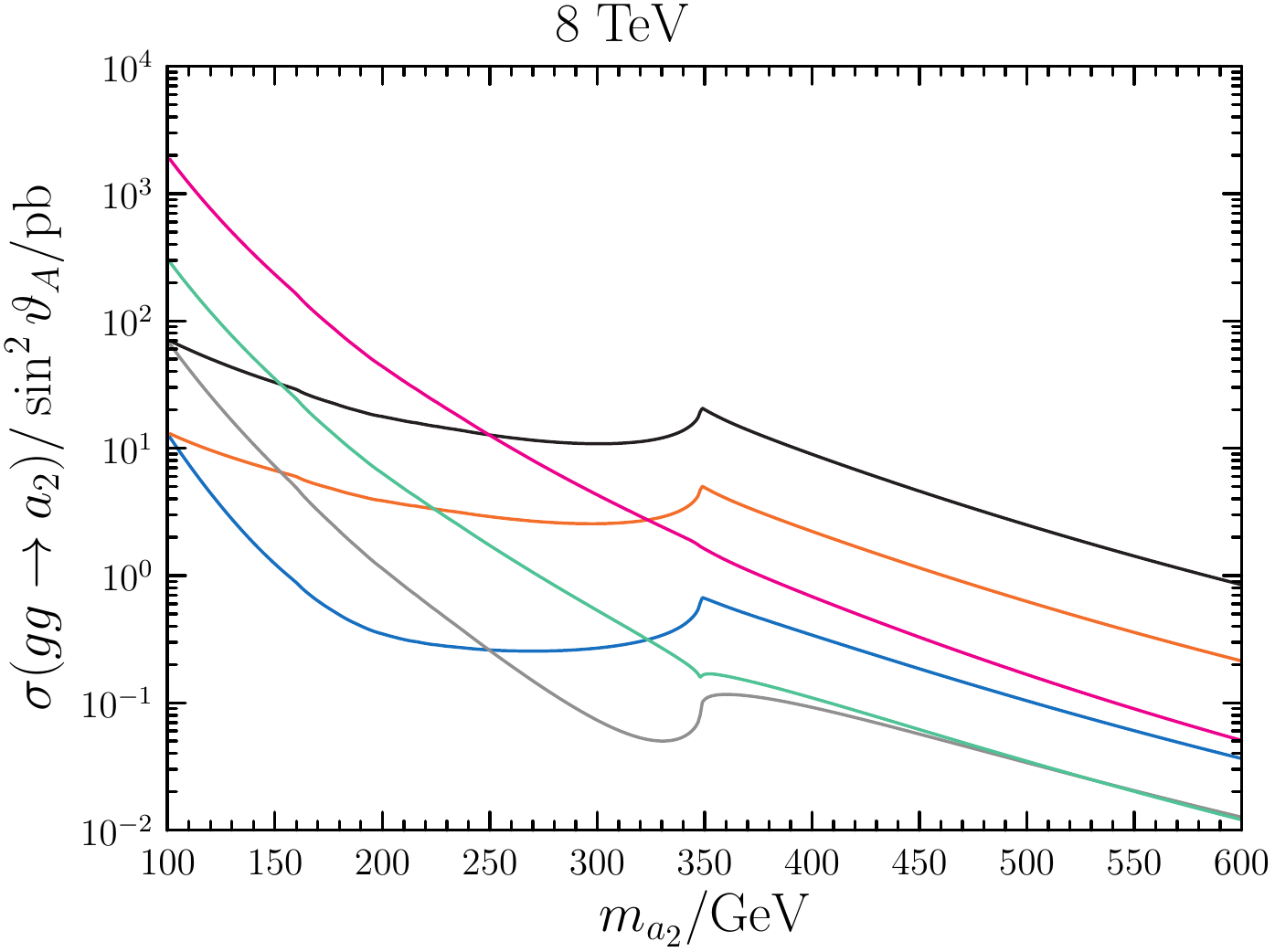}
\includegraphics[width=0.49\linewidth]{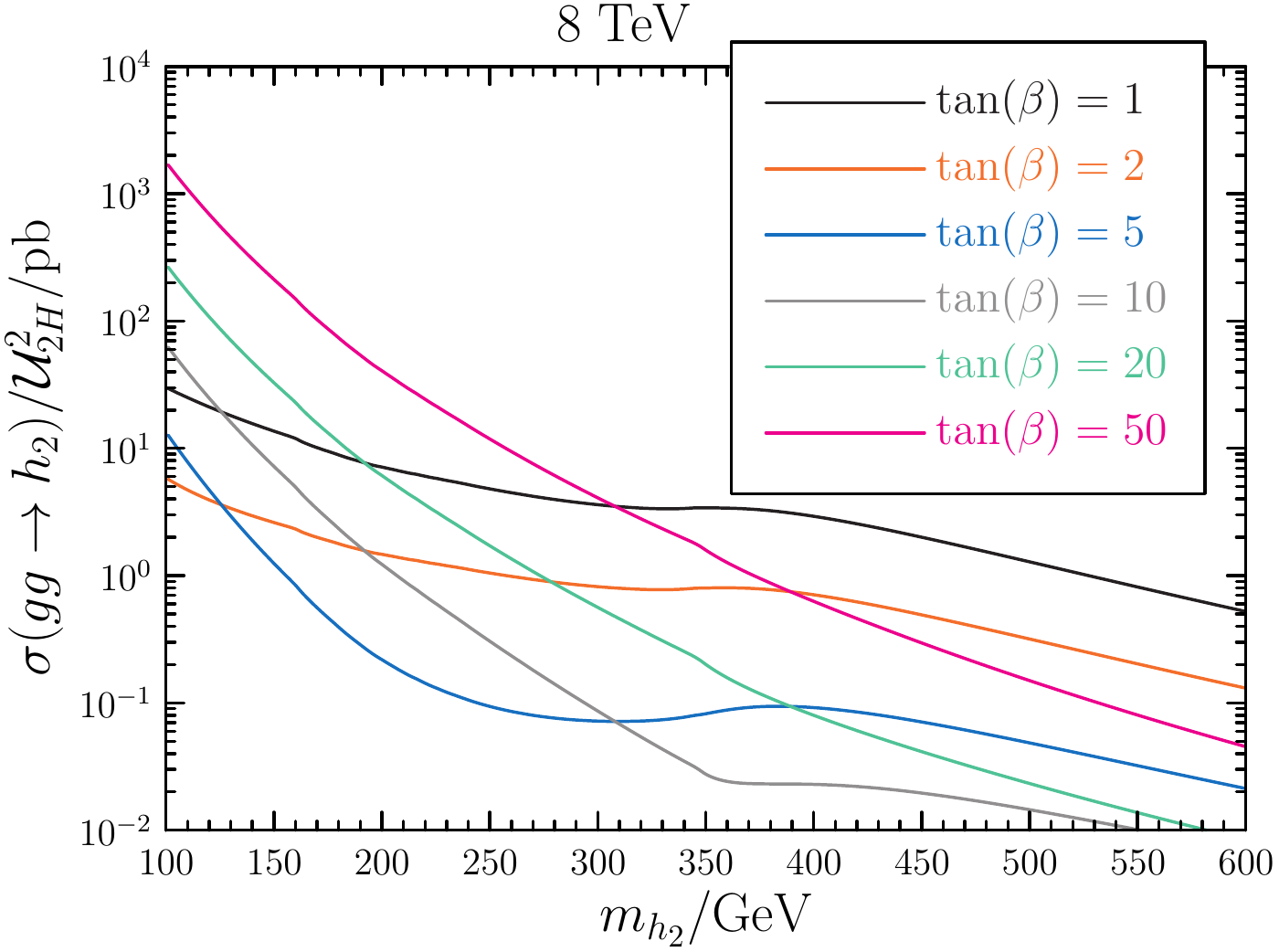}
\vskip11mm
\includegraphics[width=0.49\linewidth]{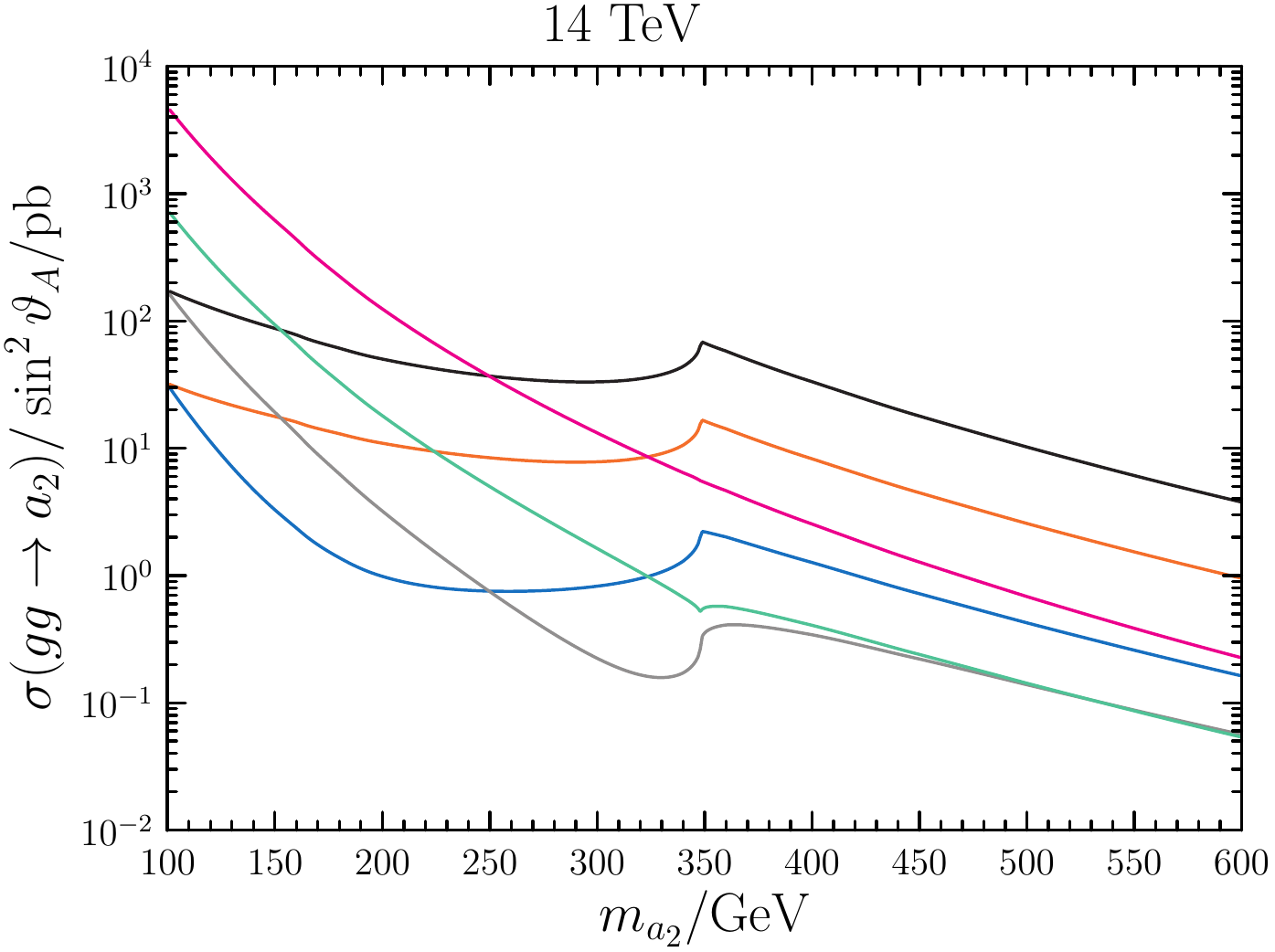}
\includegraphics[width=0.49\linewidth]{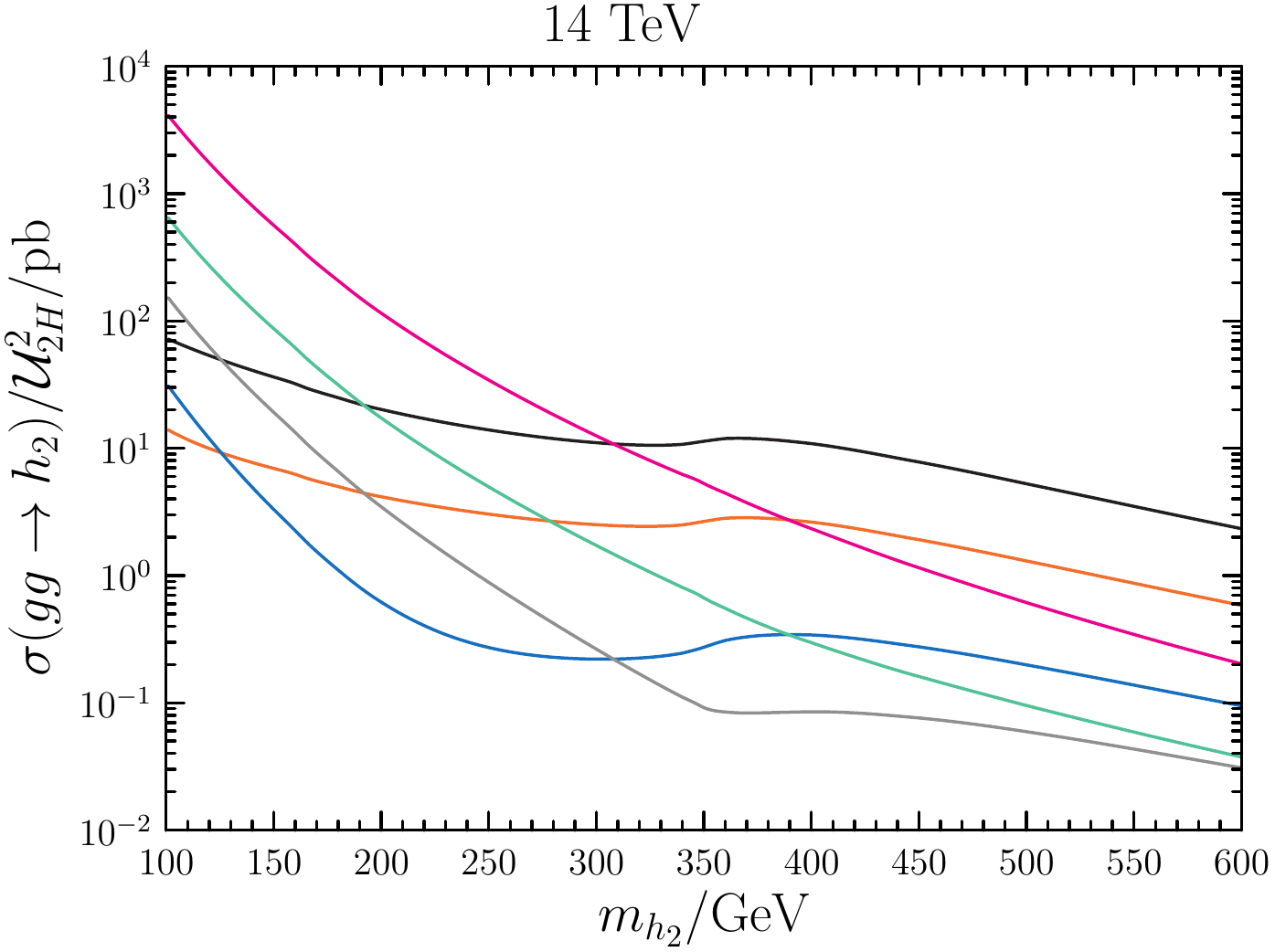}
\vskip11mm
\includegraphics[width=0.49\linewidth]{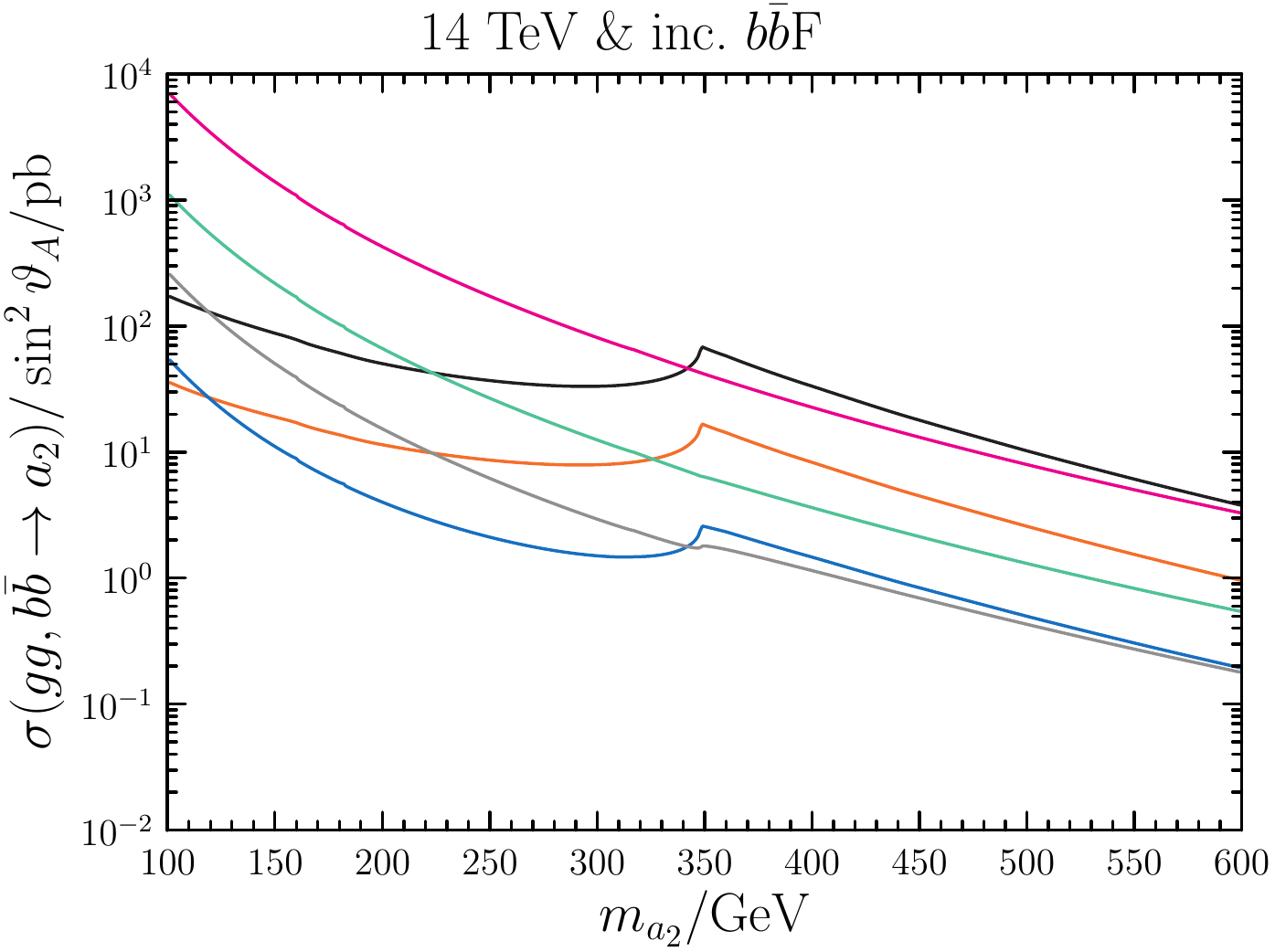}
\includegraphics[width=0.49\linewidth]{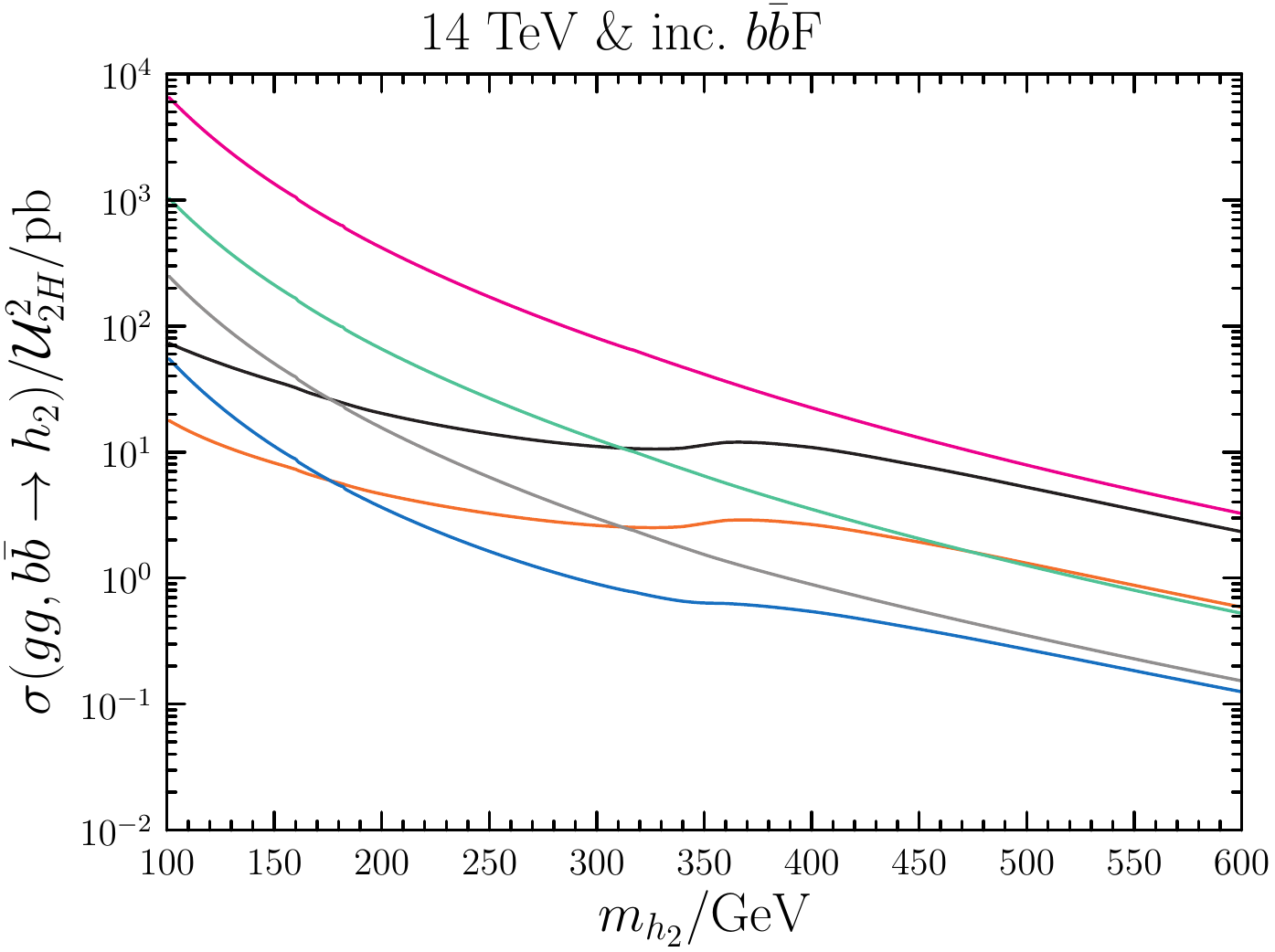}
\caption{
The LHC $gg$F production cross-sections at (above) 8~TeV and (central) 14~TeV
for (left) the $CP$-odd state of the type-II 2HDM $A_H$ and (right) the $CP$-even state of the type-II 2HDM orthogonal to the SM-like state $H$,
for various masses and values of $\tan\beta$. Below: the cross-sections at 14~TeV summing the contributions from $gg$F and $b\bar b$F.
\label{fig:prod}}
\end{figure}
%------------------------------

At leading order the $gg$F production cross-section for a scalar or pseudoscalar $\phi$ is proportional to
\dis{
S^\phi_0 &= \left|\frac{3}{4}\sum_q g^\phi_q A^\phi_{1/2} \left(\frac{m_\phi^2}{4m_q^2}\right)\right|^2,
}
where $g^\phi$ is the relative coupling to the quark $q$ (relative to that of the SM Higgs) and $m_q$ is the quark pole mass. The form factors $A^\phi_{1/2}$ are equal to
\begin{eqnarray}
A^{\mathcal{H}}_{1/2}(\tau) &=& 2[\tau+(\tau-1)f(\tau)]/\tau^2\quad\mbox{and}\\
A^{\mathcal{A}}_{1/2}(\tau) &=& 2f(\tau)/\tau 
\end{eqnarray}
for scalar and pseudoscalar couplings respectively. The universal scaling function $f$ can be found, for example, in Ref.~\cite{Djouadi:2005gi,Djouadi:2005gj}. In the limit $\tau \rightarrow 0$ the functions $A^{\mathcal{H}}_{1/2}(\tau)$ and $A^{\mathcal{A}}_{1/2}(\tau)$ tend to $4 / 3$ and $2$ respectively, so the ratio squared tends to 2.25. The K-factors (the ratios of cross-sections to their leading order approximations) are typically around 1.8 and cannot be neglected. In this work, to calculate the $CP$-odd ($A_H$) and $CP$-even ($H$) doublet production we take the 8 and 14~TeV $gg$F production cross-sections recommended by the CERN Higgs Working Group~\cite{CERNHWG} (calculated at NNLL QCD and NLO EW) for a SM Higgs of the same mass $M$ and multiply by the ratio
\begin{eqnarray}
\frac{\left|\sum_q g_q A^{\mathcal{A,H}}_{1/2}
\left(\frac{M^2}{4m_q^2}\right)\right|^2}
{\left|\sum_q A^\mathcal{H}_{1/2}
\left(\frac{M^2}{4m_q^2}\right)\right|^2},
\end{eqnarray}
where $g_q=\{\tan(\beta),\cot(\beta)\}$ for \{down-, up-\} type quarks $q$. (This is also the approach taken in Ref.~\cite{Barbieri:2013hxa}.) We checked the consistency of this approach using the Fortran code HIGLU~\cite{higlu} at NNLO QCD and NLO EW level with the CTEQ6L parton distribution functions. For the cases of $a_2$ and $h_2$ the cross-section will have an additional suppression of approximately $\sin^2 \vartheta_A$ and $\mathcal{U}^2_{2H}$ respectively, since only the doublet admixture couples to quarks. These production cross-sections at 8 and 14~TeV are shown in Fig.~\ref{fig:prod}. Note that for $A_H$ there is a sharp peak around the $t\bar{t}$ threshold region for small $\tan\beta$ (where the top loop dominates) due to the pseudoscalar form factor. Below the $t\bar{t}$ threshold the shapes of the curves are highly dependent on whether the top or bottom loop dominates. This is because the form factor looks quite different depending on whether one is above threshold (bottom loop case) or below threshold (top loop case).

At moderate and large $\tan\beta$ (i.e.~$\tan\beta \gtrsim 5$) heavy neutral Higgs production in bottom fusion ($b\bar{b}$F, upper right plot in Fig.~\ref{fig:process}) can be larger than in gluon fusion ($gg$F, upper left plot in Fig.~\ref{fig:process}). In fact, although the probability to find a bottom quark in a proton is small (whereas gluons have the largest parton distribution function at LHC center-of-mass energies), this is compensated by the fact that $b\bar b$F is an electroweak tree-level process (whereas the $gg$F is one-loop suppressed). In the lower plots of Fig.~\ref{fig:prod} we show the impact of adding the $b\bar{b}$F cross-section (calculated using FeynHiggs~\cite{feynhiggs}) to the $gg$F one for $\sqrt{s} = 14~{\rm TeV}$; clearly the effect is sizable only for large values of $\tan\beta \gtrsim 10$. Note that at small $\tan\beta$ $gg$F is large and dominant and that at large $\tan\beta$ $b\bar b$F controls the cross-section; at intermediate values of $\tan\beta \sim 5$ the $gg$F suppression is not yet compensated by the $b\bar b$F enhancement and we find relatively small cross-sections.

%%%%%%%%%%%%%%%%%%%%%%%%%%%%%%%%%
\subsection{Total cross-sections}
\label{sec:total}

%-------------------------------------------------------------
\begin{figure}[h!]
\includegraphics[width=0.43\linewidth]{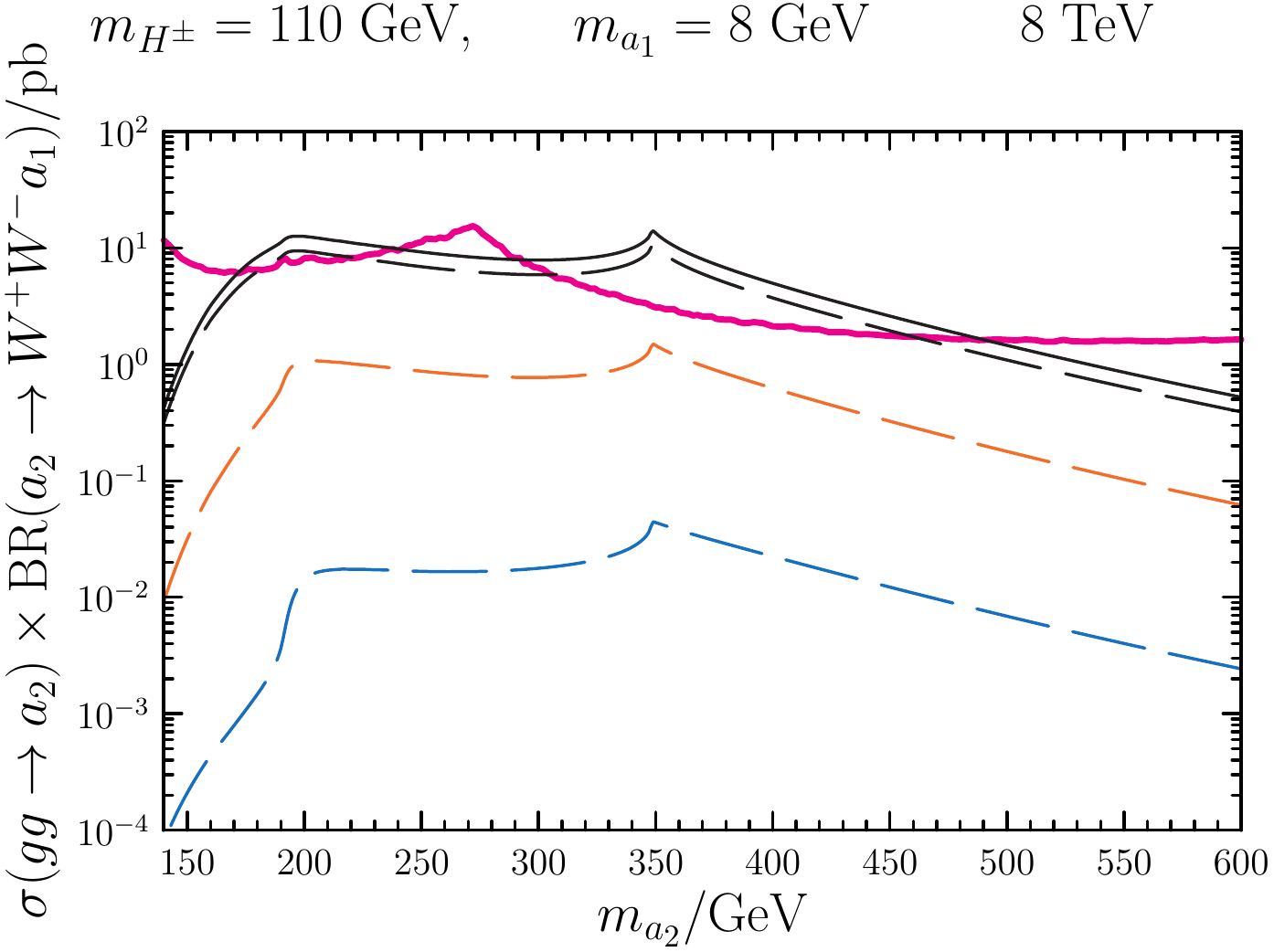}
\includegraphics[width=0.43\linewidth]{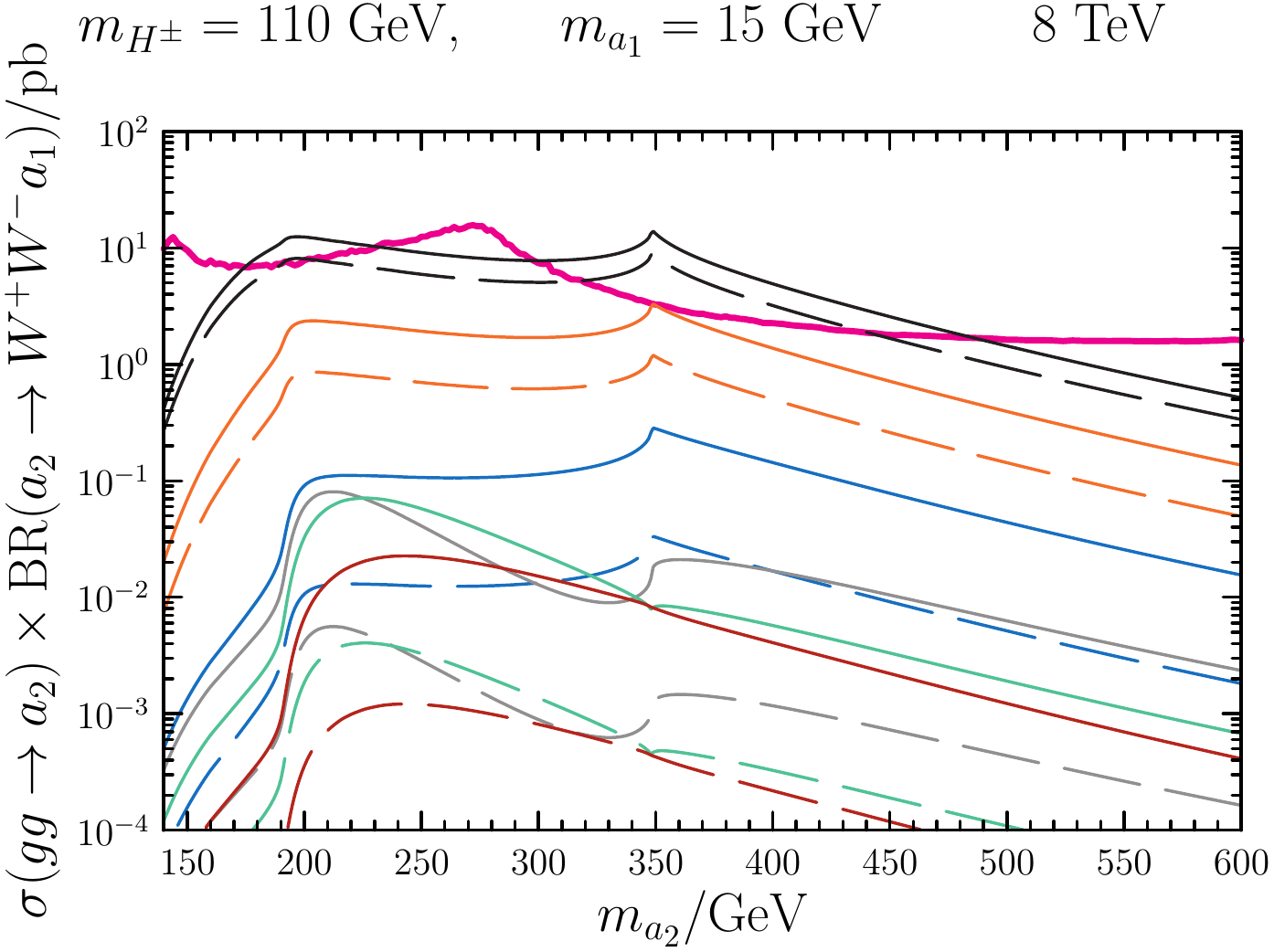}
\vskip5mm
\includegraphics[width=0.43\linewidth]{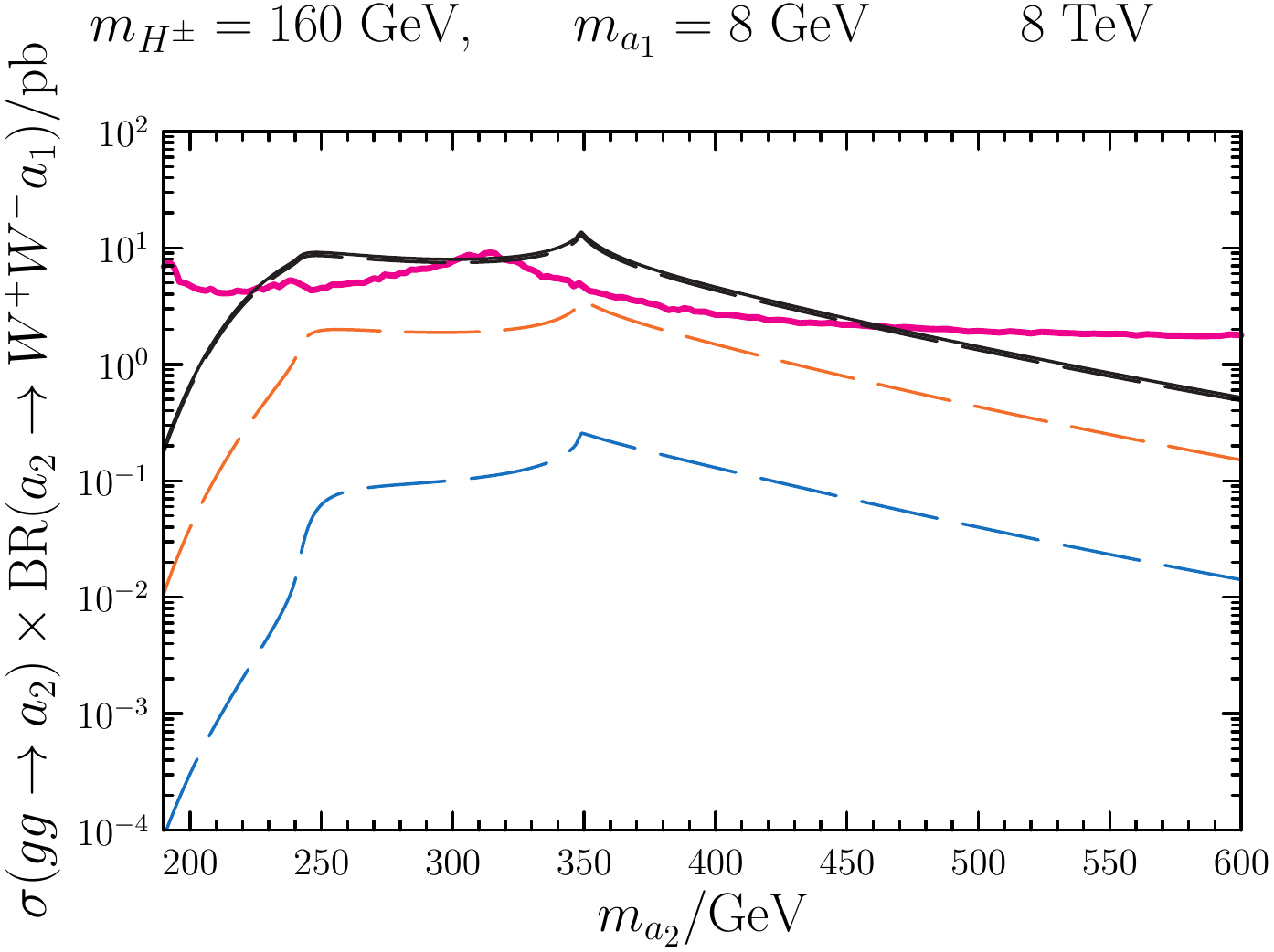}
\includegraphics[width=0.43\linewidth]{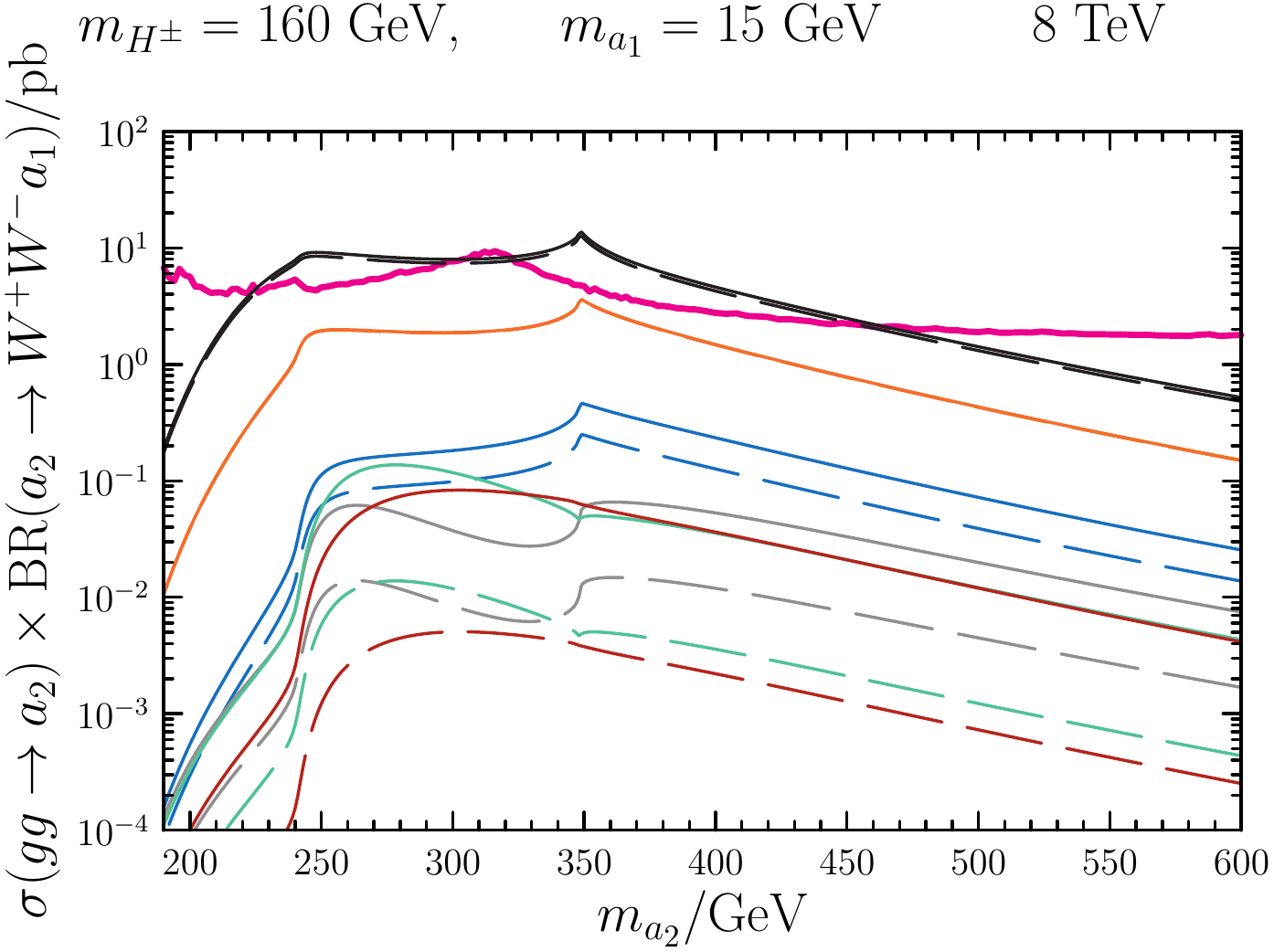}
\vskip0mm
\raisebox{-\height}{\includegraphics[width=0.43\linewidth]{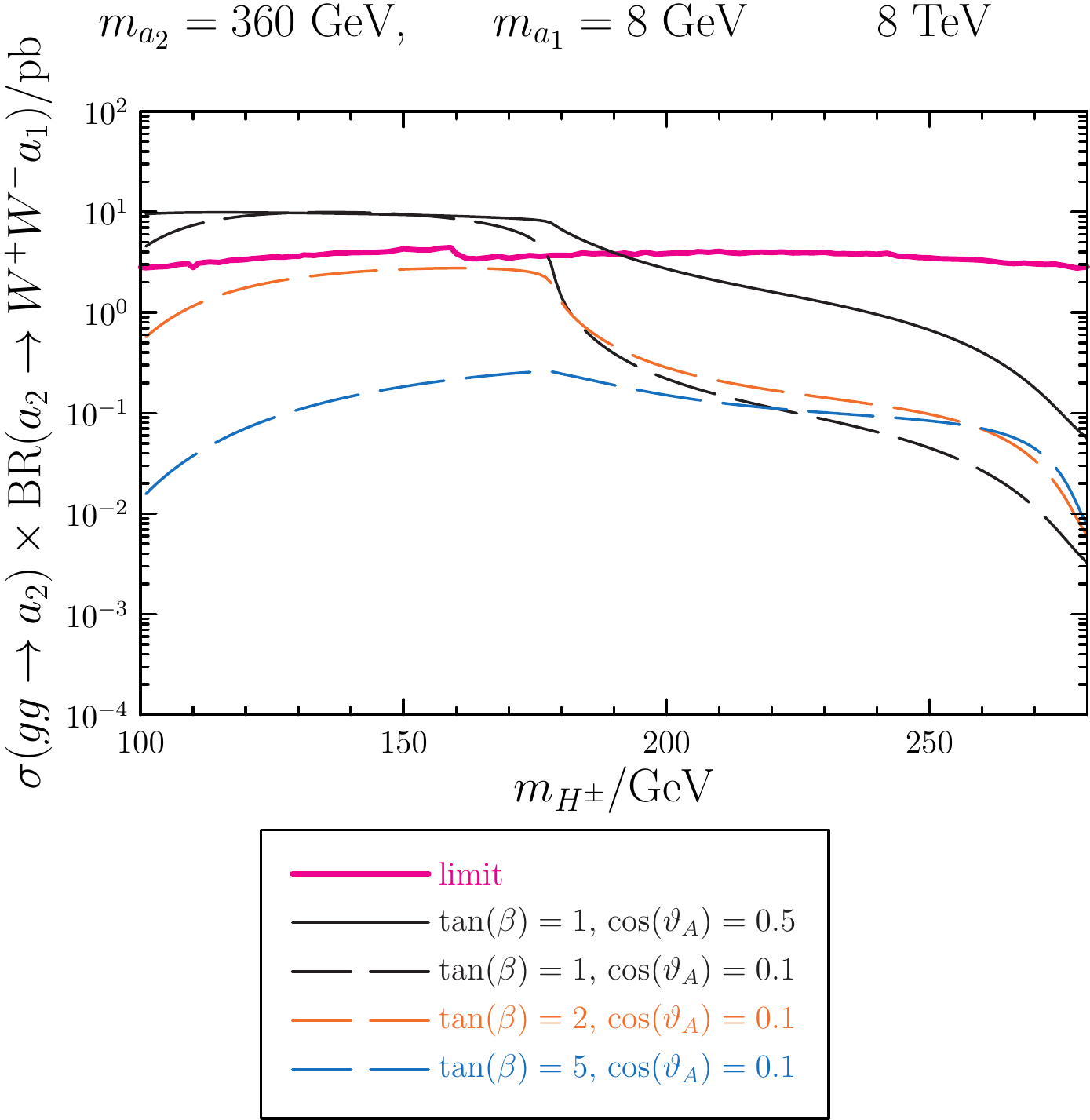}}
\raisebox{-\height}{\includegraphics[width=0.43\linewidth]{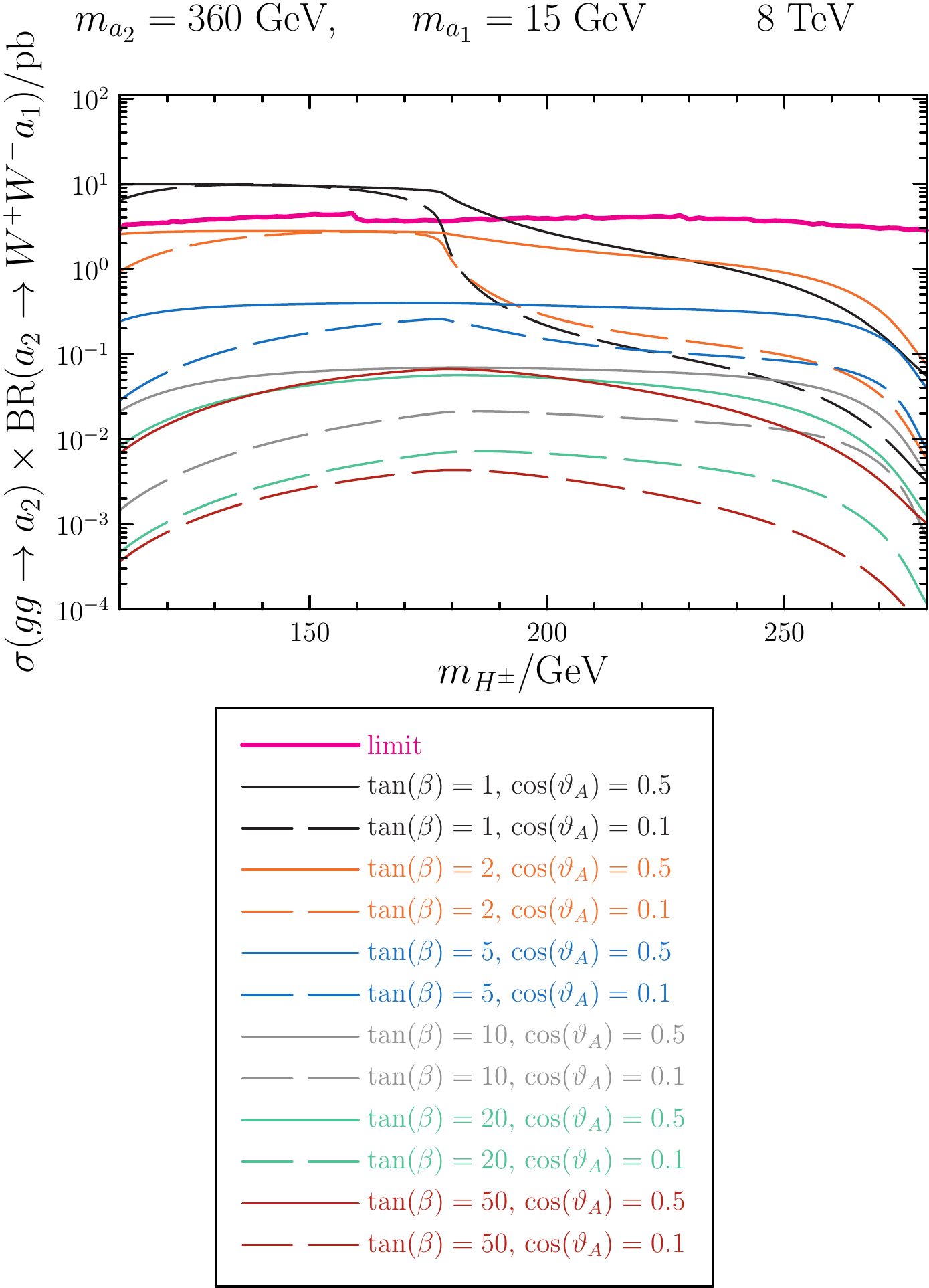}}
\caption{
The complete cross-section
$\sigma(gg\rightarrow a_2 \rightarrow W^+ W^- a_1)$ at 8~TeV.
The magenta lines are the upper limits derived in section~\ref{sec:lim}.
\label{fig:prodbrs8}}
\end{figure}
%-----------------------------------

\begin{figure}[h!]
\raisebox{-\height}{\includegraphics[width=0.49\linewidth]{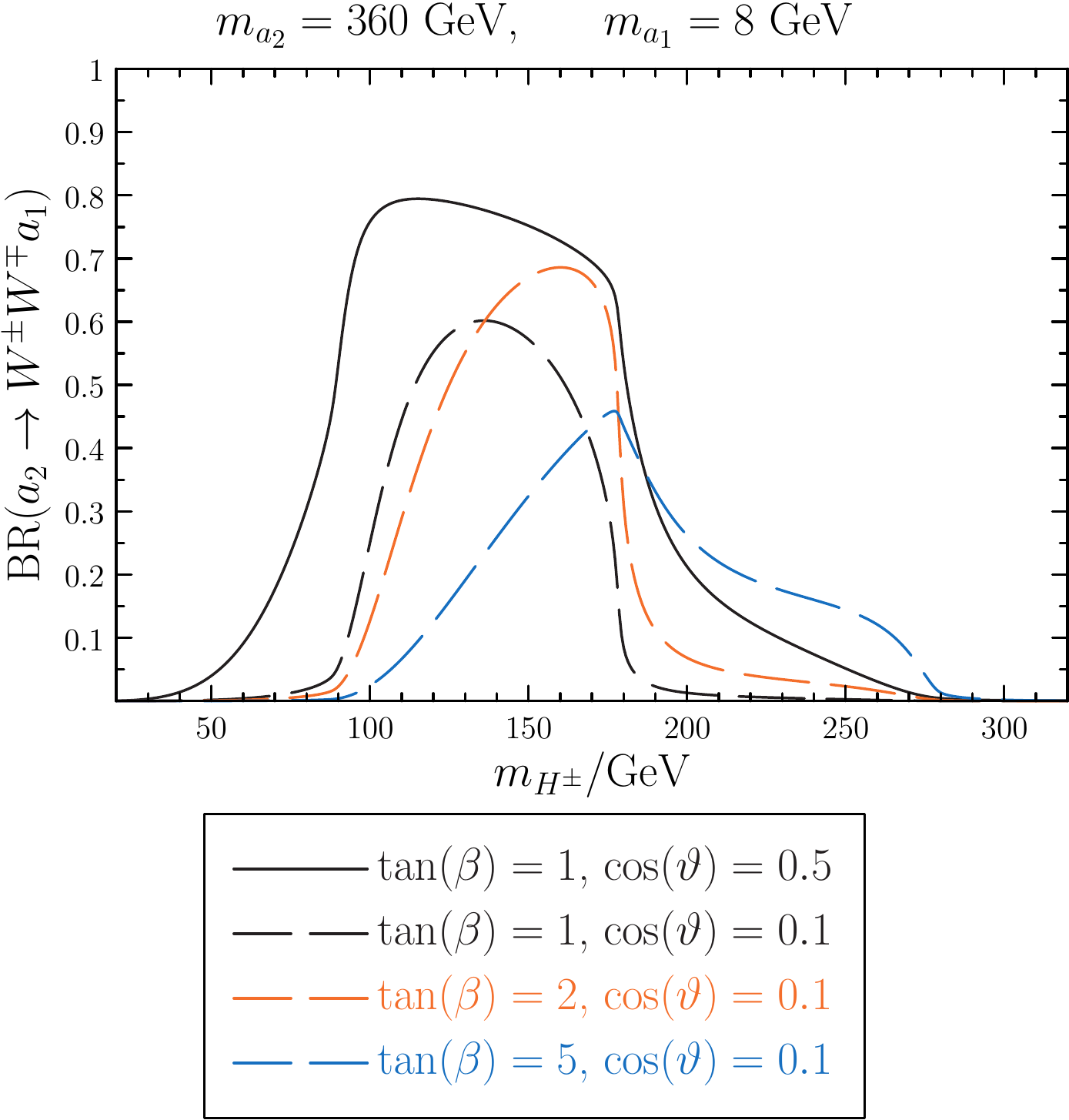}}
\raisebox{-\height}{\includegraphics[width=0.49\linewidth]{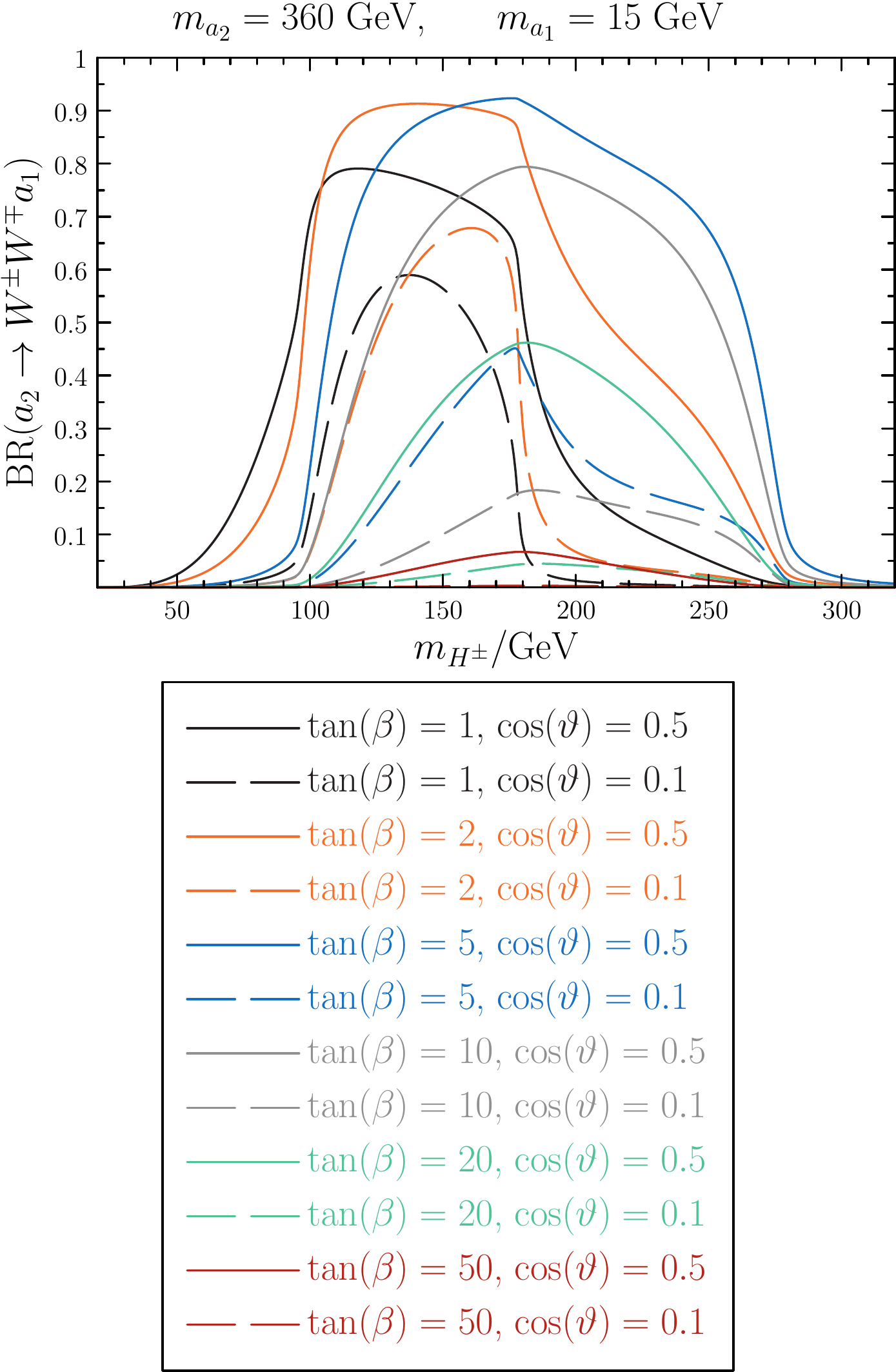}}
\caption{
The complete
BR($a_2 \rightarrow W^+ W^- a_1$).
\label{fig:brs}}
\end{figure}

Combining the previous results, we can obtain the complete cross-section times branching ratios $\sigma(gg\rightarrow a_2 \rightarrow W^+ W^- a_1)$ at 8~TeV in Fig.~\ref{fig:prodbrs8} for various masses and values of $\tan\beta$ and $\cos\vartheta_A$. For small $\tan\beta$, we can easily obtain a total cross-section times branching ratios $\mathcal{O}$(pb), which is comparable to the SM Higgs production times BR($h_{\rm SM} \rightarrow W^+ W^-$). Hence the LHC Higgs search result can constrain the maximum total cross-section of our process, as will be discussed in the next section. For very large $\tan\beta$ ($\gtrsim 20$) our study is not very sensitive because the $\tan\beta$ dependence of the $a_2$ production cross-section (responsible for the enhancement of the latter at large $\tan\beta$) is compensated by the $\tan\beta$ suppression of the branching ratio BR($a_2 \rightarrow W^+ W^- a_1)$.
The complete branching ratios BR($a_2 \rightarrow W^+ W^- a_1)$ are calculated as outlined in App.~\ref{appen:brs} and are shown in Fig.~\ref{fig:brs}.

\begin{figure}[h!]
\includegraphics[width=0.43\linewidth]{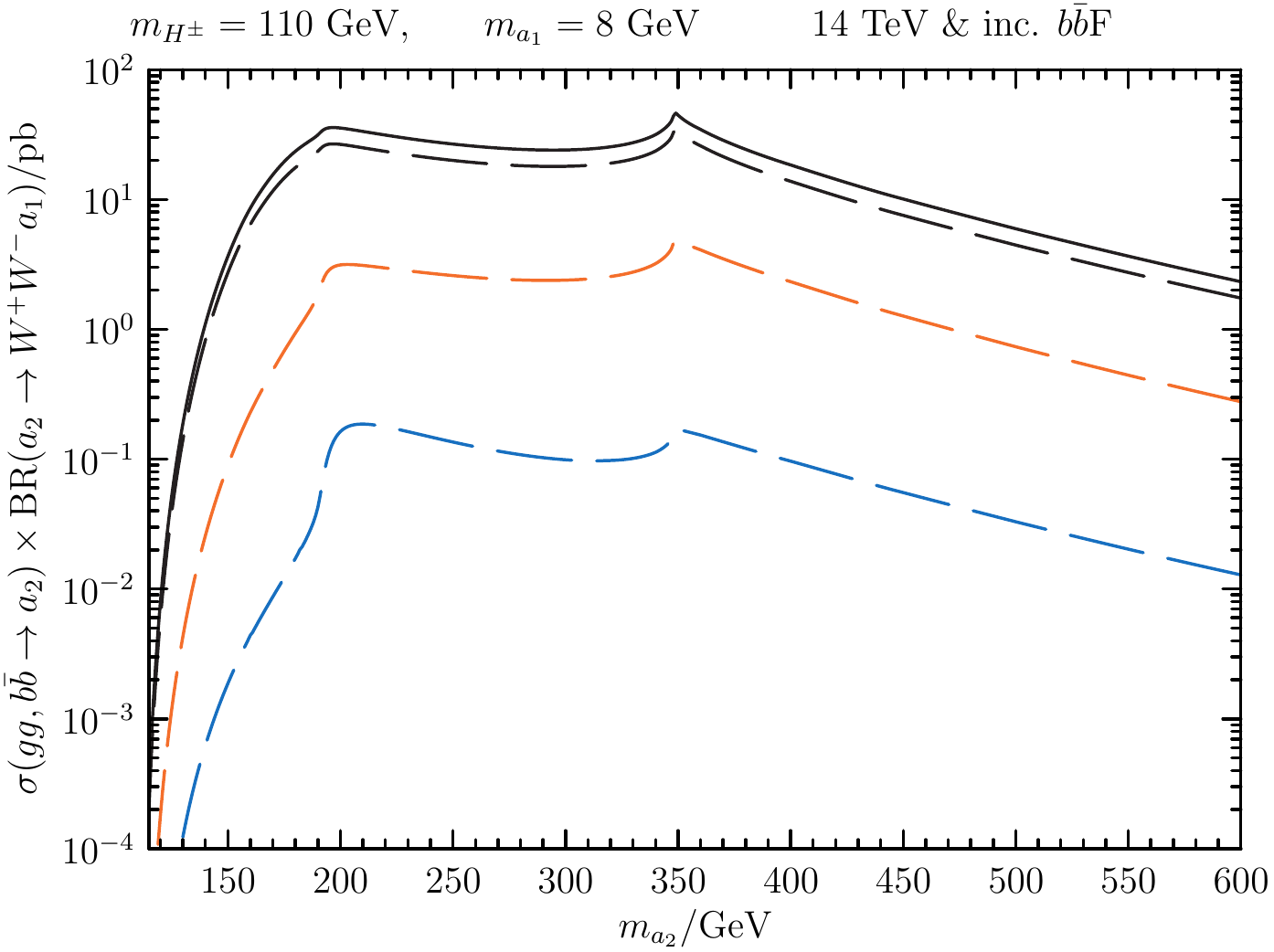}
\includegraphics[width=0.43\linewidth]{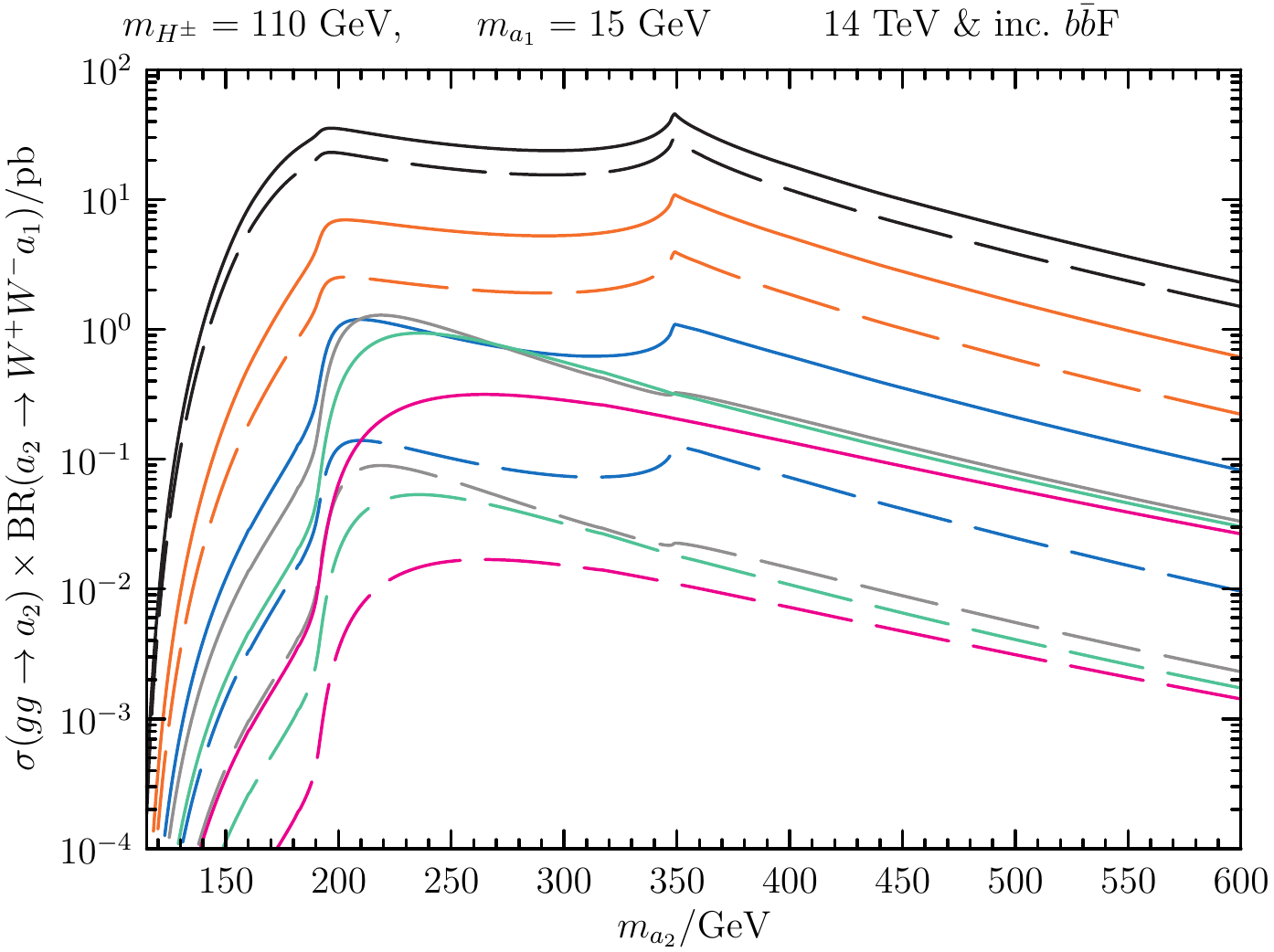}
\vskip5mm
\includegraphics[width=0.43\linewidth]{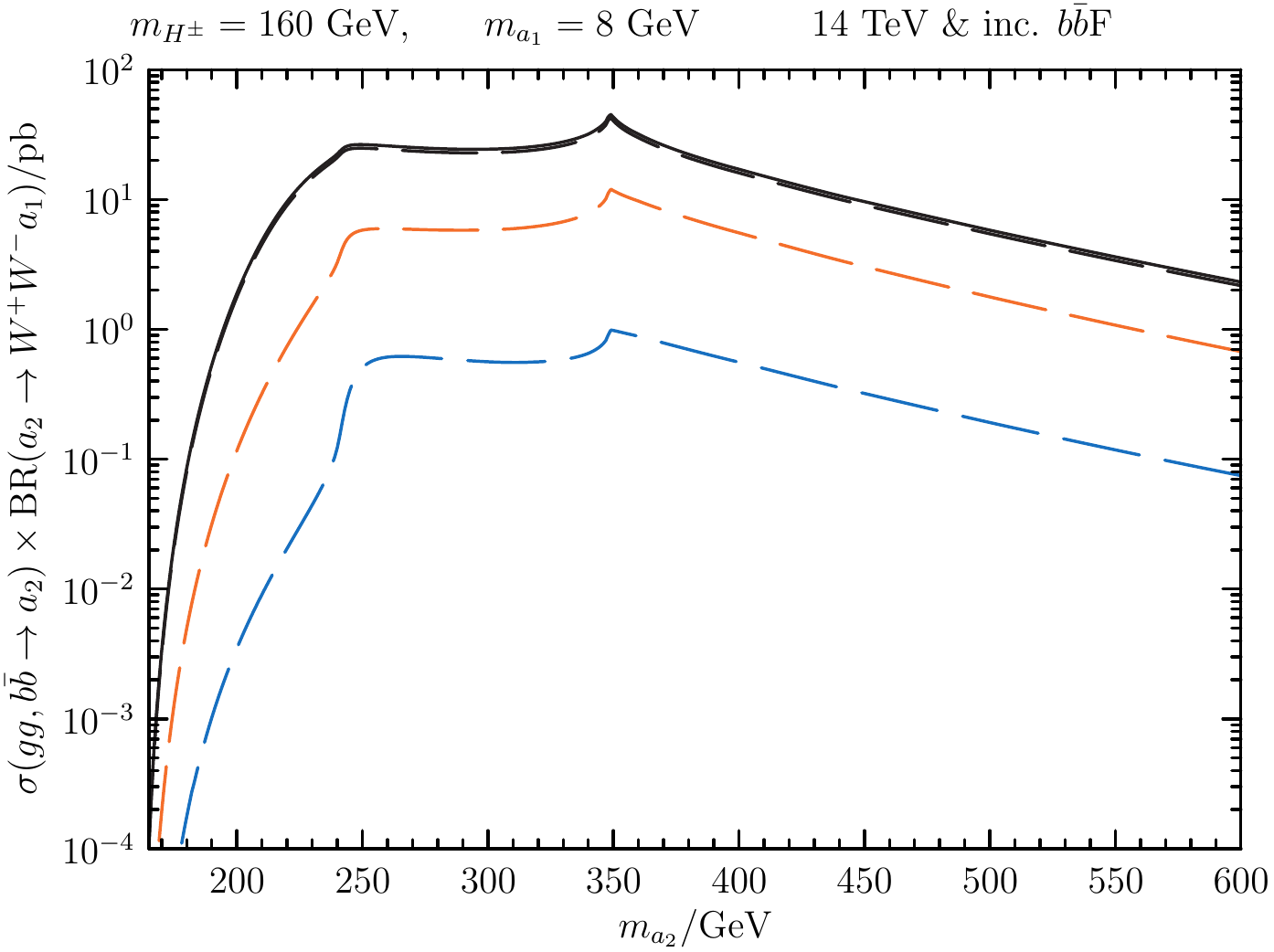}
\includegraphics[width=0.43\linewidth]{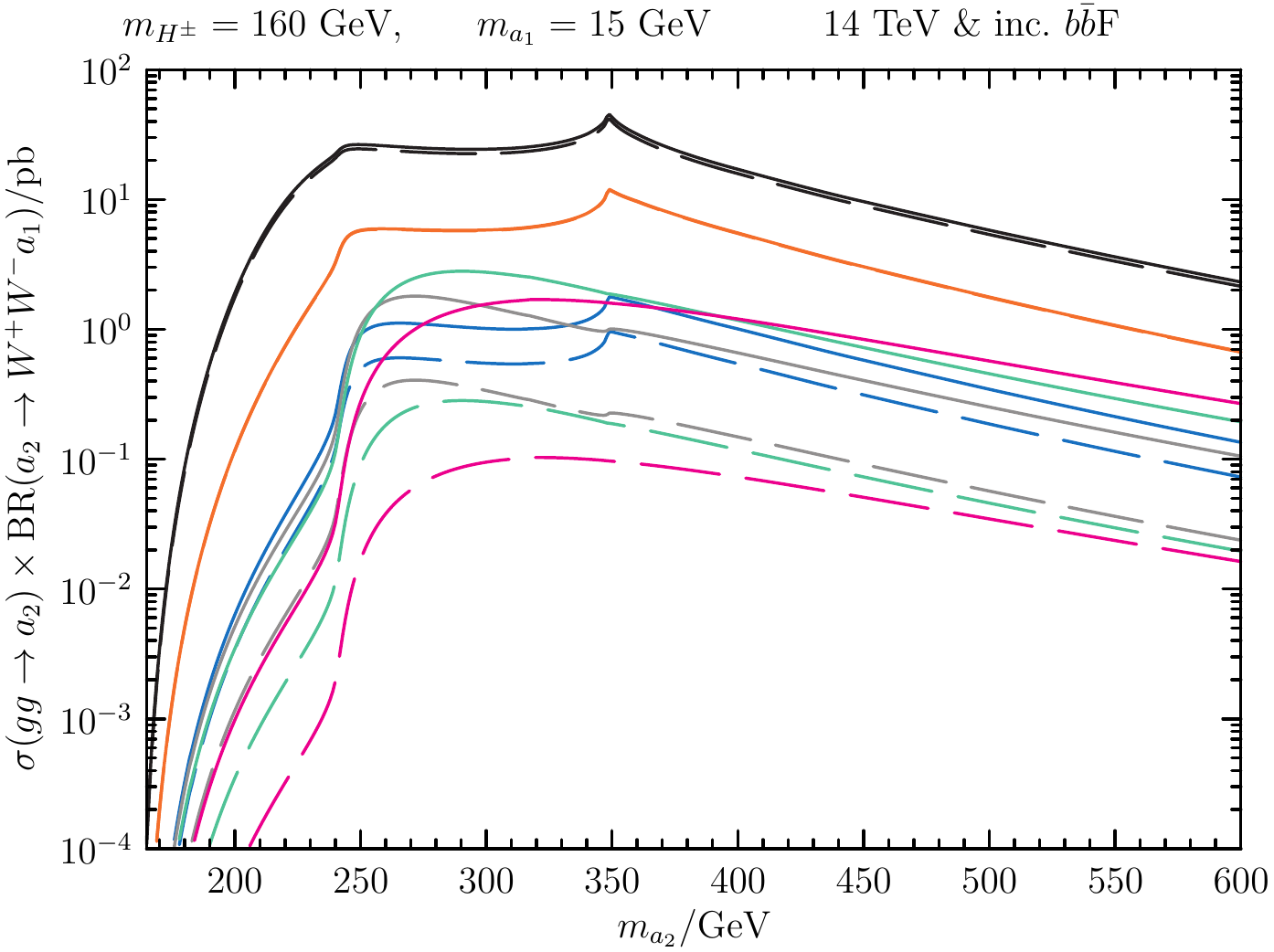}
\vskip0mm
\raisebox{-\height}{\includegraphics[width=0.43\linewidth]{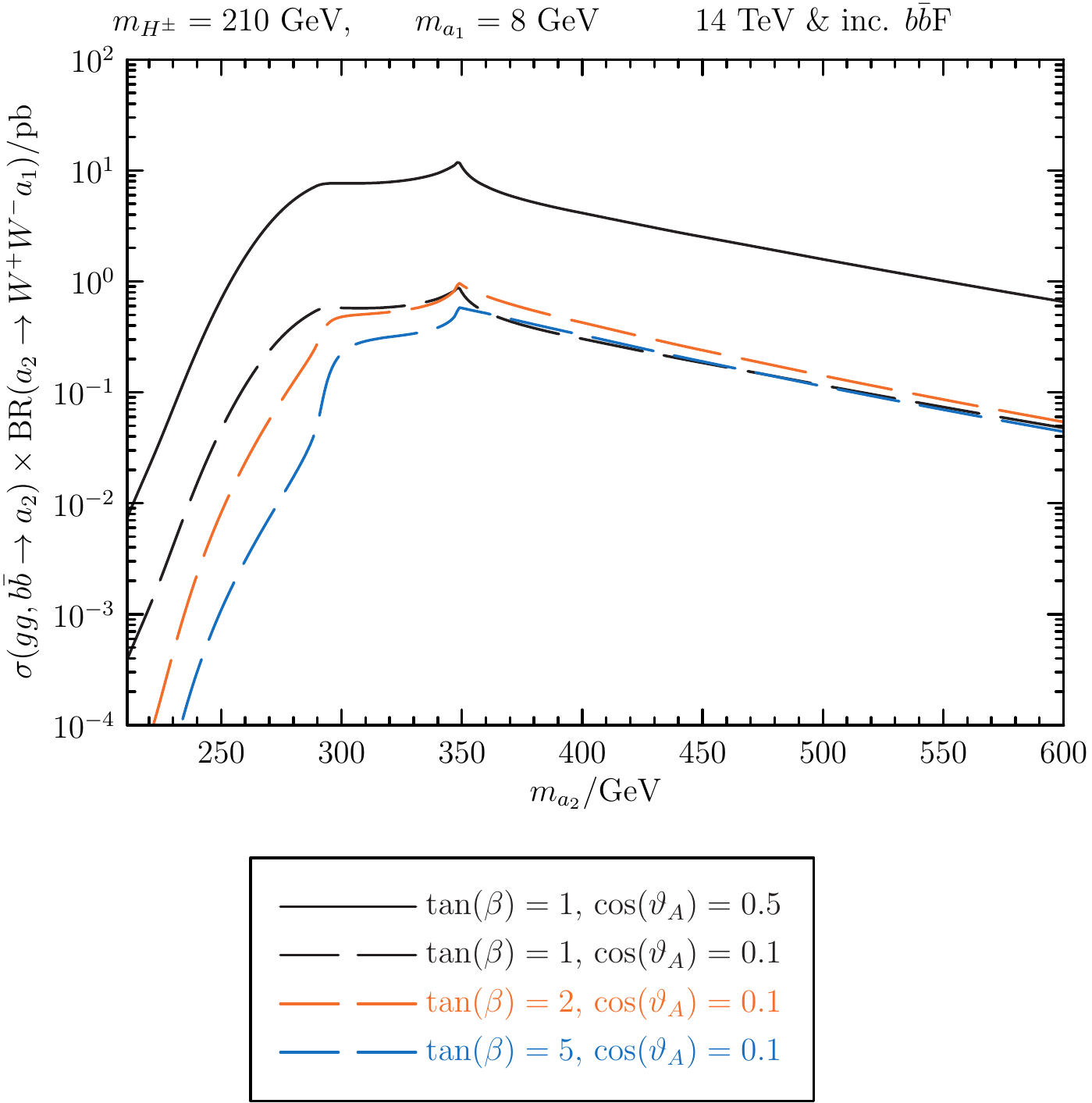}}
\raisebox{-\height}{\includegraphics[width=0.43\linewidth]{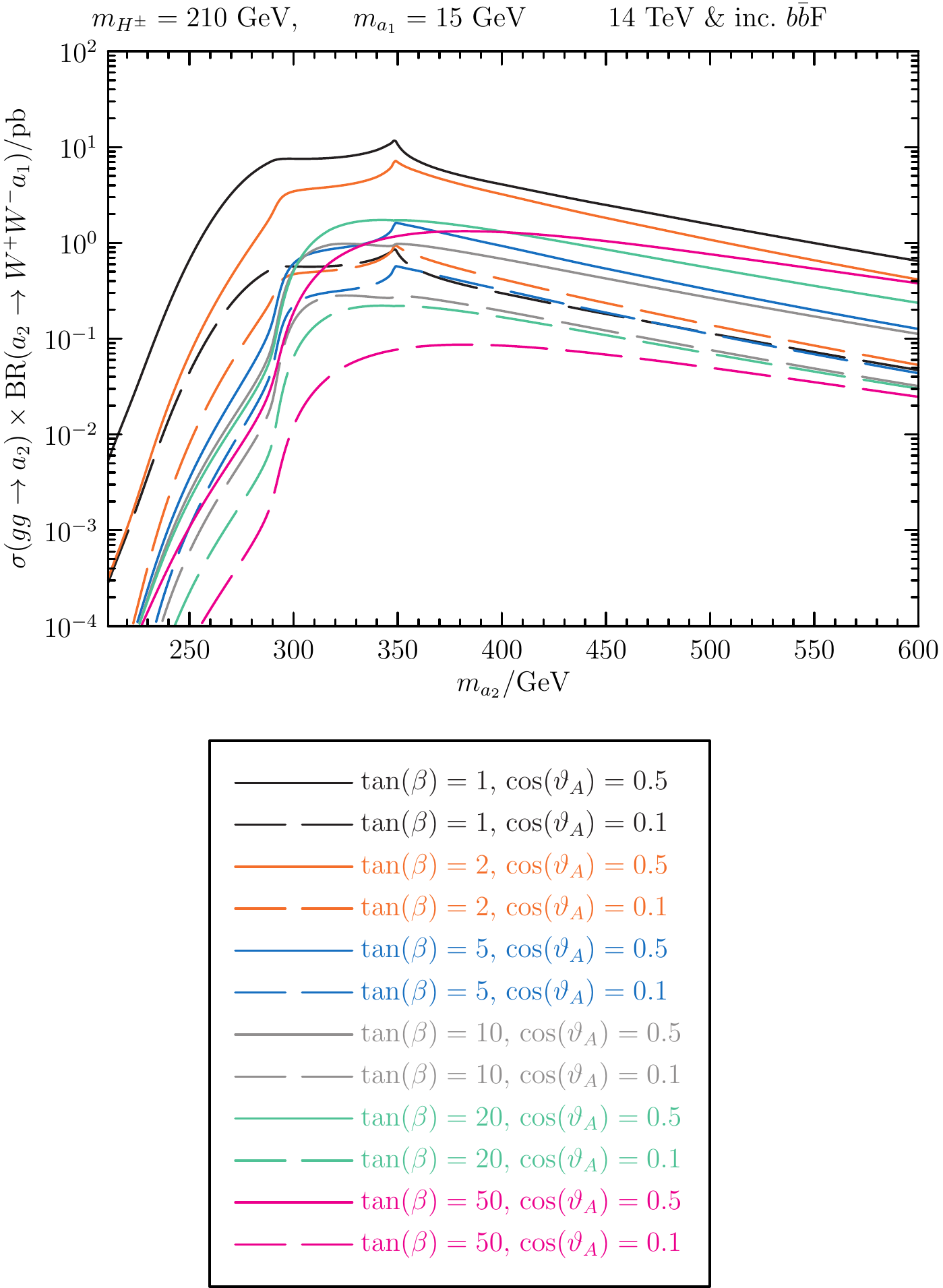}}
\caption{
The complete cross-section times branching ratios
$\sigma(gg,b\bar b\rightarrow a_2 \rightarrow W^+ W^- a_1)$ at 14~TeV. The contribution from $b\bar b$F is added to the contribution from $gg$F.
\label{fig:prodbrs14}}
\end{figure}

\begin{figure}
\includegraphics[width=0.43\linewidth]{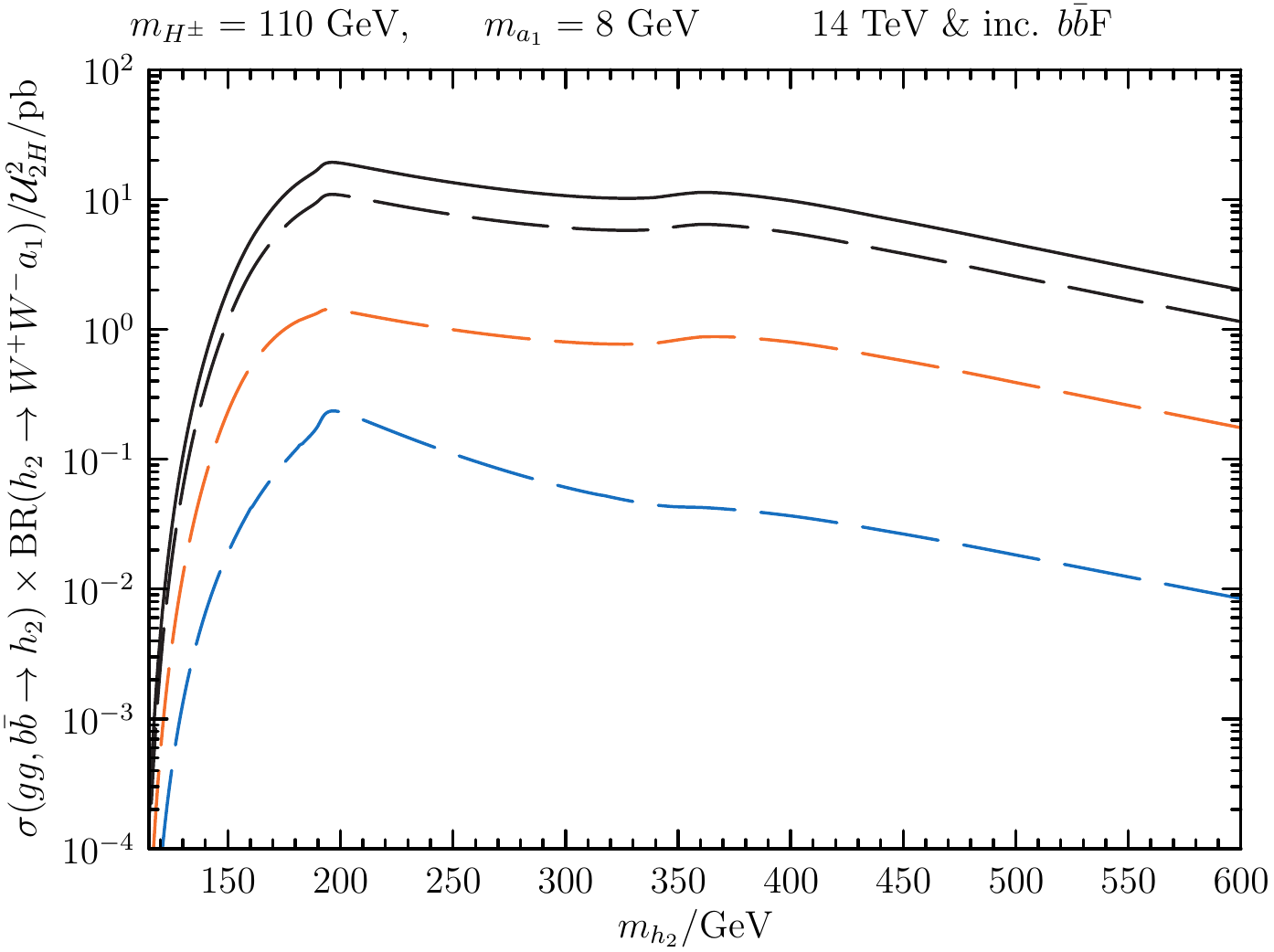}
\includegraphics[width=0.43\linewidth]{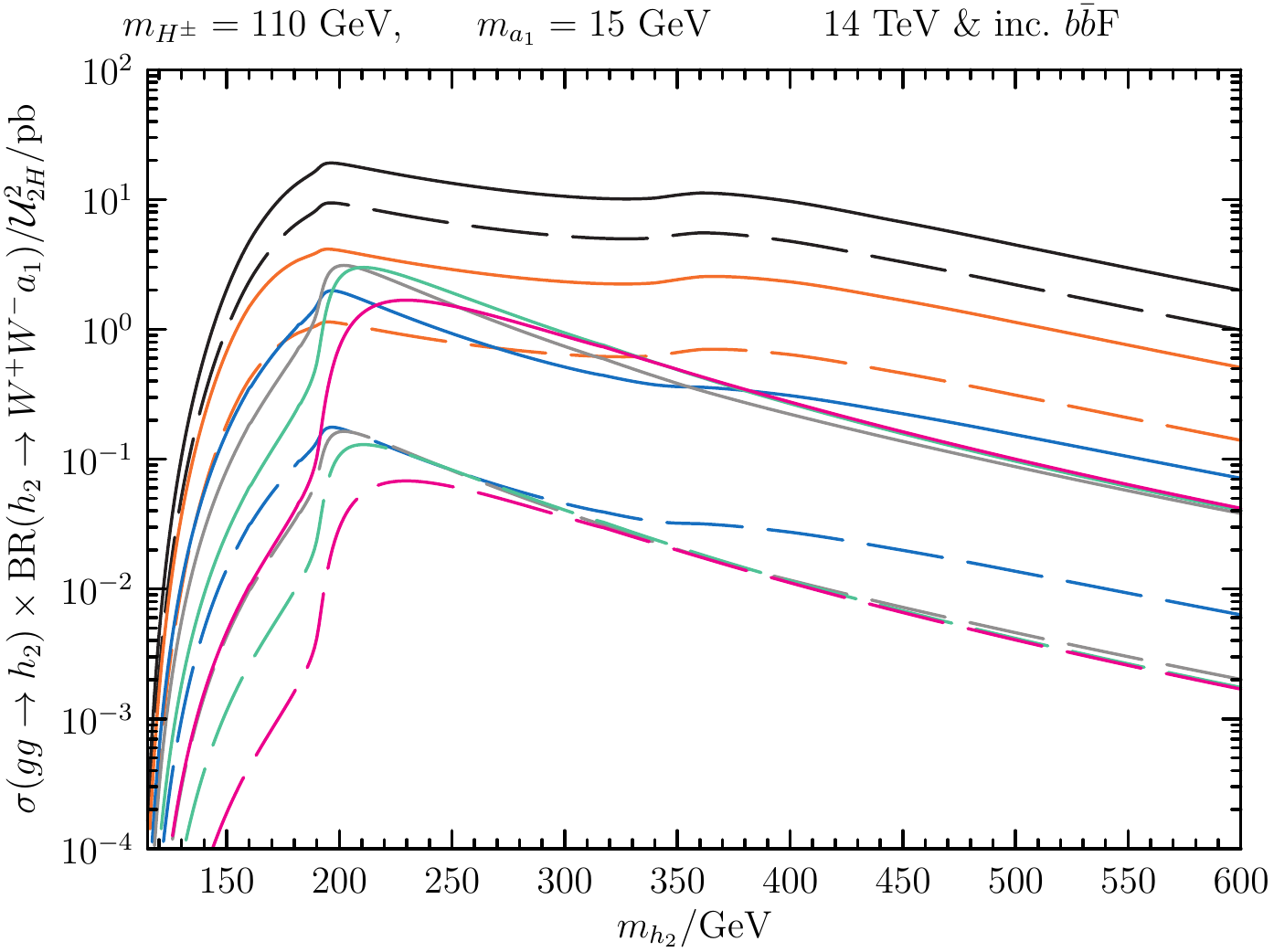}
\vskip5mm
\includegraphics[width=0.43\linewidth]{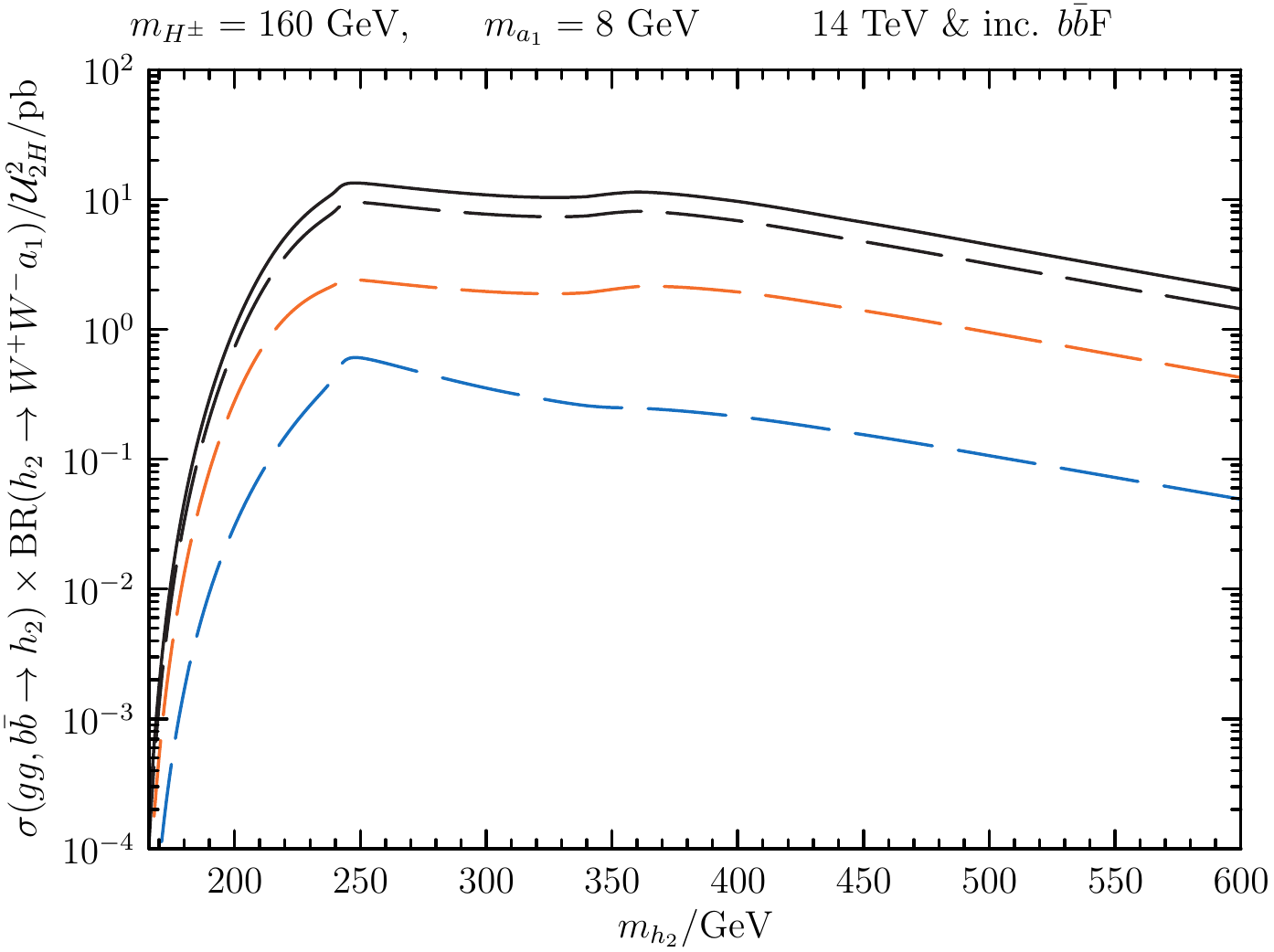}
\includegraphics[width=0.43\linewidth]{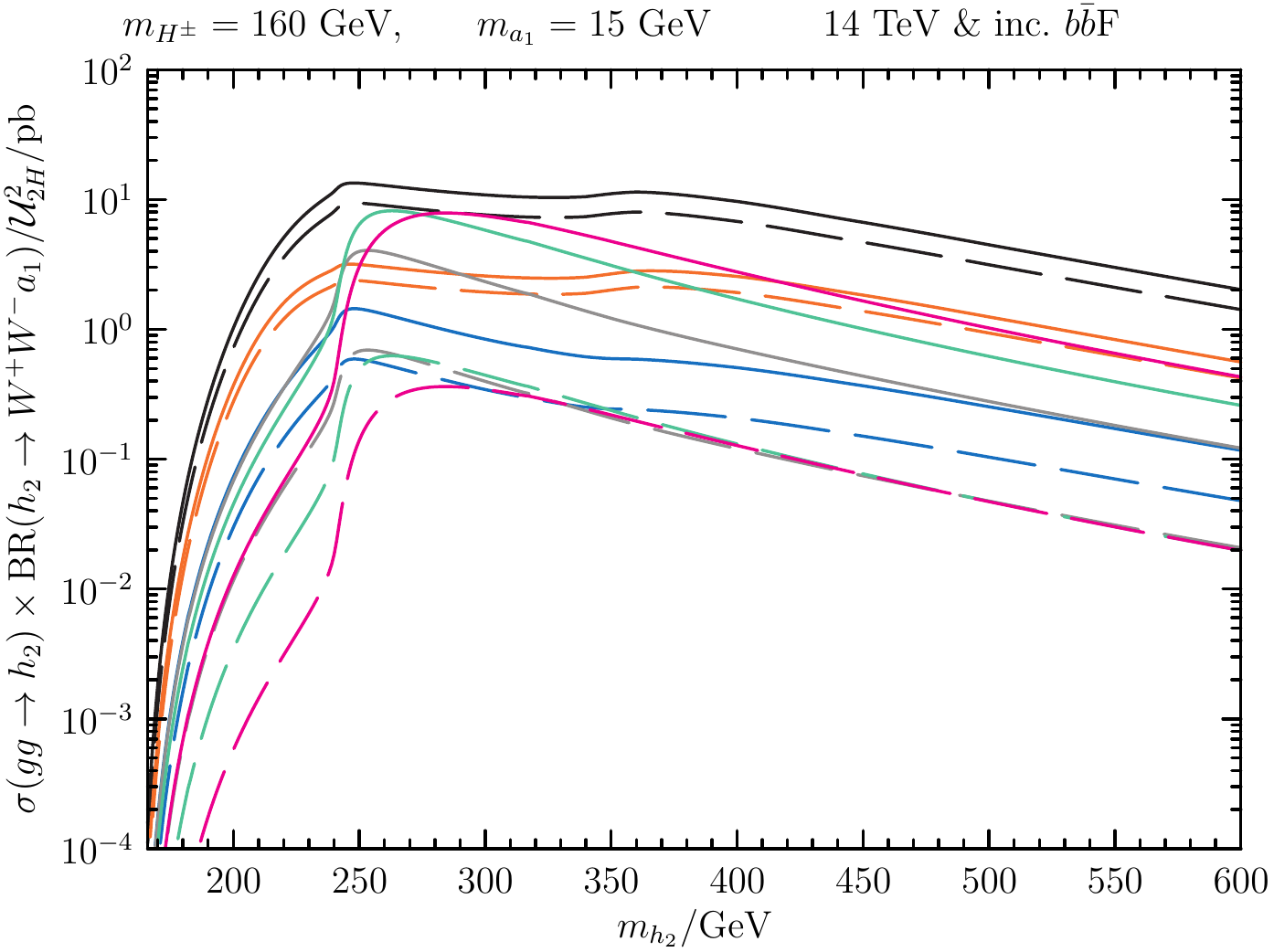}
\vskip0mm
\raisebox{-\height}{\includegraphics[width=0.43\linewidth]{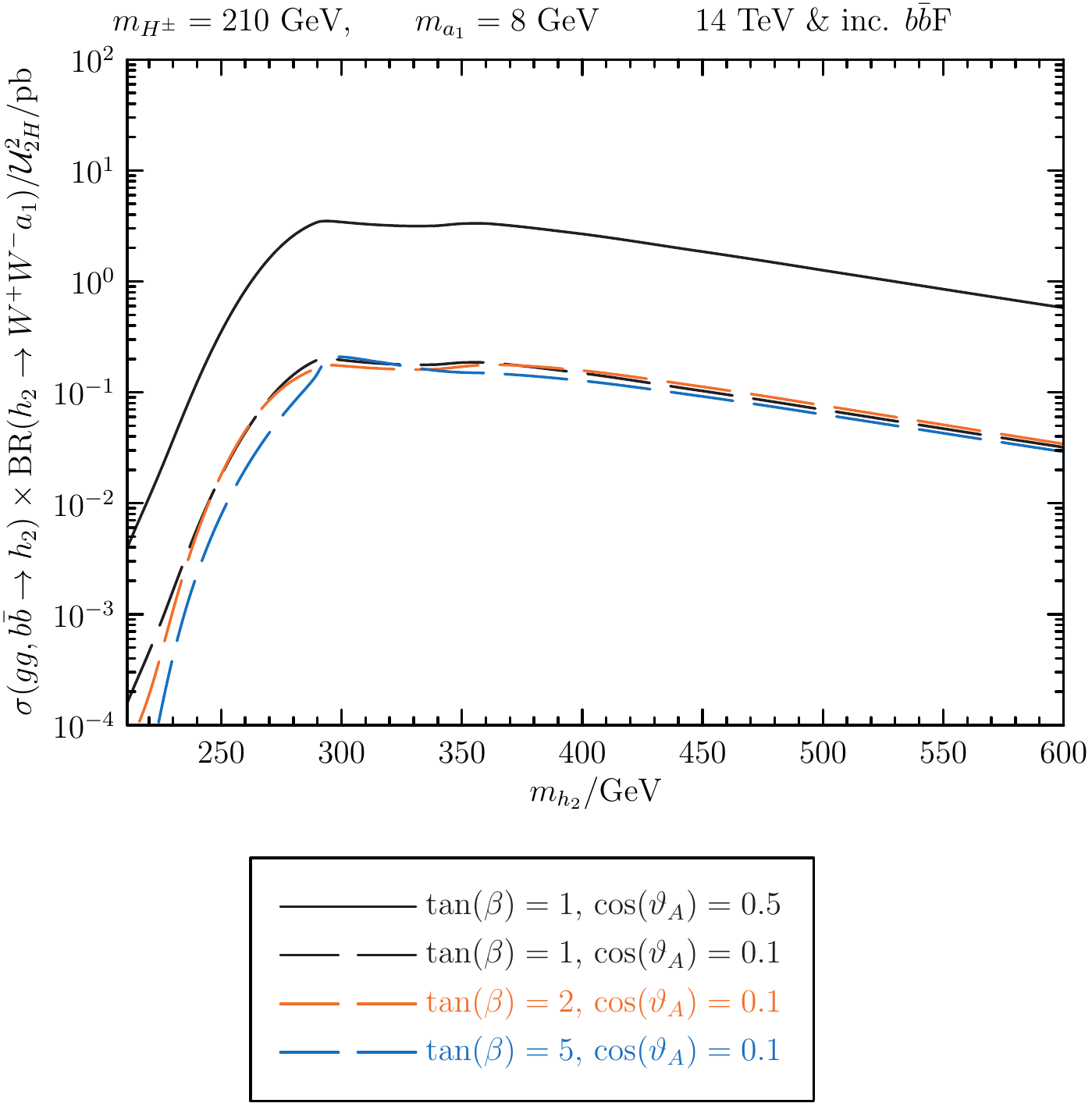}}
\raisebox{-\height}{\includegraphics[width=0.43\linewidth]{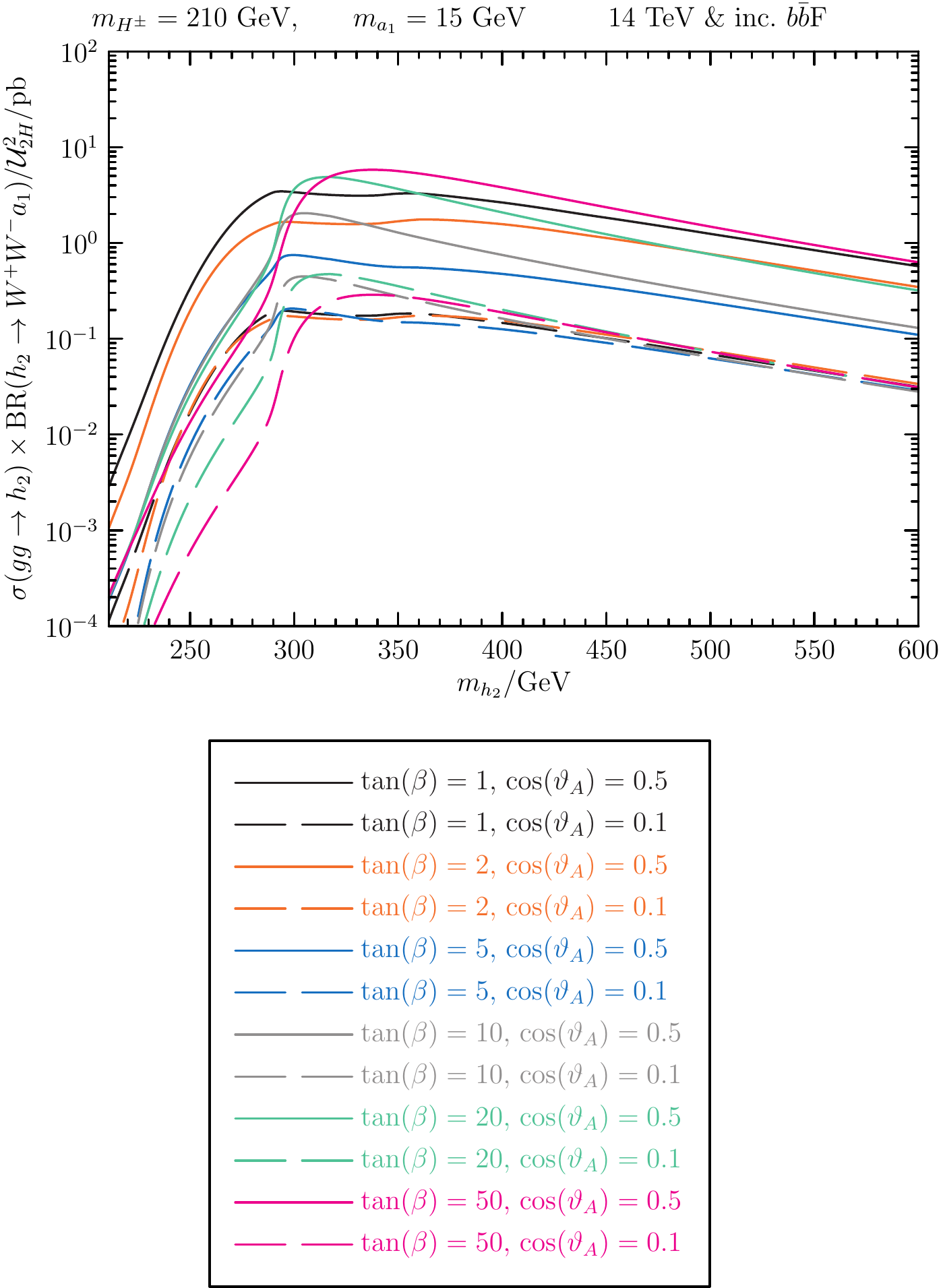}}
\caption{
The complete cross-section times branching ratios
$\sigma(gg,b\bar b\rightarrow h_2 \rightarrow W^+ W^- a_1)/\mathcal{U}^2_{2H}$ at 14~TeV. The contribution from $b\bar b$F is added to the contribution from $gg$F.
$\mathcal{U}_{2H}$ is the $H$ amplitude in the $CP$-even state $h_2$. This mixing
element suppresses the production of $h_2$, but
(given the assumptions outlined in subsection~\ref{sec:heavy})
cancels out of the branching ratios.
\label{fig:prodbrseven14}}
\end{figure}

For comparison, we also show the expected total cross-section at 14~TeV for both $\Phi=a_2$ and $\Phi=h_2$ in Figs.~\ref{fig:prodbrs14} and \ref{fig:prodbrseven14} respectively. Note that in these plots we add the $gg$F and $b\bar{b}$F production cross-sections. The most important effect of adding the latter is the flattening of the total cross-section for $\tan\beta \gtrsim 5$; therefore, once we achieve sensitivity to $\tan\beta \sim 5$ we expect to be sensitive to all values of $\tan\beta$ (depending on $\cos\theta_A$).
When $\Phi = h_2$, the complete cross-section $\sigma(gg\rightarrow h_2 \rightarrow W^+ W^- a_1)$ divided by the mixing-element-squared $\mathcal{U}^2_{2H}$ is shown. We show that it is possible to have total cross-sections $\mathcal{O}$(10 pb) in some regions of the parameter space. In these figures we include also a heavier charged Higgs masses, above the $tb$ threshold.

In this method of estimating the total cross-section, multiplying the production cross-section by the branching ratios, the non-zero width of the heavy state $\Phi$ is neglected. We check that for $m_\Phi$ above the $H^\pm W^\mp$ threshold, going beyond the zero-width approximation for $\Phi$ is a numerically small effect in the parameter space we consider. Below the $H^\pm W^\mp$ threshold the finite width effects can be important if the width of $\Phi$ is already comparable to the widths of $H^\pm$ and $W^\mp$ (the dominant contribution can come from $\Phi$ going off-shell rather than $H^\pm$ or $W^\mp$). We find that this can only occur at extreme values of $\tan\beta$ ($\gg 20$ or $\sim 1$ if $m_\Phi>2m_t$). In these cases our method can underestimate the below threshold (off-shell) total cross-section. See App.~\ref{appen:wPhi} for more details of the $\Phi$ width. Our zero-width approximation for the heavy state $\Phi$ does not affect the limits that we derive. (For the kinematics the $\Phi$ finite width effects are included.)

%%%[[[---]]]}}}-----------------------------------------------------------------
%%%[[[---]]]}}}-----------------------------------------------------------------

%---------------------------------
%------------------------------------------

%--------------------------------------
%-------------------------------------------------------------

%%%%%%%%%%%%%%%%%%%%%%%%%%%%%%%%%%%%%%%%%%%%%%%%%%%%%%%%%%%%%%%%%%%%%%%%%%
\section{The constraint from Standard Model $h \to W^+ W^-$ searches}
\label{sec:lim2hdm}

\subsection{Model independent study}
\label{sec:lim}

The CMS collaboration observed a SM Higgs signal in the $W^+ W^- \to \ell^+ \ell^- \nu \bar{\nu}$ channel (final states with zero jets or one jet were included) with a mass of approximatively 125~GeV~\cite{cmshww} at a significance of 4$\sigma$. CMS also provides an exclusion bound for a SM Higgs bosons in the mass range 128--600~GeV at 95~\% confidence level (C.L.). The process that we are considering ($pp\to W^+W^- A$) leads to a very similar final state, the only difference being the light Higgs $A$ decay products that lead to extra jets or leptons. We, therefore, expect this search to provide strong constraints on the charged Higgs production mechanism we consider and potentially to offer an avenue to discover a charged Higgs. However, due to the presence of the light Higgs $A$, the distributions of kinematic variables that we obtain are different from those expected in the SM Higgs search. In order to apply the results in Ref.~\cite{cmshww} we need to calculate how the efficiency of the various cuts adopted in that analysis are affected by the presence of the light Higgs $A$.

The constraints that we derive are valid for a light Higgs $A$ whose mass is just above the $b\bar{b}$ threshold and which decays dominantly to a pair of bottom quarks. In the CMS analysis the number of jets (for the purposes of separating the events into channels; 0, 1, or more) is defined as the number of reconstructed jets with $p_T>30$~GeV (and $|\eta|<4.7$), reconstructed using the anti-$k_T$ clustering algorithm with distance parameter $\Delta R=0.5$. For purely kinematic reasons, when the $p_T$ of the $A$ is as large as 30~GeV the angular separation ($\Delta R$) between the two $b$ quarks is going to be small (compared to 0.5) for the $A$ masses that we consider and therefore any $A$ final state with high enough $p_T$ to count as a jet will in fact have its final state $b$ quarks cluster into a single jet most of the time. This has been explicitly checked in Ref.~\cite{Dermisek:2012cn} (for $A\to \mu^+\mu^-, \tau^+\tau^-$) and in Ref.~\cite{Rathsman:2012dp} (for $A\to b\bar b$---see Fig.~6 therein).
Using MadGraph we checked that for $m_A$ up to around 15~GeV, the $\Delta R$ angular opening of the two $b$ quarks is small enough to treat the $b\bar b$ system as a single fat jet (obviously for a low enough $p_T$ cut and/or a large enough $m_A$, the two final state $b$ quarks can look like two distinct jets).
 
For $m_A$ below the $b\bar{b}$ threshold $A$ will decay mainly to $\tau$ lepton pairs or maybe to charm quark pairs. For example, for $A=a_1$, decaying via its $A_H$ admixture, decays to $\tau$ pairs will dominate until very low $\tan\beta \approx 1.3$, where decays to charm pairs begin to overtake~\cite{Dermisek:2010mg}. For such decays into charm quarks the opening angle cannot exceed 0.5 and the decay products will mostly be clustered into a single jet. For the decays into $\tau$ leptons, the decay products will also mostly be clustered into a single jet and give no additional isolated leptons; the exception is when both $\tau$ leptons decay leptonically (about 13~\% of the time). In this case there will be no jet and quite possibly extra isolated leptons that would lead to the event not passing the selection criteria in the CMS analysis. This small effect should not much affect our results.

The CMS collaboration presented exclusion bounds obtained using two different techniques to isolate the signal from the background. The first is a cut-based analysis in which separate sets of kinematic cuts are applied for each different Higgs mass hypothesis. The second is a shape-based analysis applied to the distribution of events in the two-dimensional $(m_T,m_{\ell\ell})$ plane. In this paper, we apply the cut based analysis of Ref.~\cite{cmshww} to our signal; at this time, we cannot proceed with the shape-based analysis since the CMS note does not provide enough detail.

All of the CMS data are split into four channels depending on whether the two leptons have different or the same flavor (DF, SF) and whether there is zero or one high $p_T$ ($>30$~GeV) jet ($0j$, $1j$). In each channel the expected background, expected signal, and observed data are given for several SM Higgs mass hypotheses. For each of these hypotheses a different set of cuts is applied. The cuts used for SM Higgs searches with mass hypotheses 120, 125, 130, 160, 200, and 400~GeV are presented in Tab.~1 of Ref.~\cite{cmshww}. Extra cuts are also applied for the SF channels in order to suppress background from Drell-Yan processes.

In this paper, we analyze the 19.5~fb$^{-1}$ of data collected at $\sqrt{s} = 8$ TeV and presented in Tab.~4 of Ref.~\cite{cmshww}. To obtain the observed upper limit from applying the cuts in each channel and corresponding to each SM Higgs mass hypothesis we adopt a modified frequentist construction~\cite{cls}. (A brief summary of the CL$_{\rm s}$ method is presented in App.~\ref{appen:cls}.) The 95~\% C.L. upper limit (on the number of events) that we obtain from our analysis is indicated with $\ell^{\mathcal{H}}_{\mathcal{FJ}}$, where $\mathcal{H}$ refers to each of the SM Higgs mass hypotheses and $\mathcal{FJ}$ to the channel considered ($\mathcal{F} \in \{{\rm DF, SF}\}$ and $\mathcal{J} \in \{ 0j,1j\}$). The value of $\ell^{\mathcal{H}}_{\mathcal{FJ}}$ has to be compared to the expected signal $E^{\cc{HP}}_{\cc{FJ}}$, where $\cc{P}$ stands for the considered theory and point in parameter space. (For the type-II 2HDM + singlet scenario $\cc{P}$ stands for the relevant Higgs boson masses, $\cos\vartheta_A$, and $\tan\beta$.)

In the type-II 2HDM + singlet reference scenario the expected signal in the 0j channel is then
\dis{
{\rm E}_{\mathcal{F} 0}^{\mathcal{HP}} &= \frac{s^{\mathcal{H}}_{\mathcal{F} 0} (1 - x_{\mathcal{F}}^{\mathcal{HP}})}{\mathcal{B}^{\mathcal{H}} \sigma^{\mathcal{H}}}
\underbrace{\sin^2\vartheta_A \sigma^{\mathcal{P}}}
_{\sigma(gg\to a_2)}
\mathcal{B}_{a_2}^{\mathcal{P}} 
\underbrace{\frac{A^{\mathcal{P}}\cos^2\vartheta_A }{A^{\mathcal{P}} \cos^2\vartheta_A + B^{\mathcal{P}}}}_{{\rm Br}(H^\pm \to a_1 W^{\pm})} a_{\mathcal{F}, {\rm rel}}^{\mathcal{HP}}~,
}
where the exact $\vartheta_A$ dependence has been factored out. Here $s_{\mathcal{F} 0}^{\mathcal{H}}$ is the number of expected events for each of the six SM Higgs mass hypotheses $\mathcal{H}$ in each channel $\mathcal{F}0$ in Tab.~4 of the CMS note~\cite{cmshww}. $\mathcal{B}^{\mathcal{H}} \sigma^{\mathcal{H}}$ is the production cross-section times branching ratio for that SM Higgs.
The production cross-section times branching ratio for $gg\to a_2 \to H^\pm W^\mp$ is given by $\sin^2\vartheta_A \sigma^{\mathcal{P}} \mathcal{B}_{a_2}^{\mathcal{P}}$. In the branching ratio for $H^\pm \to a_1 W^{\pm}$ we factor out the  $\cos^2\vartheta_A$ dependence and define $A^{\mathcal{P}} = \Gamma(H^\pm \to A_H^1 W^\pm)$ and $B^{\mathcal{P}} = \Gamma(H^\pm \to\!\!\!\!\!\!/ \ \  a_1 W^\pm)$, where $A_H^i$ is the pure $A_H$ interaction state with the mass of $a_i$. $x_{\mathcal{F}}^{\mathcal{HP}}$ is the fraction of events that have one more jet (in addition to those from initial or final state QCD radiation) passing the jet selection due to the decay of $a_1$. Here these events are therefore removed from the $0j$ channel and appear in the $1j$ channel. $a_{\mathcal{F}, {\rm rel}}^{\mathcal{HP}}$ is the relative acceptance for our signal and, for each Higgs mass hypothesis $\mathcal{H}$, is defined as the ratio of the fraction of $p p \to a_2 \to WW a_1$ events that survive a given cut $\mathcal{H}$ to the fraction of SM Higgs events that survive the same cut. Both of these numbers depend on $\mathcal{F}$ since extra cuts are applied in the SF channels. The exact definition of this relative acceptance is
\dis{
a_{\mathcal{F}, {\rm rel}}^{\mathcal{HP}} &= \frac{(\#{\rm ~of~events~passing~the~cut~/~total~\#~of~events~before~the~cut})_{{\rm NP}(\cc{P})}}{(\#{\rm ~of~events~passing~the~cut~/~total~\#~of~events~before~the~cut})_{{\rm SM}(\cc{H})}}~. \label{releff}
}

For the expected signal in the $1j$ channels, we obtain
\begin{align}
{\rm E}_{\mathcal{F} 1}^{\mathcal{HP}} &= \frac{s^{\mathcal{H}}_{\mathcal{F} 0} x_{\mathcal{F}}^{\mathcal{HP}} + s^{\mathcal{H}}_{\mathcal{F} 1} (1 - x_{\mathcal{F}}^{\mathcal{HP}})}{\mathcal{B}^{\mathcal{H}} \sigma^{\mathcal{H}}}\sin^2\vartheta_A \sigma^{\mathcal{P}} \mathcal{B}_{a_2}^{\mathcal{P}} \frac{A^{\mathcal{P}}\cos^2\vartheta_A }{A^{\mathcal{P}} \cos^2\vartheta_A + B^{\mathcal{P}}} a_{\mathcal{F}, {\rm rel}}^{\mathcal{HP}}~.
\label{ef1}
\end{align}
To obtain the values of $a_{\mathcal{F}, {\rm rel}}^{\mathcal{HP}}$ and $x_{\mathcal{F}}^{\mathcal{HP}}$, we used MadGraph 5~\cite{madgraph5} where the dominant $g g \rightarrow \Phi$ production is written from FeynRules~\cite{feynrules}, and obtained consistent results with SHERPA 1.4.0~\cite{sherpa}. Since the kinematic cuts are independent of the interaction couplings (and thus $\tan\beta$ or $\cos\vartheta_A$), the $a_{\mathcal{F}, {\rm rel}}^{\mathcal{HP}}$ and $x_{\mathcal{F}}^{\mathcal{HP}}$ parameters depend only on the $\Phi$, $H^\pm$, and $A$ masses.
The width of $\Phi$ does technically depend on $\tan\beta$, but the $\tan\beta$-independent contribution from $\Phi\to H^\pm W^\mp$ is dominant whenever it is important (see App.~\ref{appen:wPhi}). The effects of the cuts do not depend strongly depend on the width of $H^\pm$.
Our simulated events do not include any jets from initial or final state radiation, which is the main source of $1j$ events in the SM case; using values for $x^{\cc{HP}}_{\cc{F}}$ and $a^{\cc{HP}}_{\cc{F},\rr{rel}}$ extracted from these simulations, especially in the $s^{\cc{C}}_{\cc{F}1}(1-x^{\cc{CP}}_{\cc{F}})a^{\cc{CP}}_{\cc{F},\rr{rel}}$ part of the above $1j$ channel equation, is therefore just a reasonable approximation, since the kinematic effects of initial and final state radiation are neglected. The ratio $x^{\cc{HP}}_{\cc{F}}$ measures the fraction of events with $n$ jets (due to initial or final state QCD radiation) that end up in the (n+1)--jets bin due to hadronic decay of the pseudoscalar Higgs into high-$p_T$ b--hadrons: in principle we expect the numerical value of this ratio to be different for the cases $n=0$ and $n=1$. Taking into account that we did not observe a strong sensitivity of the bounds we extract to the precise value of this ratio and that jet isolation requirements would imply a further reduction of the ratio for $n=1$ (thus increasing the number of expected signal events and strengthening the exclusion bounds), we believe that Eq.~(\ref{ef1}) represents a reasonable and conservative approximation.

%----------------------------
\begin{figure}[h!]
\includegraphics[width=0.48\linewidth]{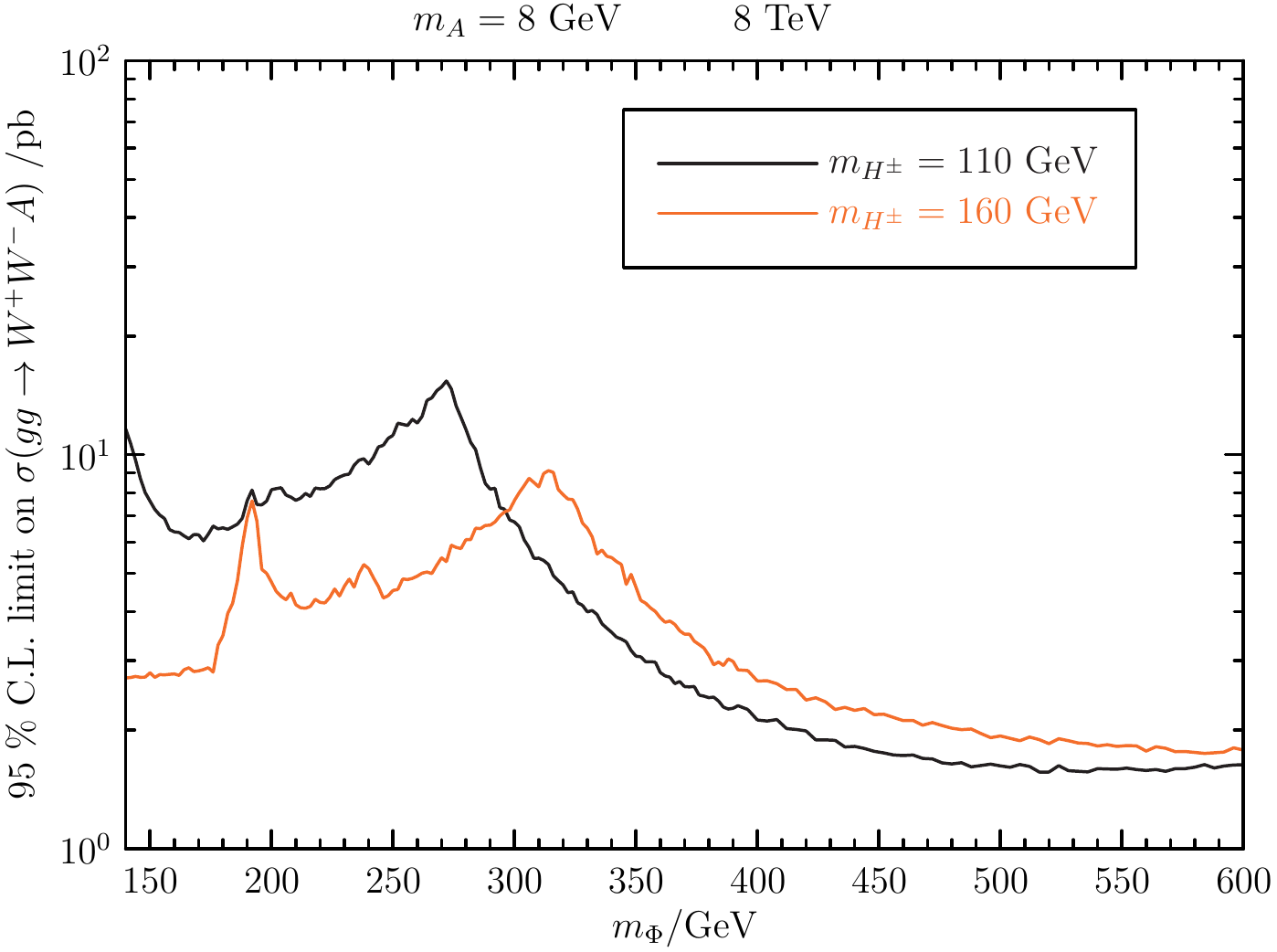}
\includegraphics[width=0.48\linewidth]{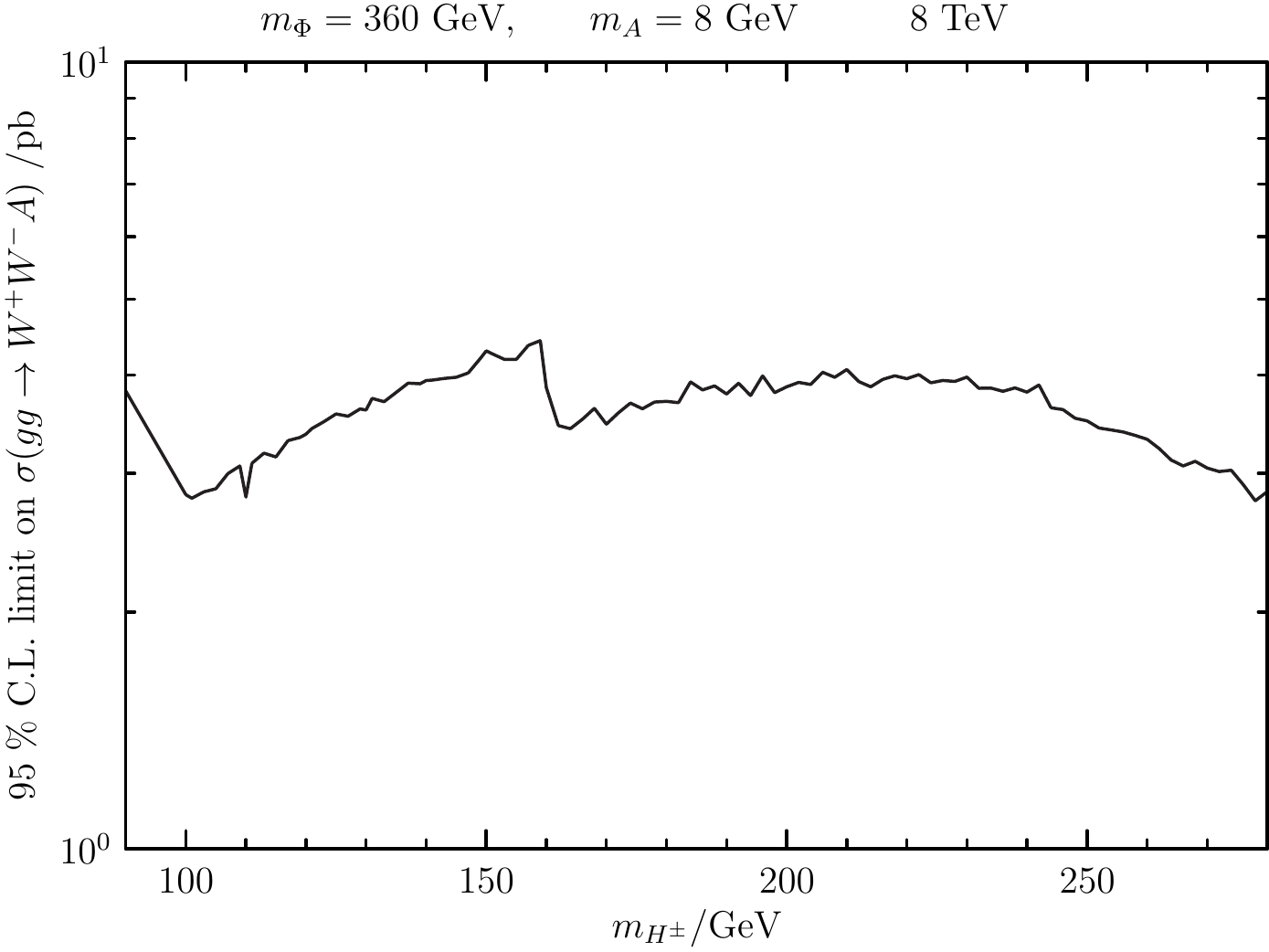}
\vskip2mm
\includegraphics[width=0.24\linewidth]{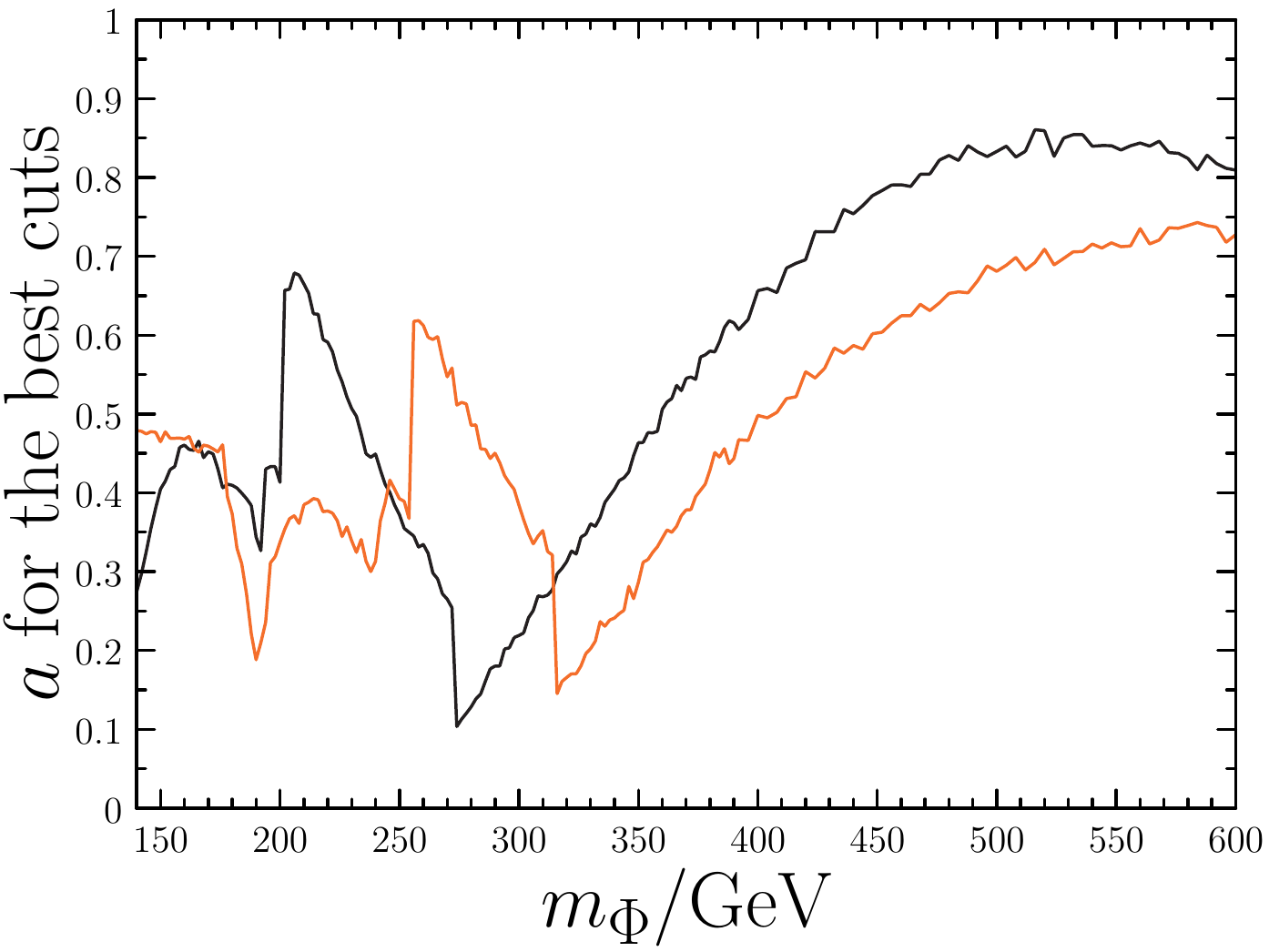}
\includegraphics[width=0.24\linewidth]{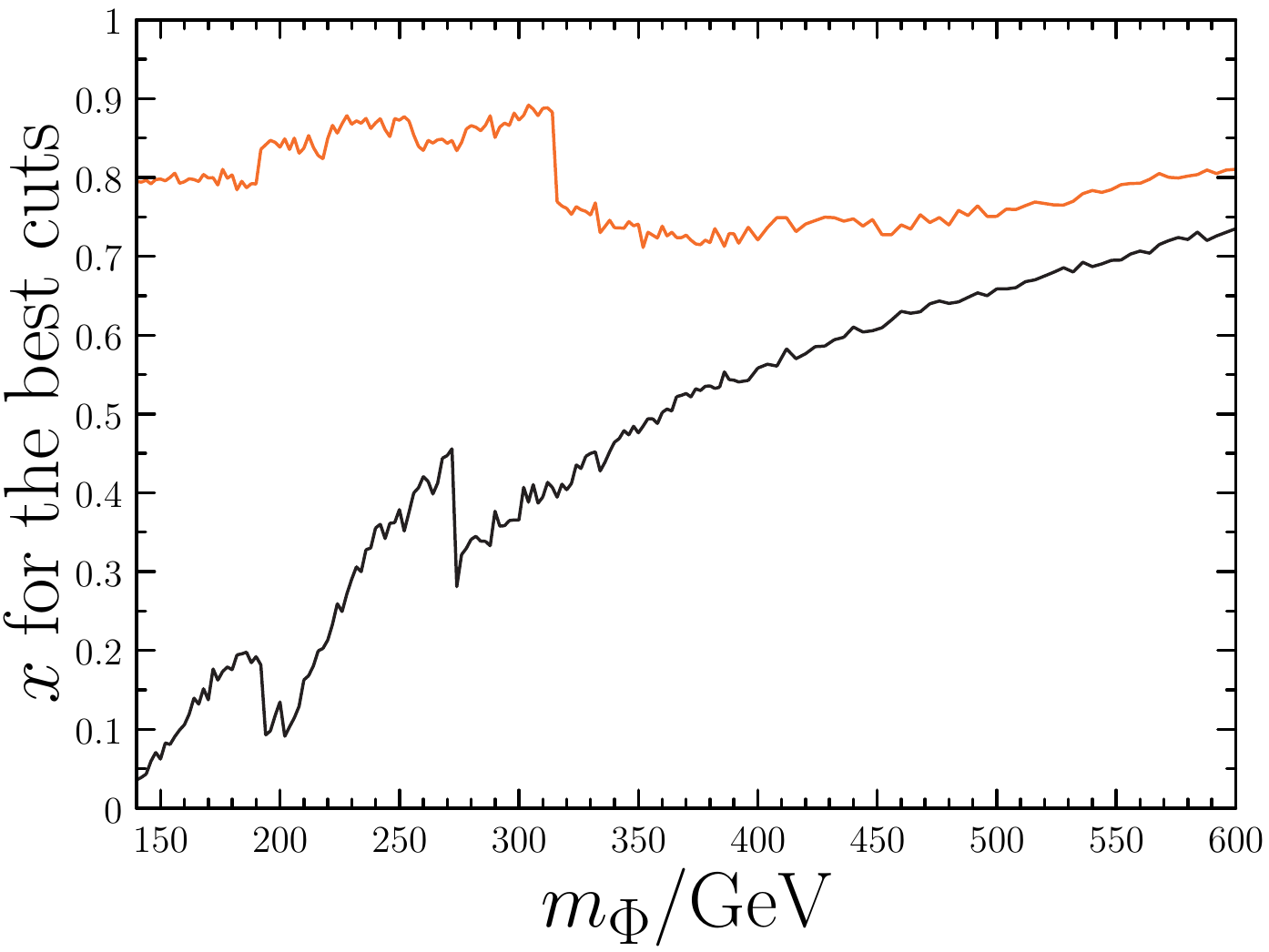}
\includegraphics[width=0.24\linewidth]{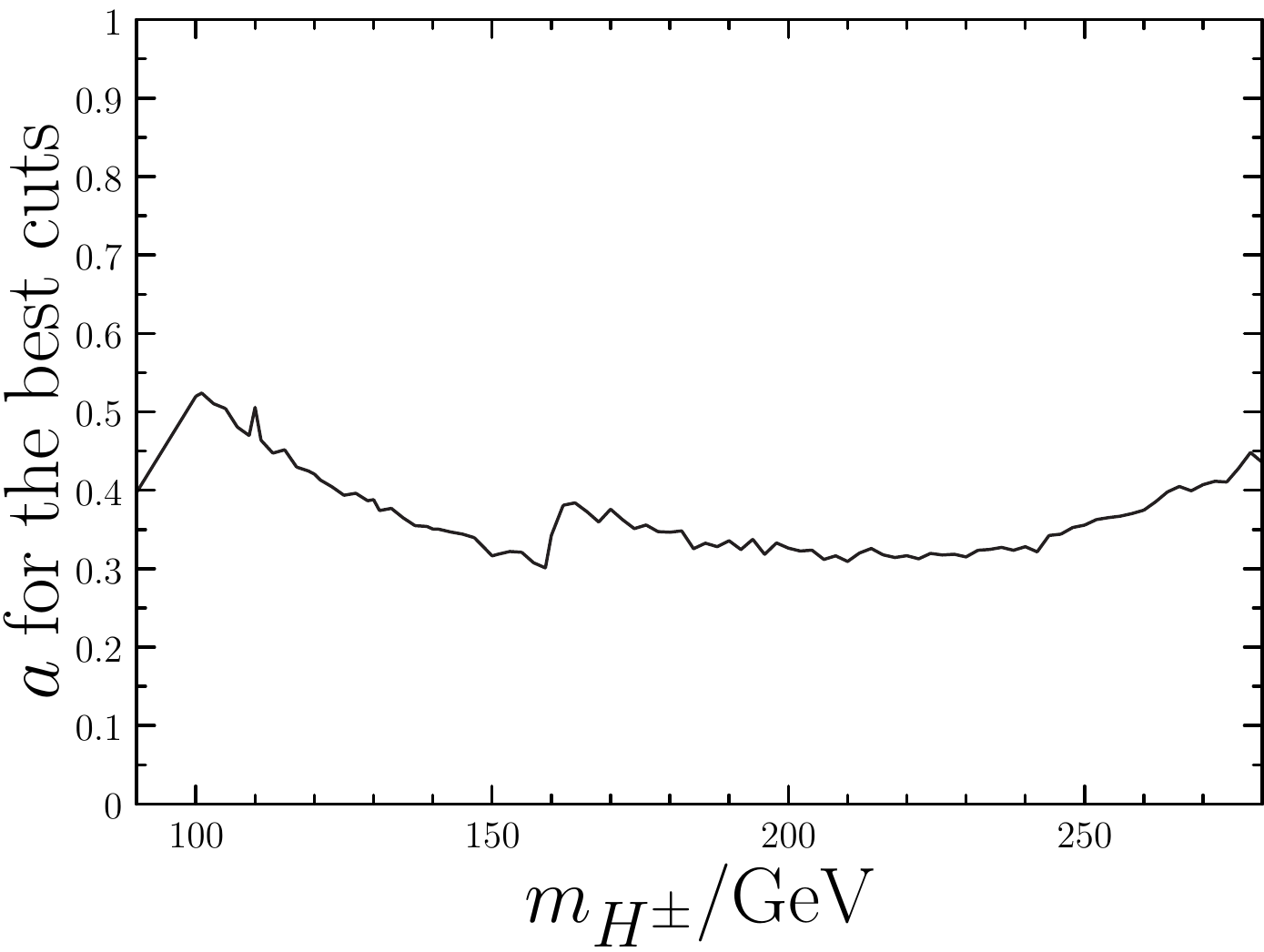}
\includegraphics[width=0.24\linewidth]{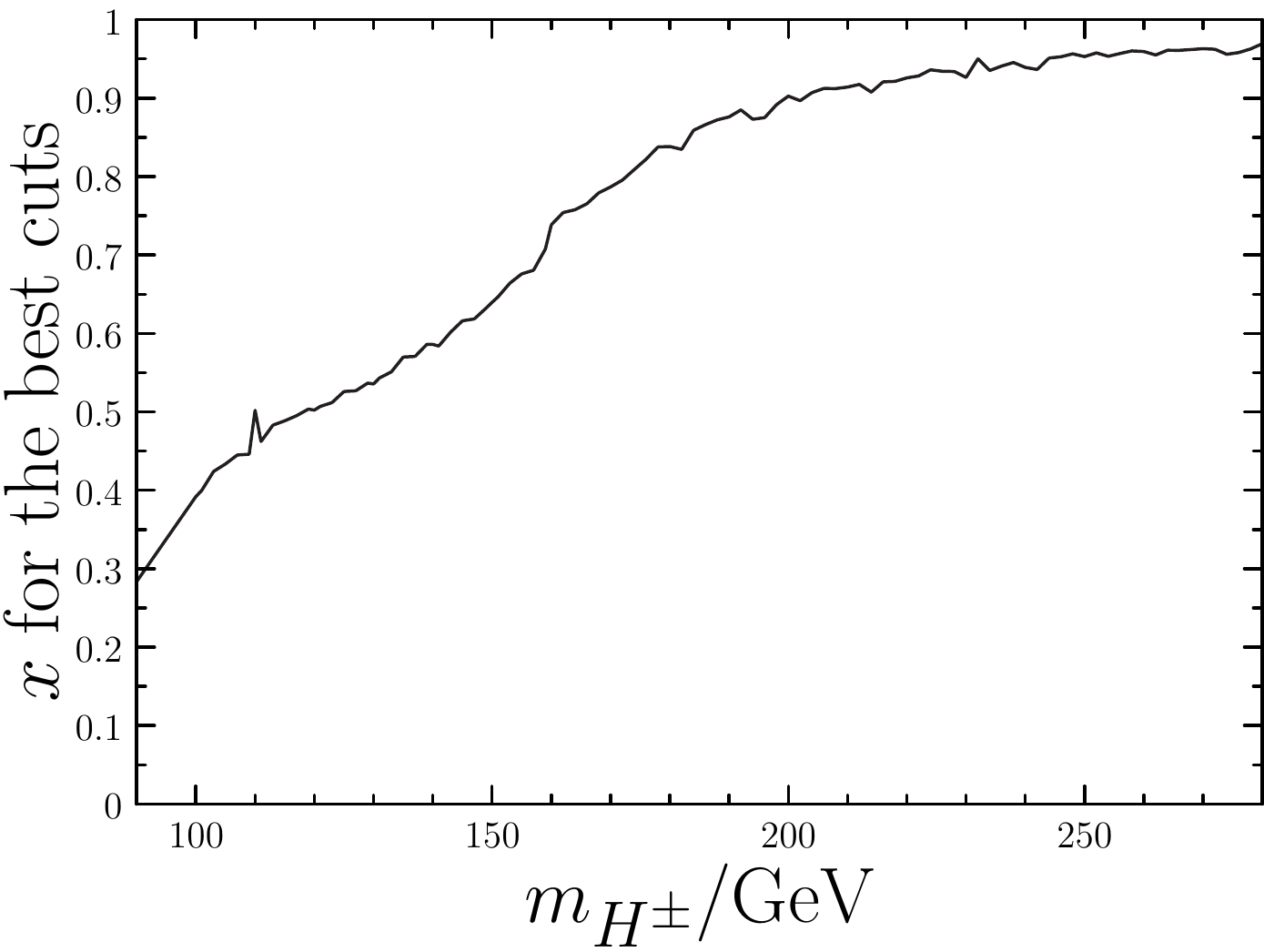}
\caption{Above: The 95\% CL$_{\rm s}$ limits on the production times branching ratios
calculated using $x^{\cc{HP}}_{\cc{F}}$ and $a^{\cc{HP}}_{\cc{F},\rr{rel}}$.
Below:~$a^{\cc{HP}}_{\cc{F},\rr{rel}}$ and $x^{\cc{HP}}_{\cc{F}}$
for the set of cuts $\cc{H}$ and channel $\cc{F}$ that sets the best limit.
The results turn out to be independent of $m_A$ to very good accuracy for the range that we consider.
\label{fig:sigmaBs}}
\end{figure}
%---------------------

For each 95~\%~CL$_{\rm s}$ limit $\ell^{\mathcal{H}}_{\mathcal{FJ}}$, derived as shown in App.~\ref{appen:cls}, we obtain the allowed parameter space by imposing
\dis{
{\rm E}_{\mathcal{FJ}}^{\mathcal{HP}} < \ell^{\mathcal{H}}_{\mathcal{FJ}}.
}
We apply whichever of these conditions leads to the best upper limit on the production cross-section times branching ratios for our signal.
These limits on cross-section times branching ratios are model independent in the sense that they apply to any model containing $\Phi$, $H^\pm$, and $A$ particles and depend only on the masses of these particles. Moreover, they do not depend on the $CP$ nature of the $\Phi$ and $A$ Higgs bosons because the $\Phi$ is produced on-shell and the structure of the $\phi\rightarrow \phi' V$ decay (where $\phi^{(\prime)}$ are spin-0) does not depend on the $CP$ nature of the $\phi^{(\prime)}$ (see App.~\ref{appen:width}). These cross-section limits are shown in the upper plots in Fig.~\ref{fig:sigmaBs} and they are superimposed on our reference scenario in Fig.~\ref{fig:prodbrs8}.
When deriving these limits $l$ we assume a fractional systematic error for the expected signal appearing in each channel of 30~\%, which we consider to be conservative (see App.~\ref{appen:cls}).
We find that the limits hardly vary with $m_A$ at all for the range that we consider. The peaks that appear in the left plot are due to us only having data for discrete values of the SM Higgs mass hypothesis. For instance, the most prominent peak corresponds to the $\Phi$ mass at which the 400~GeV cuts take over the 200 GeV cuts in providing the best upper limit. Currently only very low values of $\tan\beta~(\lesssim 2)$ can be constrained in our reference scenario. The strongest constraint is obtained near the $t\bar{t}$ threshold region, for this reason we choose $m_{a_2} = 360$~GeV as a reference point in the detailed parameter space study presented in the next subsection.

If the analysis were to be performed again using a more appropriate set of cuts for each set of masses the suppression due to the relative acceptance (see Eq.~(\ref{releff})) could certainly be reduced. In fact, since the SM Higgs to $WW$ signal and our signal are very similar, it is reasonable to presume that optimized cuts would lead to relative acceptances closer to unity. This would remove the peaks and slightly lower the baseline in the plot in Fig.~\ref{fig:sigmaBs}, leading to an order of magnitude improvement on the upper limit in some parts of the parameter space. Existing 8~TeV data could, therefore, be used to probe more moderate values of $\tan\beta$.
Estimating the possible sensitivity of a dedicated search at $\sqrt{s} = 14$~TeV is not simple, nonetheless the problem is one of distinguishing a signal over the uncertainty of the background. Assuming that with more data the background determination continues to be statistics limited and assuming that going from 8 to 14~TeV the background cross-section roughly doubles we can very roughly predict that at 14~TeV with 100~fb$^{-1}$ (500~fb$^{-1}$) of data a dedicated analysis could be sensitive to cross-sections of order 0.6~pb (0.3~pb), to be compared with the kinds of signals predicted in Figs.~\ref{fig:prodbrs14} and \ref{fig:prodbrseven14}. A proper analysis would need to be carried out by the experimental groups after collecting more data.

It is also worth pointing out that our $x^{\cc{HP}}_{\cc{F}}$ parameter is almost always closer to unity than to zero. In the SM search the limits coming from the $0j$ and $1j$ channels are comparable. In our case, however, the best limit almost always comes from the $1j$ channels, with the $0j$ channels setting much weaker limits. Almost as many events are moved out of the $1j$ channels due to the non-zero $x^{\cc{HP}}_{\cc{F}}$ than are moved from the $0j$ into the $1j$ channels, so the large $x^{\cc{HP}}_{\cc{F}}$ does not significantly increase the limits coming from the $1j$ channels; it just weakens the limits coming from the $0j$ channels. However, if one were to look at a $2j$ channel, with the same cuts as in the $0j$ and $1j$ channels, but requiring exactly two high $p_T$ ($>30$~GeV) jets, the situation could be different. Such a channel would not be useful for the SM Higgs to $WW$ search (the $2j$ channel discussed in the CMS analysis~\cite{cmshww} has completely different cuts and is designed to single out vector boson fusion production) and is therefore not considered in SM searches. However, for our process the probability to have two high $p_T$ jets even in the $gg$F production, one coming from initial or final state radiation and another coming from the $A$ decay, is significant. Such a $2j$ channel would also likely have a smaller background and could lead to better limits than the $1j$ channels for which we have data.

If we replace the $a_2$ with one of the $CP$-even states, $\Phi = h_2$, in our type-II 2HDM + singlet scenario the analysis is similar. In this case there is, however, another independent parameter, the $H$ fraction in $h_2$, $\cc{U}_{2H}^2$. This affects the production of but not the decays of $h_2$ under the assumptions outlined in subsection~\ref{sec:heavy}.

%%%%%%%%%%%%%%%%%%%%%%%%%%%%%%%%%%%%%%%%%%%%%%%%%%%%%%%%%%%%%%%%%%%%%%%%%%
\subsection{The type-II 2HDM plus singlet case}
\label{sec:param}

%------------------
\begin{figure}
\includegraphics[width=0.44\linewidth]{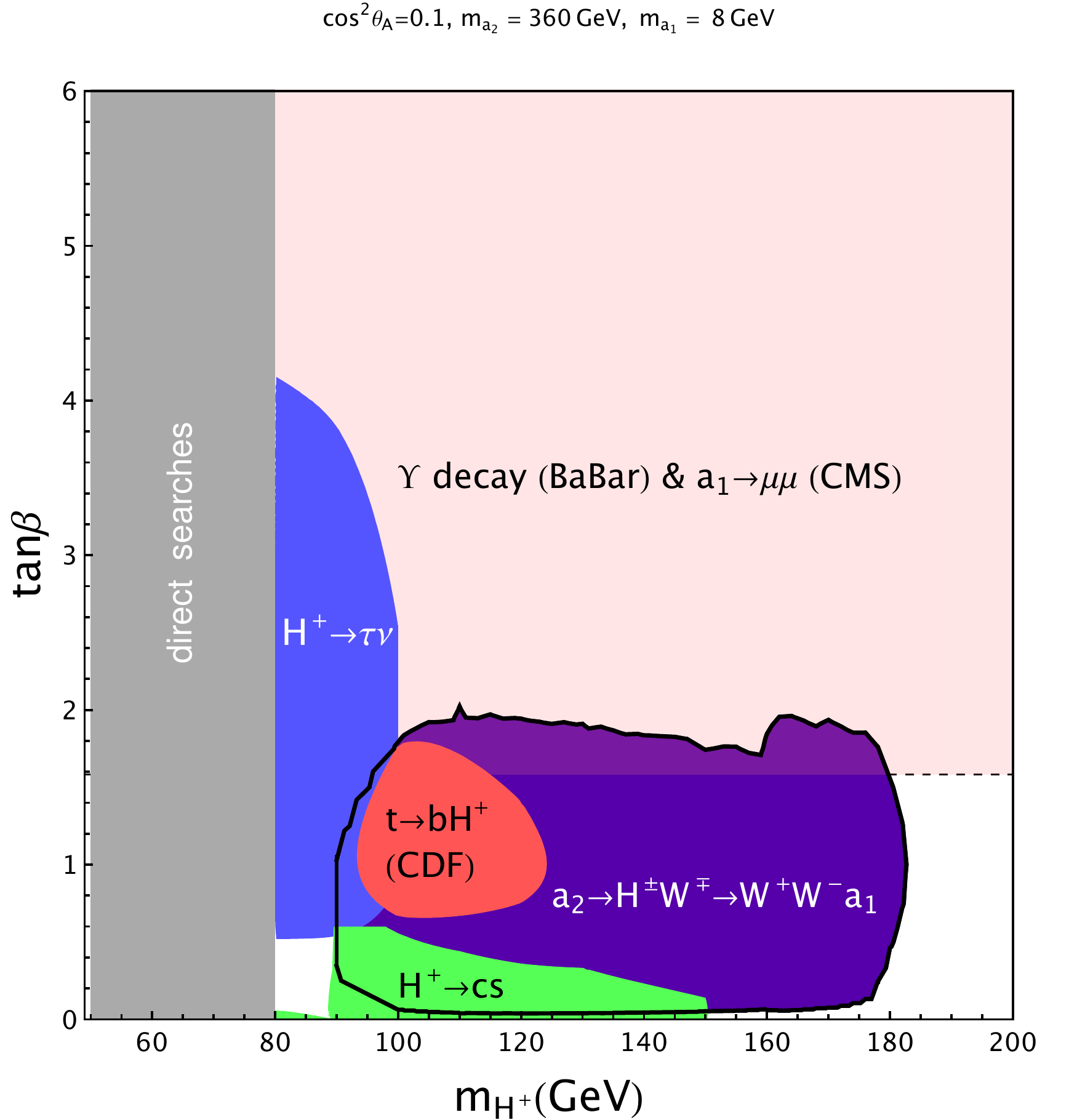}
\includegraphics[width=0.44\linewidth]{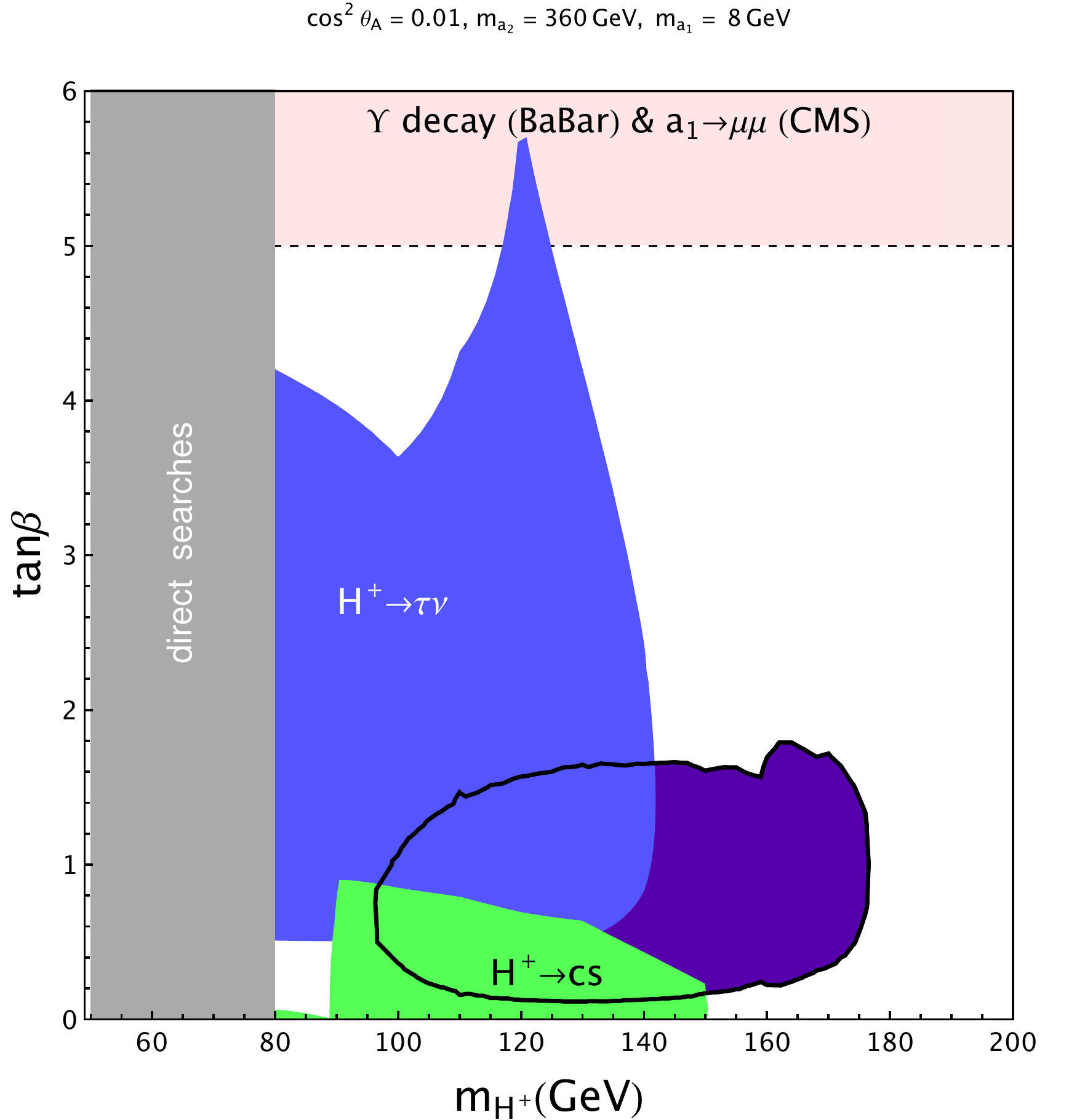}
\vskip4mm
\includegraphics[width=0.44\linewidth]{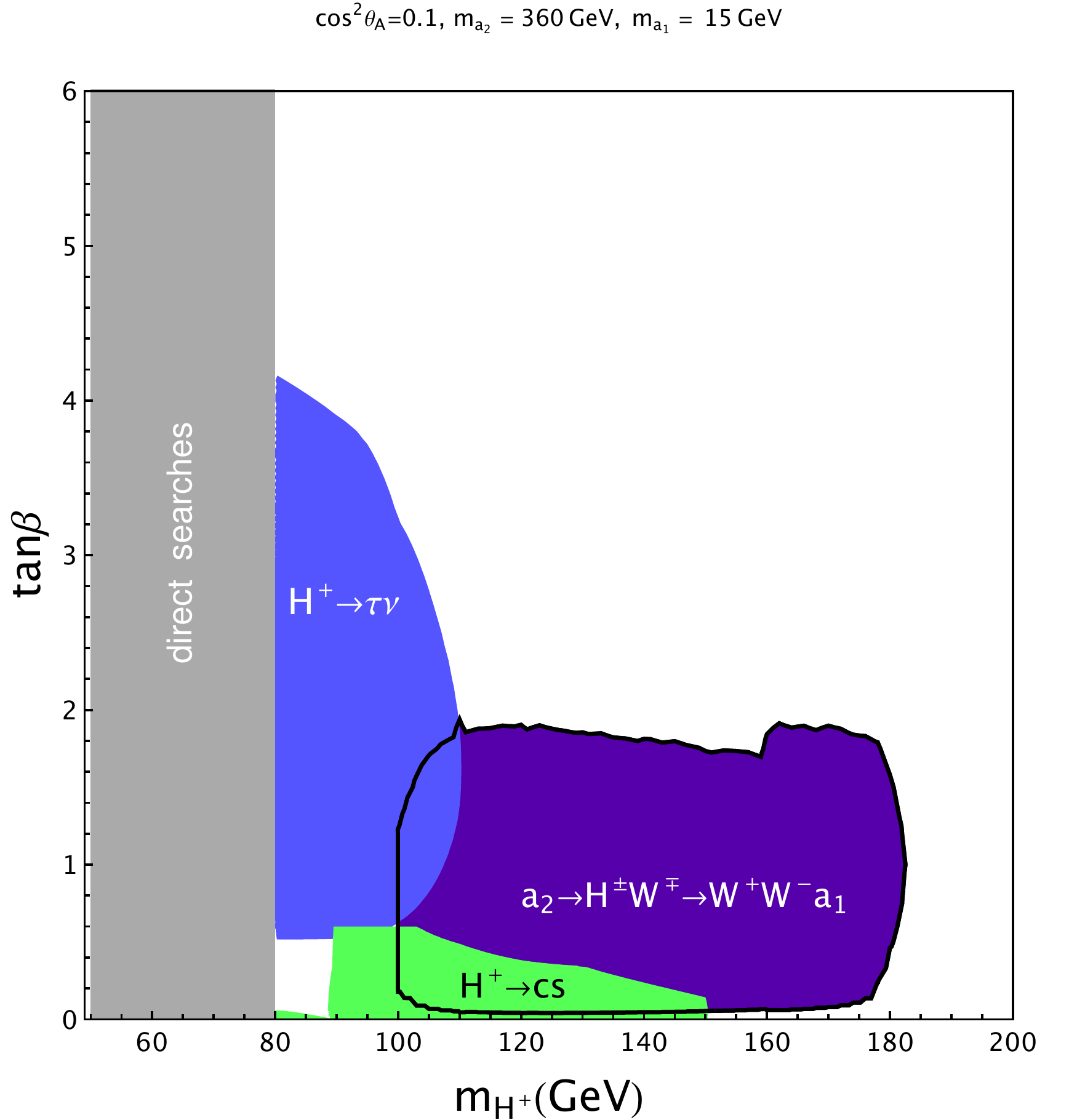}
\includegraphics[width=0.44\linewidth]{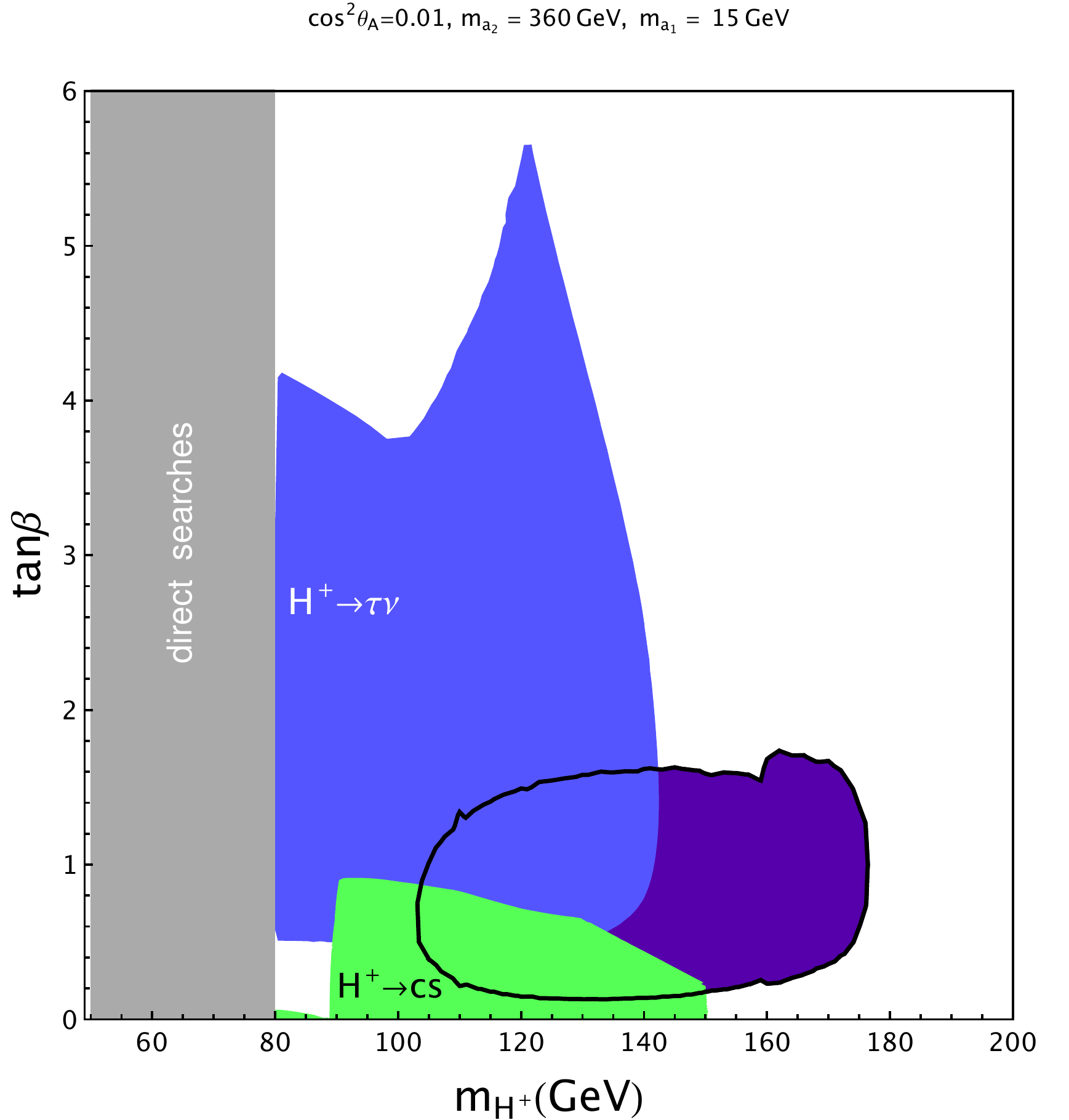}
\caption{The comparison with the result in \cite{Dermisek:2012cn} in terms of $\tan\beta$ and $m_{H^+}$ when $m_{a_2} = 360$ GeV. The additionally excluded parameter region  by our search for $gg \to a_2 \to W^+ W^- a_1$ is shown as purple area surrounded by the thick black line. We choose the mass $m_{a_2} = 360$~GeV as a reference point since the strongest constraint can be obtained nearby the $t\bar{t}$ threshold as mentioned in Sec.~\ref{sec:lim}. The red region is excluded by a direct $t \to bH^+ \to bW^+ a_1 \to bW^+ \tau^+ \tau^-$ search at CDF~\cite{Aaltonen:2011aj}. The white region is not excluded. Above: the mass of the light neutral Higgs $m_{a_1} = 8$~GeV, which is constrained by the $\Upsilon$ decay at BaBar and $a_1 \rightarrow \mu\mu$ at CMS, represented by the light pink region (above the dashed line).
The regions excluded by searching for $\tau \nu$ and $cs$ final states are shown in the blue and green respectively. Below: the mass of the light neutral Higgs $m_{a_1} = 15$~GeV, which is free from the BaBar and CMS bounds.
\label{fig:tanbefig2}}
\end{figure}
%-----------------

As explained in the previous section, SM Higgs $WW$ searches allow one to place model independent constraints on a charged Higgs produced in the decay of a heavy neutral Higgs and decaying to $W^\pm A$, where $A$ is a generic light neutral Higgs. In this section we apply the results presented in Sec.~\ref{sec:lim} to the special case of a type-II 2HDM with an extra SM singlet. In the context of this model the limits worked out in Sec.~\ref{sec:lim} apply at relatively low $\tan\beta$ ($\lesssim 2$). 

In Fig.~\ref{fig:tanbefig2} we show the limits we obtain for $m_{a_2} = 360$~GeV. As explained in the previous section we choose $m_{a_2} = 360$~GeV as a reference point because the constraints we obtain are the strongest around the resonance region $m_{a_2} \sim 2 m_t$. The figure shows the excluded regions in the $(m_{H^\pm},\tan\beta)$ plane for various values of $m_{a_1}\in\{8,15\}$~GeV and $\cos^2\vartheta_A \in \{0.1,0.01\}$. The grey region is excluded by direct searches at LEP~\cite{Heister:2002ev, Achard:2003gt, Abdallah:2003wd, Abbiendi:2008aa, Searches:2001ac}. The blue and green regions are excluded by Tevatron and LHC searches in the $\tau\nu$~\cite{Abazov:2009wy, Aad:2012tj, Chatrchyan:2012vca} and $cs$~\cite{ATLAS-cs-thesis} final states, respectively. The pink region is excluded by a combination of searches at BaBar~\cite{Aubert:2009cp, Aubert:2009cka} ($\Upsilon_{3s} \to a_1 \gamma$ channel) and at the LHC~\cite{Atlas_lightA,Chatrchyan:2012am} (direct $gg \to a_1 \to \mu\mu$ production); this pink exclusion only applies for $m_{a_1}$ just below the $b\bar{b}$ threshold and not for $m_{a_1}$ just above. The red area is excluded by a dedicated $t \to bH^+ \to bW^+ a_1 \to bW^+ \tau^+ \tau^-$ search at CDF~\cite{Aaltonen:2011aj}. The purple area surrounded by the thick black solid line is the additional region of parameter space excluded by our study in the $gg \to a_2 \to W^+ W^- a_1$ channel.

At lower values of $\cos^2\vartheta_A$ the exclusion region narrows due to the $\cos^2\vartheta_A$ dependence of BR($H^\pm \rightarrow W^\pm a_1$) (see the discussion in Sec.~\ref{sec:charged}). In particular, for $( m_{a_1}, \cos^2\vartheta_A ) =$ (8 GeV, 0.1), the light charged Higgs parameter region analyzed in Ref.~\cite{Dermisek:2012cn} is completely excluded (if a heavy Higgs with mass $m_{a_2} = 360$ GeV is present). On one hand, at low values of $\tan\beta \lesssim 0.03$ we lose sensitivity because the $a_2$ width becomes dominated by $a_2 \to t\bar t$. On the other hand, at large $\tan\beta \ge 10$ either the $a_2$ production cross-section or BR($a_2 \rightarrow W^+ W^- a_1)$ are suppressed and our search loses sensitivity. 

%----------------------------------------------
\begin{figure}
\includegraphics[width=0.49\linewidth]{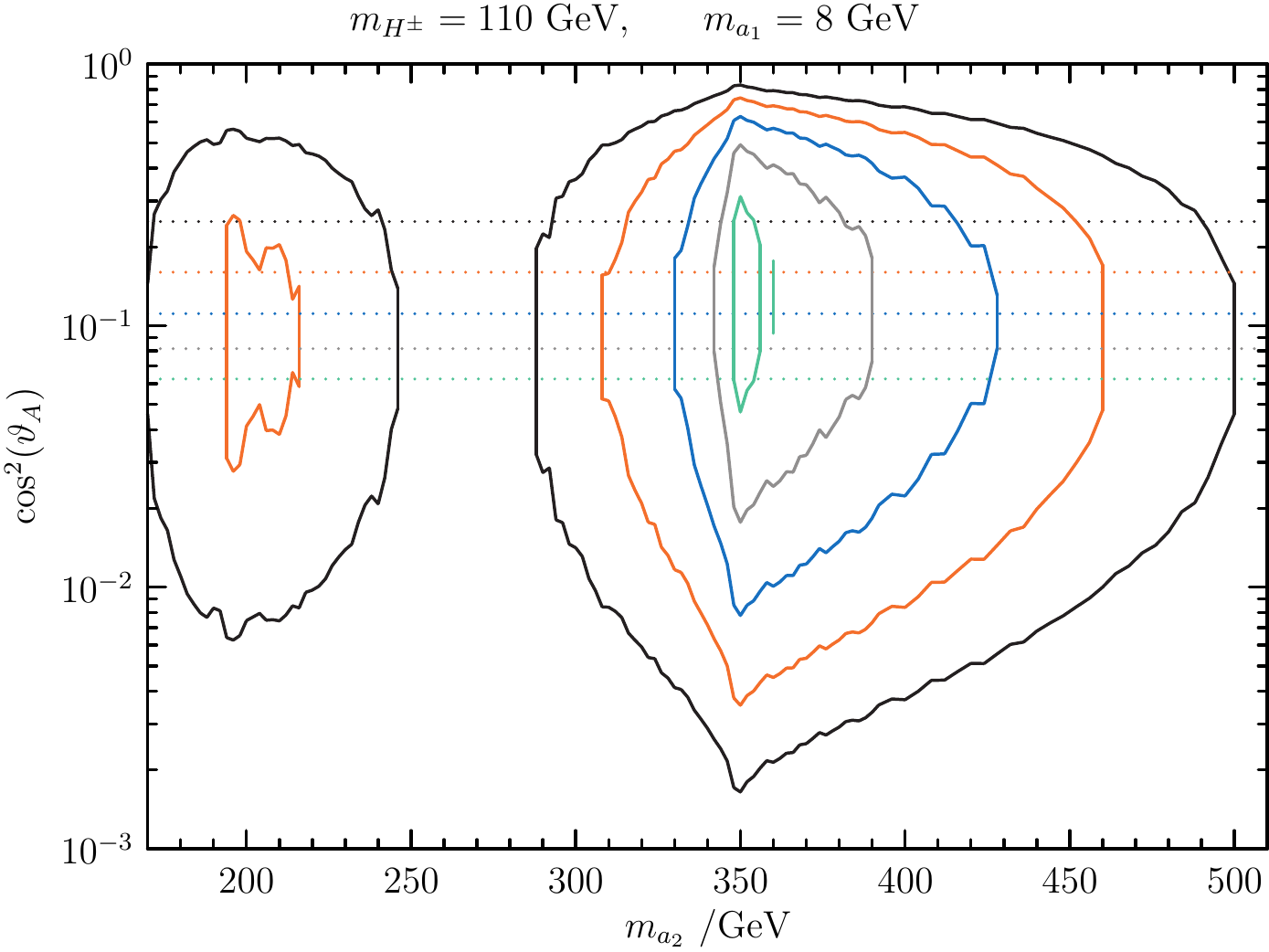}
\includegraphics[width=0.49\linewidth]{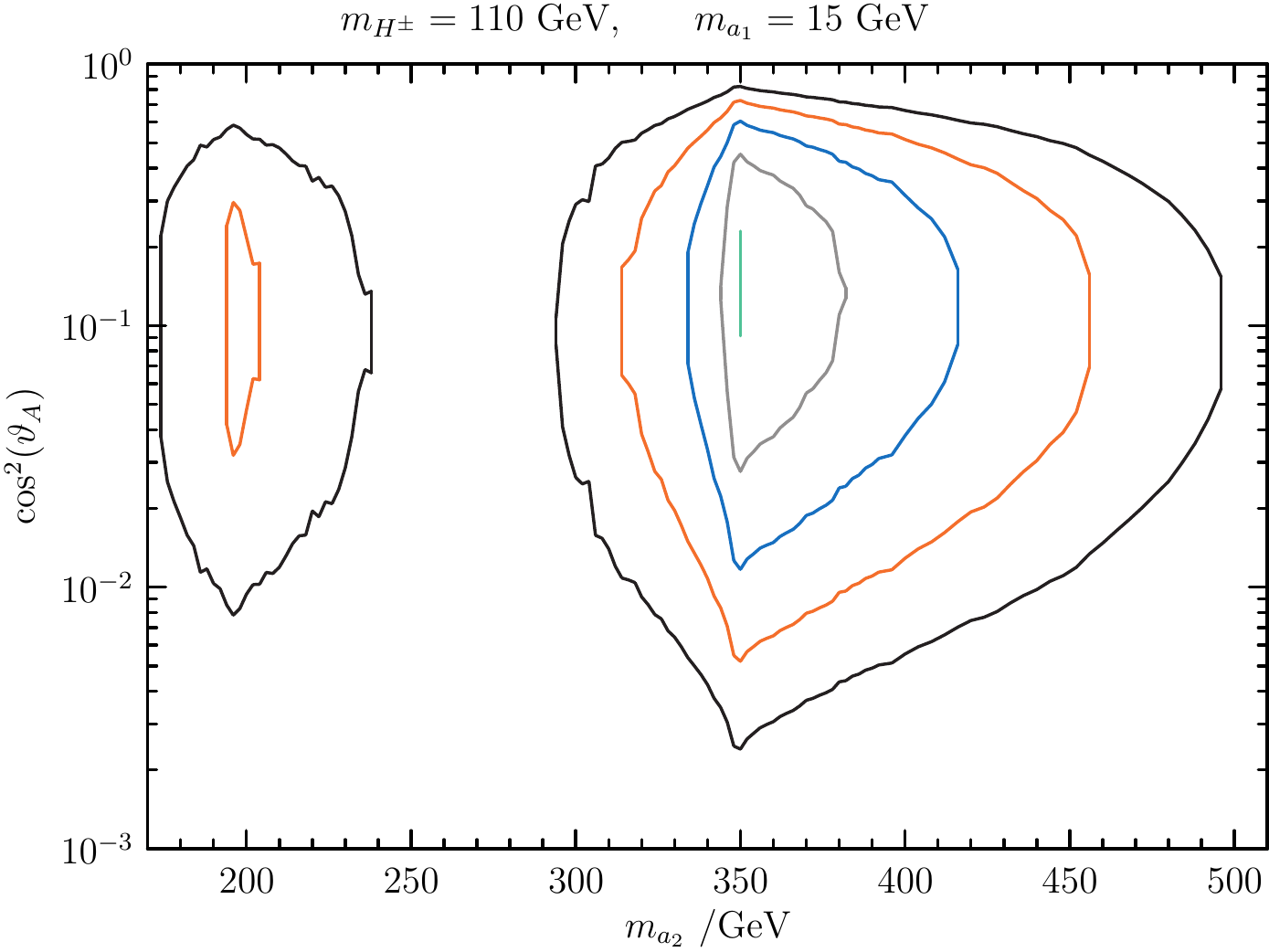}
\vskip-4mm
\raisebox{-\height}{\includegraphics[width=0.49\linewidth]{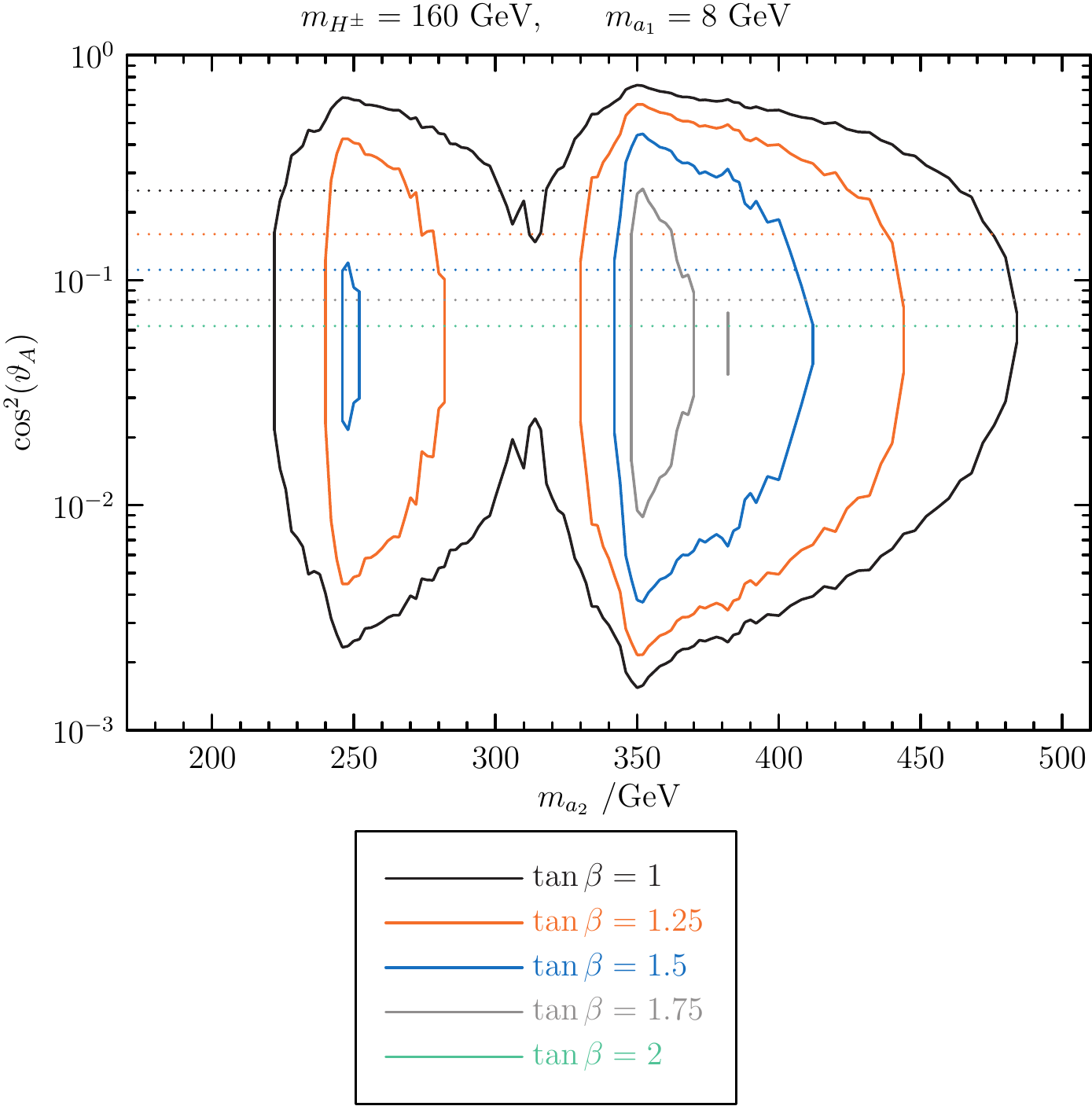}}
\raisebox{-\height}{\includegraphics[width=0.49\linewidth]{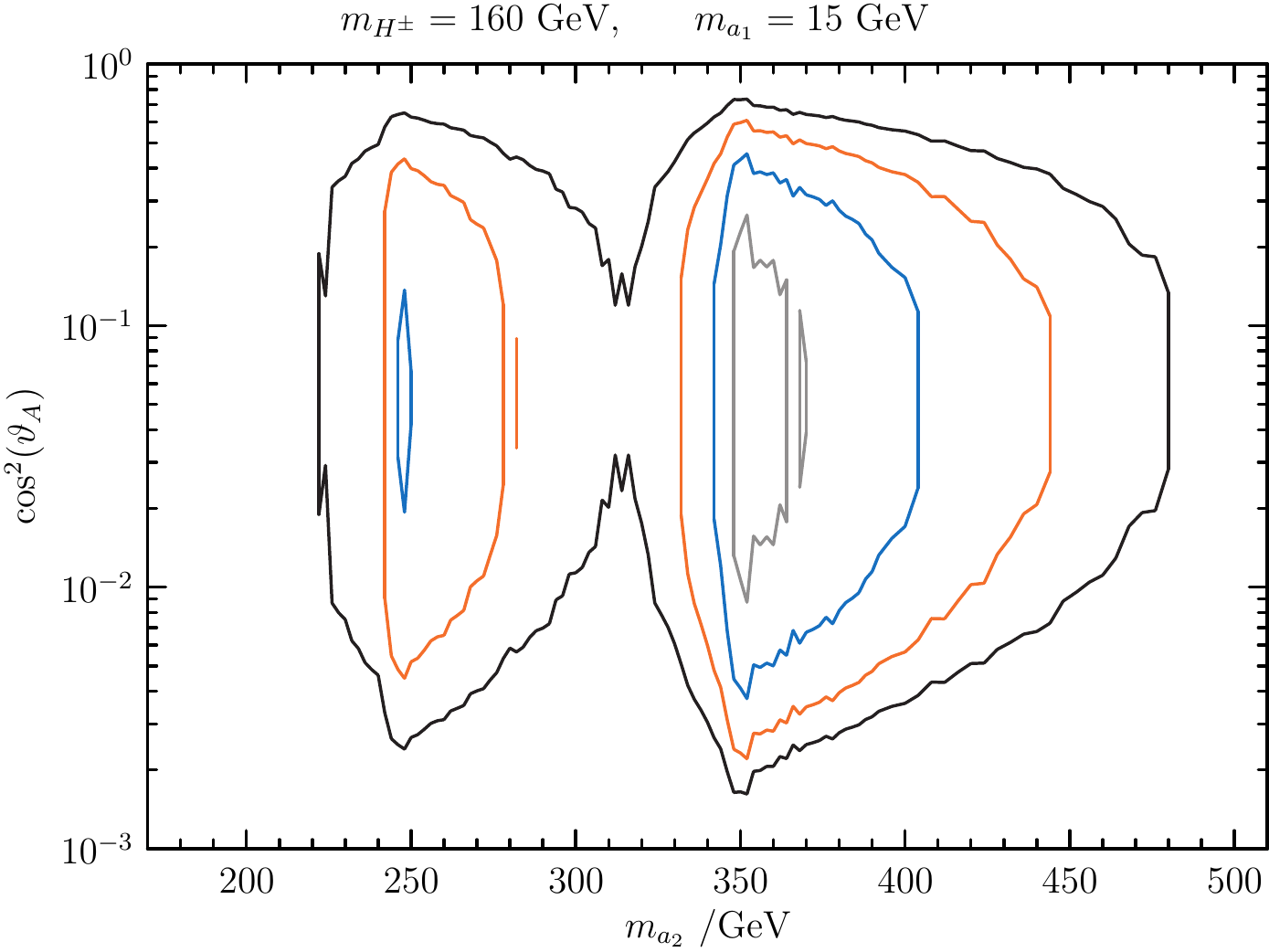}}
\caption{
For $a_1$ below the $b\bar{b}$ threshold, above the dotted lines is ruled out by negative results from
searches for $\Upsilon\rightarrow a_1\gamma\rightarrow
(\tau\tau,\mu\mu)\gamma$~\cite{Aubert:2009cp, Aubert:2009cka} and $gg\rightarrow a_1\rightarrow
\mu\mu$~\cite{Atlas_lightA,Chatrchyan:2012am},
which together roughly constrain $\tan\beta\cos\vartheta_A\lesssim 0.5$ \cite{Dermisek:2010mg,Dermisek:2012cn}.
Inside the contours (below the lines in the bottom two graphs) is ruled out
from the CMS 8~TeV SM Higgs to $WW$ search
at 95~\% C.L. by our analysis.
\label{fig:res}}
\end{figure}
%---------------------------------------------

Our study extends also to charged Higgs masses above the $tb$ threshold. Unfortunately, sensitivity in this region is not currently very strong for the following two reasons. First, in this region the $H^\pm \rightarrow W^\pm a_1$ branching ratio is suppressed at low $\tan\beta \lesssim 2$ and very large $\tan\beta \gg 10$ unless the charged Higgs mass is fairly large (see Fig.~\ref{fig:brh}).
Second, as the charged Higgs mass increases, the phase space for the $a_2 \to H^\pm W^\mp$ decay shrinks; this can be compensated by raising the $a_2$ mass at the price of a reduced production cross-section. In conclusion, we do not currently find appreciable constraints for $m_{H^\pm} \gtrsim 180$~GeV. This heavy charged Higgs parameter space could be constrained in the future with more data.

In Fig.~\ref{fig:res} we show regions that we exclude in the $(m_{a_2},\cos^2\vartheta_A)$ plane at fixed values of $m_{H^\pm}\in\{110,160\}$~GeV, $m_{a_1} \in \{8,15\}$~GeV, and $\tan\beta$. The region above the dotted line is excluded by direct $a_1$ searches at BaBar and at the LHC ($\tan\beta\cos\vartheta_A\lesssim 0.5$~\cite{Dermisek:2010mg,Dermisek:2012cn}) when $m_{a_1}$ is just below the $b\bar{b}$ threshold. The reason for the weakening of the limits for intermediate $a_2$ masses in Fig.~\ref{fig:res} is purely due to the fact that we have data for the cuts corresponding to SM Higgs mass hypotheses of 200~GeV and 400~GeV, but nothing in between. This then causes the peaks of weakening limits in Fig.~\ref{fig:sigmaBs} and the effects can be seen in Fig.~\ref{fig:res}. (See also Fig.~\ref{fig:prodbrs8}.)

%%%%%%%%%%%%%%%%%%%%%%%%%%%%%%%%%%%%%%%%%%%%%%%%%%%%%%%%%%%%%%%%%%%%%%%%%%
\section{Conclusions}
\label{sec:conclusions}

The experimental discovery at the LHC of a particle compatible with the SM Higgs boson is the first step towards a full understanding of the electroweak symmetry breaking mechanism. Assuming that the particle discovered at the LHC is a fundamental scalar, it becomes imperative to figure out what exactly the Higgs sector is. Many beyond-the-SM scenarios contain a second Higgs doublet and predict the existence of at least one charged Higgs and several neutral $CP$-even and -odd Higgs bosons. Most experimental searches have been conducted under the rather traditional assumption that the charged Higgs dominantly decays into $\tau\nu$ or $c\bar s$ pairs at low-mass ($m_{H^\pm} \lesssim m_t$) and into $t\bar b$ otherwise. The existence of a light neutral Higgs $A$ opens the decay channel $H^+ \to W^+ A$ and offers new discovery venues.

In this paper, we study a charged Higgs whose production mechanism relies on a heavy neutral Higgs ($\Phi$) and whose dominant decay is into a light neutral Higgs ($A$)
\be
pp \to \Phi \to W^\mp H^\pm \to W^+ W^- A. \label{proc}
\ee
For $m_A \gtrsim 2 m_b$, this particle decays dominantly to pairs of $b$ quarks that are detected, at sufficiently high $p_T$, as a single jet. Under these conditions, the final state is simply $W^+ W^-$ plus jets and is, therefore, constrained by SM Higgs searches in the $WW$ channel (this is also mostly true for $m_A$ below the $b\bar{b}$ threshold). For $m_A \lesssim 2 m_b$, the $A$ dominantly decays into $\tau$ pairs, whose decay products will also mostly be clustered into a singlet jet unless both $\tau$'s decay leptonically. (The latter case provides no extra jets and extra isolated leptons that would lead to the event not passing the selection criteria in the CMS analysis. This may however be another useful signal to search for.) Using existing data on searches for a SM Higgs in the range 128--600~GeV we are able to place constraints on this new physics process. In particular, we find that the upper limit on the production cross-section times branching ratios for the process in (\ref{proc}) are in the $\mathcal{O}$(1--10~pb) range for a wide range of $\Phi$, $H^\pm$ and $A$ masses. The results (presented at the top of Fig.~\ref{fig:sigmaBs}) depend very loosely on the details of a given model and will be useful to constrain a vast array of theories that contain three such particles. In particular the limits depend only on the masses of the three particles and not on the $CP$ nature of $\Phi$ and $A$. For the sake of definiteness we specialize our results to an explicit type-II 2HDM plus singlet reference scenario and show that our results are able, at low $\tan\beta$, to exclude previously open regions of parameter space.

The constraints we derive are shown in Figs.~\ref{fig:tanbefig2} and \ref{fig:res}. They are limited both because we only have partial access to the relevant data and because the cuts used for the SM Higgs search are not quite optimized for the process we consider. We point out that a slight modification of the search strategy, using more appropriate cuts that depend on the hypothesized masses of $\Phi$ and $H^\pm$, would lead to better limits and would be sensitive to more moderate values of $\tan\beta$. Our analysis extends, in principle, to arbitrarily large charged Higgs masses. In practice, the parametric dependence of the production cross-section and branching ratios on the charged Higgs mass limits our present sensitivity to $m_{H^\pm} \lesssim 180$~GeV. However, the parameter space with a heavier charged Higgs could be constrained in the future at the 14~TeV LHC. We point out that once the contribution to production from $b \bar b$ fusion is taken into account alongside $gg$ fusion, sensitivity to all values of $\tan\beta$ in our reference scenario should be achieved at the 14~TeV LHC. With 100~fb$^{-1}$ of data we very roughly estimate that sensitivity to cross-sections of order 0.6~pb would be achieved, to be compared to the kinds of cross-sections predicted in Figs.~\ref{fig:prodbrs14} and \ref{fig:prodbrseven14}. A search for the process where the charged Higgs is produced in the same way but goes to $tb$ is also being considered~\cite{heavytb}.

Finally, let us comment on the possibility that our process might contribute sizably to the total $pp \to W^+W^-$ cross-section. A recent CMS measurement with $3.54 \; {\rm fb}^{-1}$ of integrated luminosity at 8 TeV, found a slight excess in this channel: $69.9 \pm 2.8 \pm 5.6 \pm 3.1$~pb against a SM expectation of $57.3^{+2.4}_{-1.6}$~pb without the inclusion of the SM Higgs contribution~\cite{ww}. Even after accounting for this an additional contribution of several pb seems to be required (see for instance Ref.~\cite{Curtin:2012nn} for a possible explanation of this tension in a supersymmetric framework). If this discrepancy survives, the process discussed in this paper could potentially offer a contribution of the correct order of magnitude.

%%%%%%%%%%%%%%%%%%%%%%%%%%%%%%%%%%%%%%%%%%%%%%%%%%%%%%%%%%%%%%%%%%%%%%%%%%%%%%%%%%%%%%%%%%%%%%%%%%%%%
\acknowledgments{R.D. is supported in part by the Department of Energy under grant number DE-FG02-91ER40661. S.S. is supported by the TJ Park POSCO Postdoc fellowship and NRF of Korea No. 2011-0012630. E.L. would like to thank Frank Siegert for substantial help with the Sherpa Monte Carlo.}

\vskip 0.5cm
%%%%%%%%%%%%%%%%%%%%%%%%%%%%%%%%%%%%%%%%%%%%%%%%%%%%%%%%%%%%%%%%%%%%%%%%%%%%

\appendix

%%%%%%%%%%%%%%%%%%%%%%%%%%%%%%%%%%%%%%%%%%%%%%%%%%%%%%%%%%%%%%%%%%%%%%%%%%%%%%%%%%%%%
\section{Decay rates~\footnote{
A more complete list of two- and three-body tree-level decays relevant in Higgs sector extensions containing doublets and singlets, along with accompanying \texttt{C++} code, will be presented in Ref.~\cite{jon}}}
\label{appen:width}

Let
\be
\lambda_{12}&=&(1-k_1-k_2)^2-4k_1k_2,
\ee
where $k_i = m_i^2/M^2$ and $M$ is the mass of the decaying particle, and let
\be
\beta_{1}=\sqrt{\lambda_{11}}&=&\sqrt{1-4k_1}.
\ee
Let us further define $x_i=2E_i/M$ and $\gamma_i=\Gamma_i^2/M^2$.

\subsection{$\Phi\rightarrow \phi W$}

Allowing the $W$ to be off-shell and assuming it can decay to all light fermions
(excluding tops), which we take to be massless, we can write
\be
&\Gamma(\Phi\rightarrow \phi W^*) =
\f{3G_F}{4\pi\sqrt{2}}M\f{M_W\Gamma_W}{\pi}
\int_{0}^{1-k_\phi}\rr{d}x_2
\int_{1-x_2-k_\phi}^{1-k_\phi/(1-x_2)}\rr{d}x_1&\nn\\&
\f{
(1 - x_1) (1 - x_2) - k_\phi
 }{
(1 - x_1 - x_2 - k_\phi + k_W)^2 + k_W \gamma_W}.&
\ee
Here 1 and 2 label the fermions from the $W$ decay~\footnote{
This is for one particular charge of $W$. The equivalent formula for a $Z$ boson is obtained by replacing $W\to Z$ everywhere. The formulae in Ref.~\cite{Djouadi:2005gj} (2.20) and Ref.~\cite{DKZ} (41,58,59) are a factor of 2 too large for the $W$ boson case, whereas the formula for the $Z$ boson case are correct. This is because $\delta Z$ (as defined in Ref.~\cite{Djouadi:2005gj}), rather than being the ratio of the $Z$ and $W$ widths times $\cos^3\vartheta_W$, contains an extra factor of $\nf{1}{2}$. This is the symmetry factor relevant for the $VV$ decays, but not the $\phi V$ decays. There is also a typo in the $\sin^4\vartheta_W$ term in $\delta Z$ in Ref.~\cite{Djouadi:2005gj}.}.
This formula is valid for
$A_H\rightarrow H^\pm W^\mp$,
$H\rightarrow H^\pm W^\mp$, and $H^\pm \rightarrow A_H W^\pm$. For
$a_2\rightarrow H^\pm W^\mp$ and $H^\pm \rightarrow a_1 W^\pm$, with the conventions defined in Sec.~\ref{sec:def}, there
is a suppression by $\sin^2(\vartheta_A)$ and $\cos^2(\vartheta_A)$
respectively.
Writing the integral in this way, the inner $x_1$ integration can be
performed analytically and
the remaining integrand
behaves well for numerical integration and the
outer integration over $x_2$ can
evaluated numerically very quickly.
For completeness, above threshold in the zero-width on-shell
approximation we can write~\footnote{
This is also for one particular charge of $W$ and the equivalent formula for a $Z$ boson is again obtained by replacing $W\to Z$. This on-shell formula in Ref.~\cite{Djouadi:2005gj} (2.18) contains a typo that makes it dimensionally inconsistent. The formulae in Ref.~\cite{DKZ} (38,39,51,57) are correct, except that (57) contains an erroneous factor of $\cos^2\vartheta_W$.}
\be
\Gamma(\Phi\rightarrow\phi W) &=&
\f{G_F}{8\pi\sqrt{2}}M^3\lambda_{\phi W}^{3/2}.
\ee
In this massless fermion approximation we can write
\be
\Gamma_W &=& \f{9G_FM_W^3}{6\pi\sqrt{2}}.
\ee

\subsection{Other $A_H$ decays}

For light quarks $q$
\be
\Gamma(A_H\rightarrow q\bar{q}) &=&
\f{3G_F}{4\pi\sqrt{2}}(g^{A_H}_q)^2Mm_q^2(1+\Delta_{qq}),
\ee
where
\be
\Delta_{qq} &=& 5.67\,\f{\alpha_{\rr{QCD}}(M)}{\pi} + (35.94-1.36\,n_f)\f{\alpha_{\rr{QCD}}(M)^2}{\pi^2}
,
\ee
$m_q$ is the running mass at the scale $M=m_{A_H}$, and $n_f$ is the QCD number of flavours at $M$. Further QCD corrections for the scalar and pseudoscalar decays to quarks are derived in Refs.~\cite{Larin:1995sq} and \cite{Chetyrkin:1995pd} and summarised in Ref.~\cite{Djouadi:2005gj}, but these are only valid in the heavy top mass limit, i.e.~when the boson is light compared to the top quark.

For charged leptons
$l$
\be
\Gamma(A_H\rightarrow l^+l^-) &=&
\f{G_F}{4\pi\sqrt{2}}\tan^2(\beta)Mm_l^2\beta_l.
\ee
\be
&\Gamma(A_H\rightarrow t\bar{t}^*\rightarrow t\bar{b}W^-) =
\f{3G_F^2}{64\pi^3}\f{M^3m^2_t}{\tan^2(\beta)}
\int\rr{d}x_{\bar{b}}
\int\rr{d}x_t\f{1}{(1-x_t)^2+k_t\gamma_t}&\nn\\&
\Big[-(1 - x_t)^2 (1 - x_t - x_{\bar{b}} - k_W + k_t)
+ 2 k_W ((1 - x_t) (1 - x_{\bar{b}}) - k_W)&\nn\\&
-   k_t ((1 - x_t) (1 - x_{\bar{b}}) - 2 (1 - x_t) - k_W - k_t)\Big].&
\ee
Here $m_b$ has been neglected in the integrand.
The leading QCD correction can be included
by using the running mass for the $m_t^2$ factor that appears out front, which
comes directly from the Yukawa coupling in the Feynman rule.
In the integrand and in the integration limits
the running mass is not used (for $k_t$ and $k_b$) so that the
threshold appears in the correct place~\footnote{
This formula is correct in Ref.~\cite{DKZ} (55,56), but the expression in Ref.~\cite{Djouadi:2005gj} (2.8) is a factor of 2 larger. The formula (2.8) as written is correct below threshold after one takes into account that either top can go off-shell, but is then a factor of 2 too large above threshold. Our approach is given in the text.}.
For three-body decays written in terms of the $x$s (energies) of two (1 and 2) of the three final states particles (1, 2, and 3) the kinematic limits are, without neglecting any masses,
\be
&2\sqrt{k_2} \leqslant x_2 \leqslant
1-k_3+k_2-k_1-2\sqrt{k_1k_3},&\\\nn\\
&x_1 \;\vphantom{A}^\geqslant_\leqslant\;
\f{
(1 - x_2 + k_1 + k_2 - k_3) \left(1 - \f{x_2}{2}\right)
\mp
\sqrt{\f{x_2^2}{4} - k_2}\sqrt{
(1 - x_2 + k_1 + k_2 - k_3)^2 - 4 k_1\left(1 - \f{x_2}{2}\right)^2
+ 4 k_1 \left(\f{x_2^2}{4} - k_2\right)}}
{1 - x_2 - k_2}.&\nn
\ee
\be
\Gamma(t\rightarrow bW^+) &\approx&
\f{G_F}{8\pi\sqrt{2}}m_t^3(1-k_W)(1+2k_W)\lambda_{bW}^{1/2}.
\ee
A formula for $\Gamma(A_H\rightarrow t^*\bar{t}^*\to bW^+\bar bW^-)$
that is valid both above and below
threshold can be obtained by doubling $\Gamma(A_H\rightarrow
t\bar{t}^*\rightarrow t\bar{b}W^-)$ and using $4\gamma_t$ in place of
$\gamma_t$.
Above threshold in the zero-width on-shell
approximation we can write
\be
\Gamma(A_H\rightarrow t\bar{t}) &=&
\f{3G_F}{4\pi\sqrt{2}}\f{Mm^2_t}{\tan^2(\beta)}\beta_t.
\ee

\subsection{Other $H$ decays}
\be
\Gamma(H\rightarrow q\bar{q}) &=&
\f{3G_F}{4\pi\sqrt{2}}(g^{A_H}_q)^2Mm_q^2(1+\Delta_{qq})\;,
\ee
where
\be
\Delta^2_{H} &=&
\f{\alpha_{\rr{QCD}}(M)^2}{\pi^2}
\left(1.57-\f{2}{3}\ln\left(\f{M^2}{m_t^2}\right)
+\nf{1}{9}\ln\left(\f{m_q^2}{M^2}\right)^2\right)\; ,
\ee
\be
\Gamma(H\rightarrow l^+l^-) &=&
\f{G_F}{4\pi\sqrt{2}}\tan^2(\beta)Mm_l^2\beta^3_l\;,
\ee
\be
&\Gamma(H\rightarrow t\bar{t}^*\rightarrow t\bar{b}W^-) =
\f{3G_F^2}{64\pi^3}\f{M^3m^2_t}{\tan^2(\beta)}
\int\rr{d}x_{\bar{b}}
\int\rr{d}x_t\f{1}{(1-x_t)^2+k_t\gamma_t}&\nn\\&
\Big[
-(1 - x_t)^2 (1 - x_t - x_{\bar{b}} - k_W + 5k_t)&\nn\\&
+ 2 k_W ((1 - x_t) (1 - x_{\bar{b}}) - k_W -2k_t(1-x_t) +4k_tk_W)&\nn\\&
- k_t (1-x_t)(1-x_{\bar{b}})
+ k_t (1 - 4k_t)(2(1-x_t)+k_W+k_t)
\Big]& \; .\footnotemark
\ee
\footnotetext[8]{Again this formula is correct in Ref.~\cite{DKZ} (48,49). The situation for Ref.~\cite{Djouadi:2005gj} (2.8) is the same as discussed in the previous footnote.}
Again, a formula for $\Gamma(H\rightarrow t^*\bar{t}^*)$
that is valid both above and below
threshold can be obtained by doubling $\Gamma(H\rightarrow
t\bar{t}^*\rightarrow t\bar{b}W^-)$ and using $4\gamma_t$ in place of
$\gamma_t$.
Above threshold in the zero-width on-shell
approximation we can write
\be
\Gamma(H\rightarrow t\bar{t}) &=&
\f{3G_F}{4\pi\sqrt{2}}\f{Mm^2_t}{\tan^2(\beta)}\beta^3_t.
\ee

\subsection{Other $H^\pm$ decays}
For light up- and down-type quarks $u$ and $d$, assuming $k_u,k_d \ll 1$,
\be
\Gamma(H^+\rightarrow u\bar{d}) &=&
\f{3G_F}{4\pi\sqrt{2}}M
(m_u^2\cot^2(\beta)+m_d^2\tan^2(\beta))
(1+\Delta_{qq}),
\ee
where $m_u$ and $m_d$ are running masses. For charged leptons $l$
\be
\Gamma(H^+\rightarrow l^+\nu_l) &=&
\f{G_F}{4\pi\sqrt{2}}\tan^2(\beta)Mm_l^2(1-k_l)^2.
\ee
One of the $(1-k_l)$ factors comes from the matrix-element-squared and the other is the phase-space factor $\sqrt{\lambda_{l\nu}}$.
\be
&\Gamma(H^+\rightarrow t^*\bar{b}\rightarrow b\bar{b}W^+) =
\f{3G_F^2}{32\pi^3}M^3
\int\rr{d}x_b
\int\rr{d}x_{\bar{b}}&\nn\\&
\Bigg[m^2_t\cot^2(\beta)\f{
-k_W^2 (1 + x_{\bar{b}}) - (1 - x_{\bar{b}})^2 (1 - x_{\bar{b}} - x_b) + 
 k_W (1 - x_{\bar{b}}) (3 - x_{\bar{b}} - 2 x_b)
 }{
(1 - x_{\bar{b}} - k_t)^2 + k_t \gamma_t}&\nn\\&
+m^2_bk_t\tan^2(\beta)\f{
(1-x_{\bar{b}}-k_W)(1-x_b-k_W)-k_W(1-x_{\bar{b}}-x_b-k_W)
 }{
(1 - x_{\bar{b}} - k_t)^2 + k_t \gamma_t}&\nn\\&
-2m_bm_t\sqrt{k_bk_t}\f{
(1-x_{\bar{b}}-k_W)(1-x_{\bar{b}}+2k_W)
 }{
(1 - x_{\bar{b}} - k_t)^2 + k_t \gamma_t}\Bigg].&
\ee
The leading QCD correction can be included
by using the running masses for the $m_t^2$, $m_b^2$, and $m_bm_t$ factors
that appears out front for each of the three terms, which
come directly from the Yukawa couplings in the Feynman rule.
Elsewhere in the integrand $m_b$ has been neglected~\footnote{
The $m_t^2$ term given in Ref.~\cite{DKZ} (63) seems to be incorrect, producing a different shape to our formula below threshold and not agreeing with the on-shell formula above threshold. All our formulae are checked to make sure that they reproduce the on-shell zero-width approximation formulae sufficiently above threshold, up to finite width effects.}.
Elsewhere in the integrand and in the integration limits
the running masses are not used (for $k_t$ and $k_b$) so that the
threshold appears
in the correct place.
Above threshold in the zero-width on-shell
approximation
\be
\Gamma(H^+\rightarrow t\bar{b}) &=&
\f{3G_F}{4\pi\sqrt{2}}M\sqrt{\lambda_{t\bar{b}}}\nn\\
&&\quad\left[(1-k_t-k_b)(m^2_b\tan^2(\beta)+m^2_t\cot^2(\beta))
-4m_bm_t\sqrt{k_tk_b}\right].
\ee

\subsection{Off-shell $H^\pm$}

%--------------------------------------------------
\begin{figure}
\raisebox{-\height}{\includegraphics[width=0.49\linewidth]{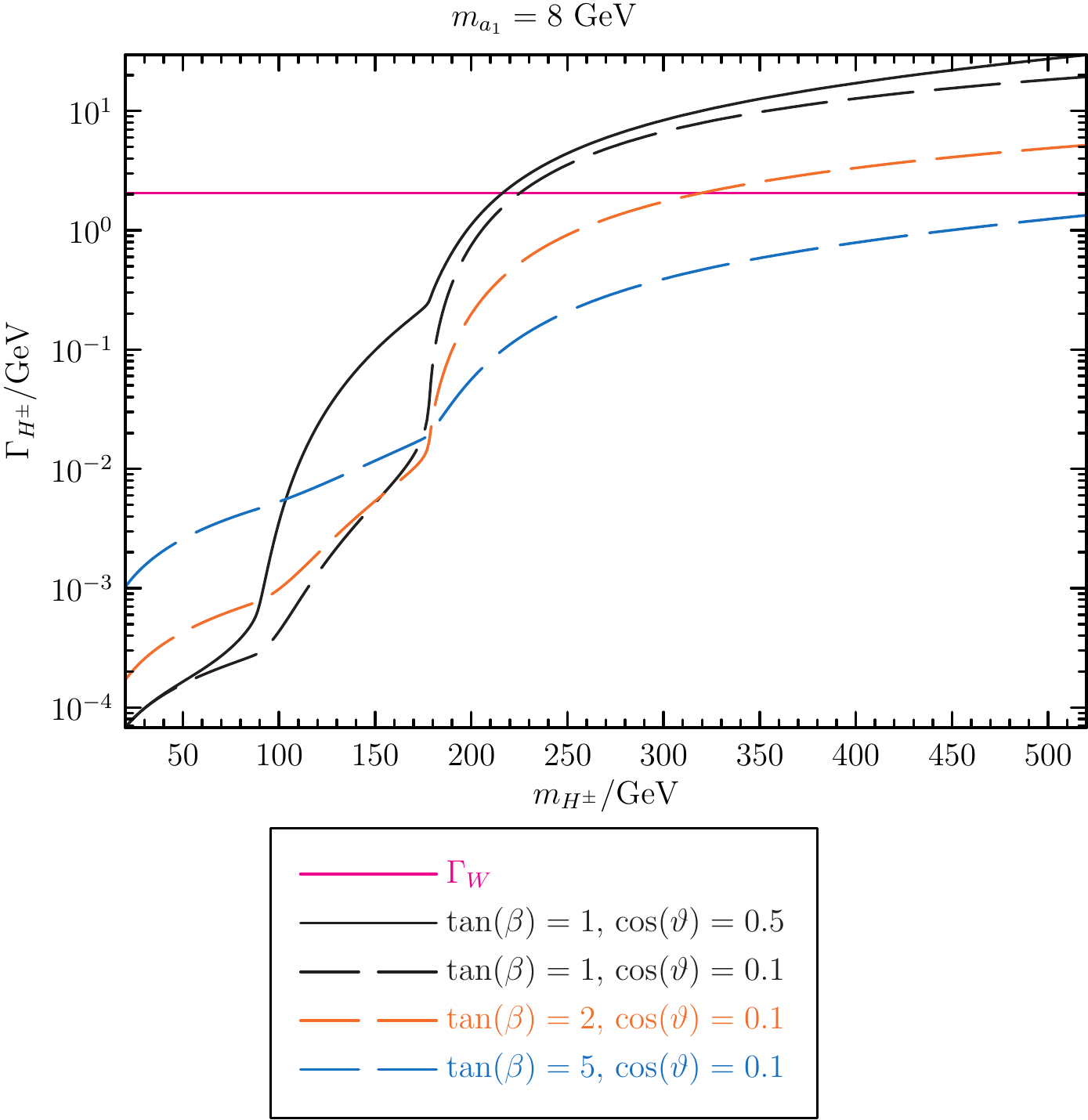}}
\raisebox{-\height}{\includegraphics[width=0.49\linewidth]{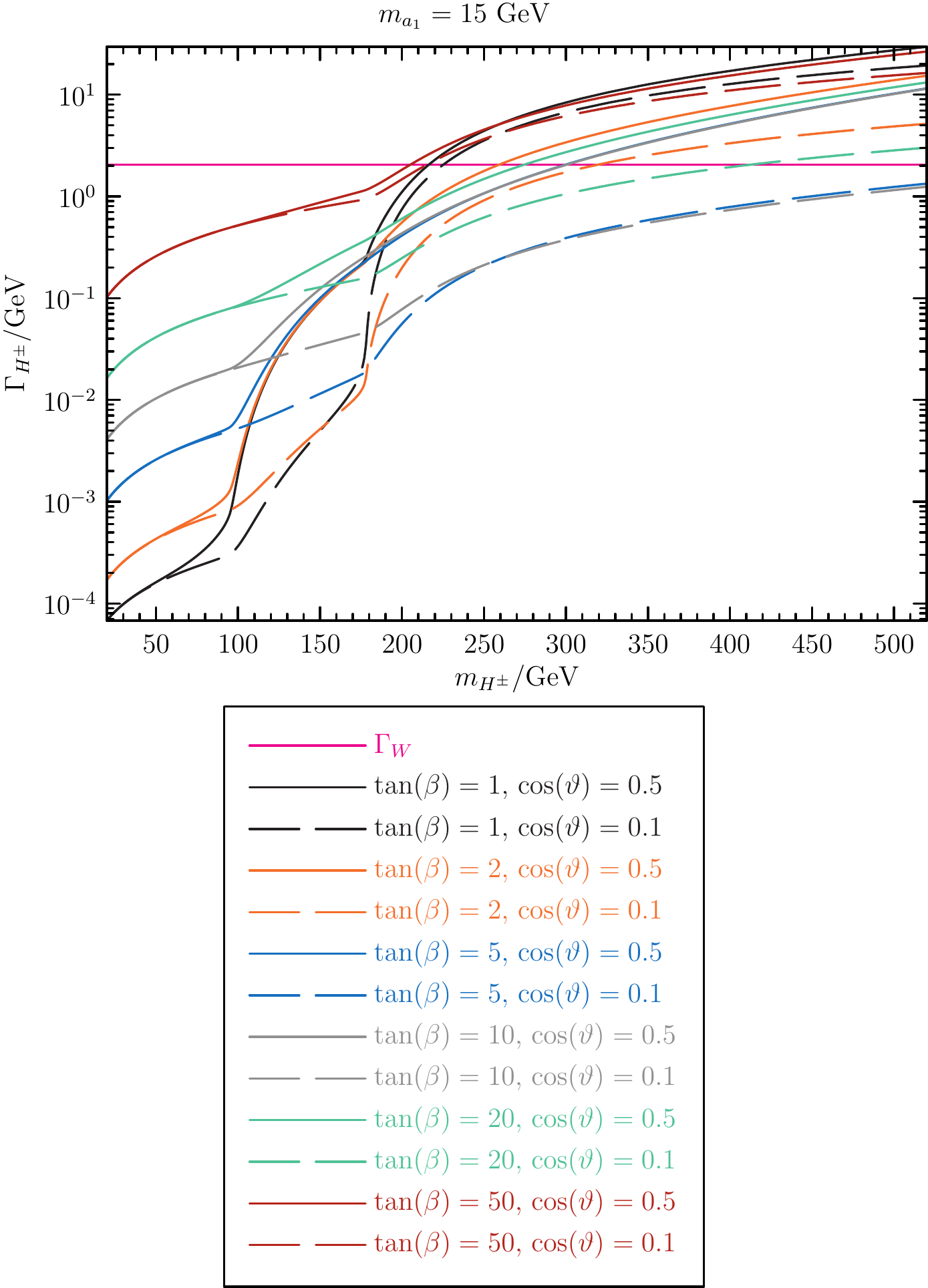}}
\caption{The full width of $H^\pm$. When the heavy neutral Higgs $\Phi$ decays into $H^\pm W^\mp$ off-shell this determines which one preferably goes off-shell.}
\label{fig:hpdecaywidth}
\end{figure}
%----------------------------------------------------

In the off-shell decay $\Phi \rightarrow W^{\mp*}H^{\pm*} \rightarrow W^{\mp *} W^\pm A$, the decay widths $\Gamma_W$ and $\Gamma_{H^\pm}$ roughly decide which one is preferred to be off-shell. The full decay width of $H^\pm$ in our type-II 2HDM + singlet reference scenario is shown in comparison with $\Gamma_W$ in Fig.~\ref{fig:hpdecaywidth}. For an $H^\pm$ with a mass much above the $tb$ threshold the possible three-body decay of $\Phi$ through an off-shell $H^\pm$ needs to be considered.

\be
&\Gamma(\Phi\rightarrow W^+H^{-*}\rightarrow W^+\bar{t}b) =
\f{3G_F^2}{64\pi^3}M^3
\int\rr{d}x_t
\int\rr{d}x_b&\nn\\&
\Bigg[\left(m^2_t\cot^2(\beta)+m^2_b\tan^2(\beta)\right)\f{
(-1+x_t+x_b+k_W-k_t-k_b)\left((2-x_t-x_b+2k_W)^2+4k_W(1-x_t-x_b-k_W)\right)
 }{
(1 - x_t-x_b - k_W + k_H)^2 + k_H \gamma_H}&\nn\\&
-4m_bm_t\sqrt{k_bk_t}\f{
(2-x_t-x_b+2k_W)^2+4k_W(1-x_t-x_b-k_W)
 }{
(1 - x_t-x_b - k_W + k_H)^2 + k_H \gamma_H}\Bigg].&
\ee

\be
&\Gamma(\Phi\rightarrow W^+H^{-*}\rightarrow W^+W^-\phi) =
\f{G_F^2}{128\pi^3}M^5
\int\rr{d}x_2
\int\rr{d}x_1&\nn\\&\f{
\left[(x_1-2k_W)^2-4k_W(1-x_1+k_W)\right]
\left[(1-x_1-k_\phi)^2-4k_Wk_\phi\right]}{
(1 - x_1 + k_W - k_H)^2 + k_H \gamma_H}.&
\ee
Here 1 and 2 label the two $W$s.
These formulae are valid for $\Phi,\phi\in\{A_H,H\}$ and are suppressed
by mixing angles for other mass eigenstates that are not completely doublet.

\section{Branching Ratios}
\label{appen:brs}

Only two and three-body decay rates are calculated to allow for fast numerical
integration. Neglecting the $H^\pm$ width the branching ratio for
$\Phi\rightarrow W^\pm W^\mp A$ can be expressed as the product of the
two branching ratios
\be
\rr{BR}(\Phi\rightarrow W^\pm W^\mp A) &=&
2\rr{BR}(\Phi\rightarrow W^+H^-)\rr{BR}(H^-\rightarrow W^-A).
\ee
These individual branching ratios can be calculated using off-shell $W$s and
tops.

Alternatively, allowing the $H^\pm$ to go off-shell, we can write 
\be
&&\rr{BR}(\Phi\rightarrow W^\pm W^\mp A) =\\
&&\quad\f{
2\Gamma(\Phi\rightarrow W^+H^{-*}\rightarrow W^+W^-A)_y
+2\Gamma(\Phi\rightarrow W^{+*}H^-)_y\rr{BR}(H^-\rightarrow W^-A)
}{
2\Gamma(\Phi\rightarrow W^+H^{-*}\rightarrow W^+X)_y
+2\Gamma(\Phi\rightarrow W^{+*}H^-)_y
+\Gamma(\Phi\nrightarrow W^{\pm *}H^{\mp *})\nn
},
\ee
where $X$ means $W^-A$ or fermions and
the subscript $y$ indicates that the width
\be
y&=&\Gamma_{H^\pm}+\Gamma_W
\ee
should be used in place of the actual off-shell particle width
in the integrand denominator ($\gamma\rightarrow y^2/M^2$). Here tops
and $W$s coming from the $H^\pm$
are on-shell. This formula is really only needed for
$H^\pm$ masses above the $tb$ threshold anyway, as can be seen by looking at Fig.~\ref{fig:hpdecaywidth}. This formula provides a very good approximation to real answer calculated using four-body decay widths (allowing both the $H^\pm$ and $W^\mp$ to be off shell)---much better than just allowing the particle with the largest width to be off-shell---but is built out of three-body decay widths and can therefore be quickly evaluated using single numerical integration.

\section{The $\Phi$ Width}
\label{appen:wPhi}

%--------------------------------------------------
\begin{figure}
\raisebox{-\height}{\includegraphics[width=0.473\linewidth]{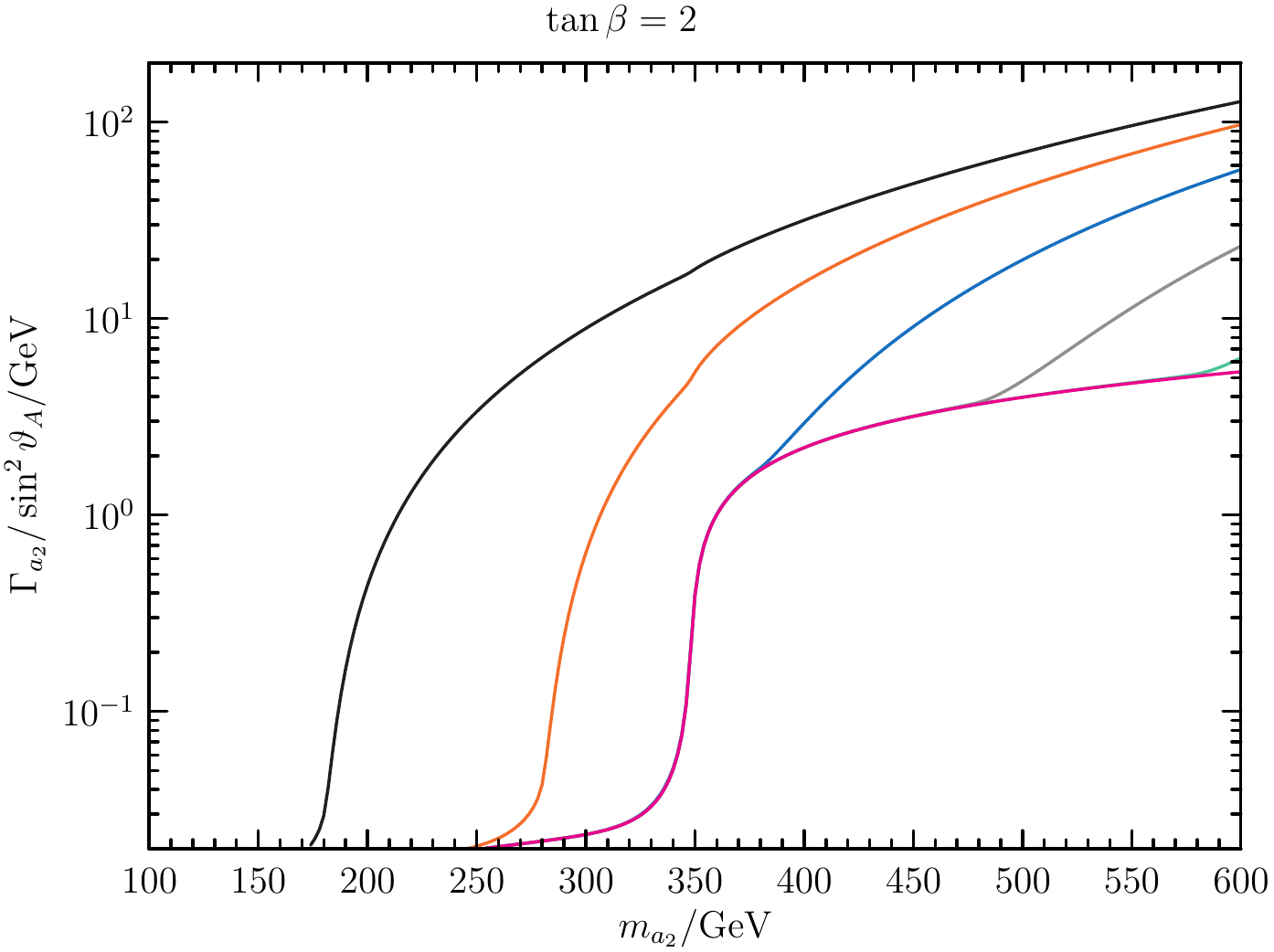}}
\raisebox{-\height}{\includegraphics[width=0.50\linewidth]{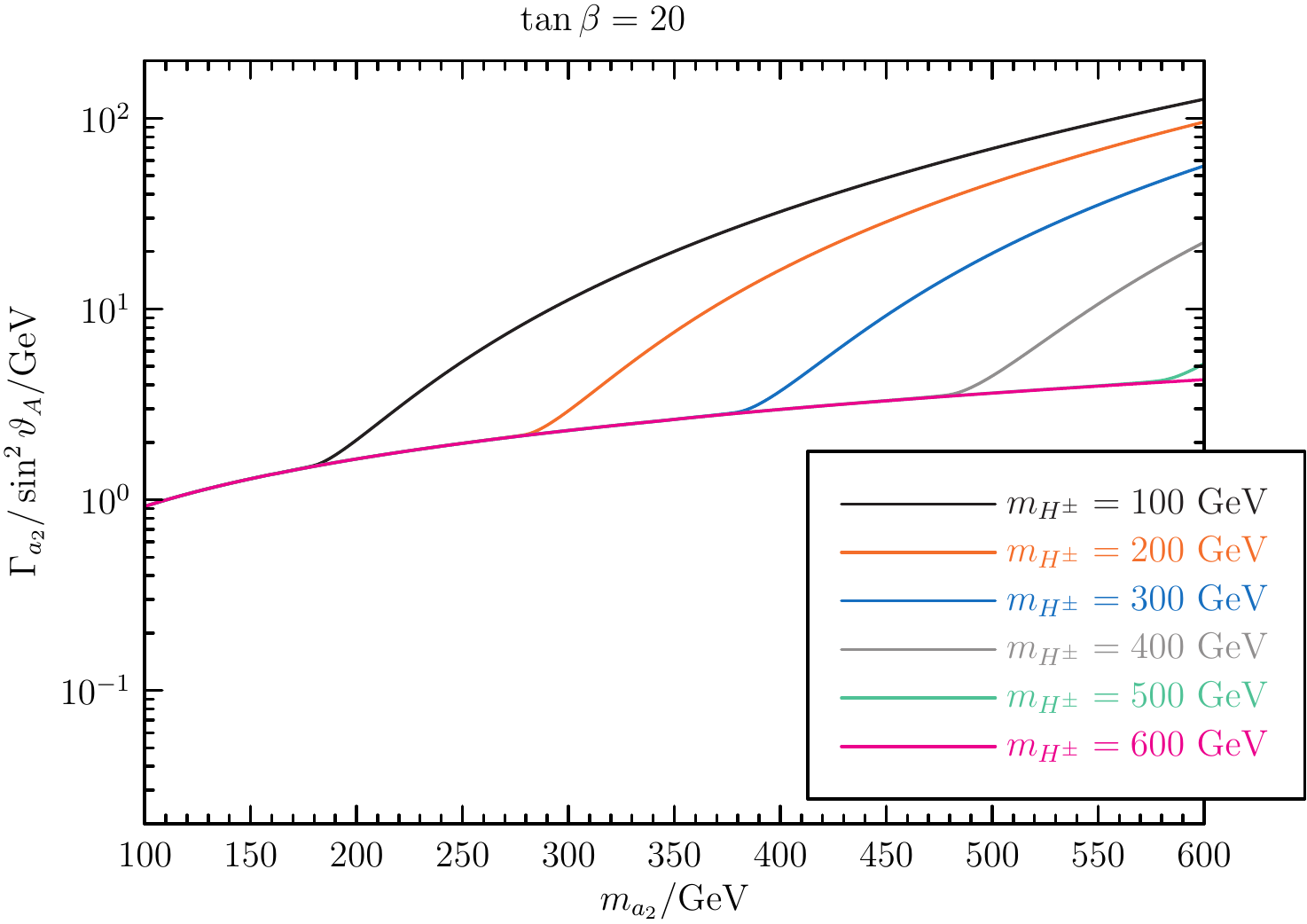}}
\caption{The full width of $a_2$ divided by $\sin^2\vartheta_A$, its doublet amplitude squared. Below the $H^\pm W^\mp$ threshold (coinciding lines) the dominant contribution comes from $t\bar{t}$ on left (proportional to $\cot^2\beta$) and $b\bar{b}$ on the right (proportional to $\tan^2\beta$).
\label{fig:a2decaywidth}}
\end{figure}
%----------------------------------------------------

Fig.~\ref{fig:a2decaywidth} shows the width of $a_2$ (divided by its doublet fraction) in the type-II 2HDM + singlet scenario. The contribution to the total cross-section from $a_2$ going off-shell (rather than $H^\pm$ or $W^\pm$) can be important at high $\tan\beta$ or at very low $\tan\beta$ if $m_{H^\pm}+M_W>2m_t$. For large $a_2$ masses the width of $a_2$ can become very large.

%%%%%%%%%%%%%%%%%%%%%%%%%%%%%%%%%%%%%%%%%%%%%%%%%%%%%%%%%%%%%%%%%%%%%%%%%%%%%%%%%%%%%
\section{CL$_{\rm s}$ limits}
\label{appen:cls}

A $1-\alpha$ confidence level CL$_s$ limit on a signal $s$ is defined by
\begin{eqnarray}
\alpha=\frac{P(D\ge\lambda|H_0)}{P(D\ge\lambda|H_1)}
\end{eqnarray}
where $D$ is the data and $\lambda$ is the expected distribution in the
signal-plus-background hypothesis ($H_0$) and in the background only
hypothesis ($H_1$).

For each channel and set of cuts we have a background $B=b\pm\sigma_b$. The
signal-plus-background may be expressed as
\begin{eqnarray}
B+S &=& b + s \pm \sqrt{\sigma_b^2+s+s^2\Sigma^2},
\end{eqnarray}
where $s$ is the expected signal, its statistical error is taken to be
$\sqrt{s}$, its fractional systematic error is taken to be $\Sigma$. In this paper, we set
$\Sigma = 30$\% as a conservative bound. 

We approximate everything as Gaussian.
We therefore take
\begin{eqnarray}
P(D\ge\lambda|H_1) &=& \Phi\left(\frac{D-b}{\sigma_b}\right),\\
P(D\ge\lambda|H_0) &=&
\Phi\left(\frac{D-b-s}{\sqrt{\sigma_b^2+s+s^2\Sigma^2}}\right),
\end{eqnarray}
Where $\Phi$ is the cumulative distribution function.
For a given $b$, $\sigma_b$, $D$, $\Sigma$, and $\alpha$
the $1-\alpha$ confidence level
limit on $s$ can therefore be found. We call this solution $s=l$. For our calculation, the parameters are obtained from Ref.~\cite{cmshww}.

%%%%%%%%%%%%%%%%%%%%%%%%%%%%%%%%%%%%%%%%%%%%%%%%%%%%%%%%%%%%%%%%%%%%%%%%%%

\end{document}